\documentclass[%
 reprint,
 amsmath,amssymb,
 aps,soul,
pre,
]{revtex4-2}

\usepackage{graphicx}% Include figure files
\usepackage{dcolumn}% Align table columns on decimal point
\usepackage{bm}% bold math
\usepackage[dvipsnames]{xcolor} % Enables more colors

\usepackage{hyperref}% add hypertext capabilities
\hypersetup{
    colorlinks=true,        % False: boxed links; True: colored links
    linkcolor=MidnightBlue, % Color of internal links (sections, pages)
    citecolor=ForestGreen,  % Color of bibliographical citations
    filecolor=magenta,      % Color of file links
    urlcolor=cyan,          % Color of external links
    breaklinks=true,        % Allows links to break over lines
    bookmarksnumbered=true, % Put section numbers in PDFs bookmarks
    pdfpagemode=UseOutlines, % Show bookmarks in PDF viewer
}
\usepackage{xcolor}
\renewcommand{\selectlanguage}[1]{}
\usepackage{lipsum}
\makeatletter
\newcommand*{\balancecolsandclearpage}{%
  \close@column@grid
  \cleardoublepage
  \twocolumngrid
}
\begin{document}

\preprint{APS/123-QED}

\title{Polar chiral active matter as a motile, disordered Josephson array: \\ Information supercurrents and Goldstone spin waves}

% The Motile Josephson Array: \\ Bridging Active Turbulence and Superconductivity

% Motile Josephson arrays: The thermodynamics of \\ information supercurrents in active chiral matter

% Chiral Active Turbulence as a Motile Disordered Josephson Array

\author{Magnus F Ivarsen}
\email{Contact: magnus.fagernes@gmail.com}
\altaffiliation[Also at ]{
The European Space Agency Centre for Earth Observation, Frascati, Italy}
\affiliation{Department of Physics and Engineering Physics, University of Saskatchewan, Saskatoon, Canada}%

\begin{abstract}
%\color{black}
We consider a minimalist model of polar chiral active matter: overdamped, self-propelled agents coupled through a localized Kuramoto--Sakaguchi interaction, which causes alignment. Intrinsic frustration drawn from a broad distribution constitutes a temperature for the ensemble. In the co-moving frame of the local order parameter, the agent dynamics reduce exactly to the Adler equation, placing each agent in a tilted washboard potential: trapped agents are phase-synchronized, and we demonstrate thereby that synchronization (phase rigidity) is maintained by information supercurrents; agents that are running in this potential form a resistive bath. The model is therefore formally isomorphic to a disordered, resistively shunted Josephson array, and a Monte-Carlo sweep over the frequency dispersion empirically recovers the disorder-broadened Adler--Ohmic crossover of the ensemble-averaged slip velocity. Lifting the dynamics from $S^1$ to $S^2$, the polar alignment torque (the Kuramoto-term) is geometrically equivalent to the Gilbert-damping term of the Landau--Lifshitz--Gilbert equation; the mapping establishes an effortless azimuthal precession, yielding a Goldstone-mode dispersion that carries an effective inertia $\propto R^2$, where $R$ is the local order parameter. This furnishes a microscopic basis for the spin-wave transport assumed in inertial-spin models of flocking. Within its regime of validity, i.e., dry, polar, chiral agents under marginal synchronization with sufficient frustration, the model is well-described as a dissipative spintronic fluid.
%\color{black}
\vspace{9pt}

\end{abstract}

\maketitle

\begin{figure*}
    \centering
    \includegraphics[width=\textwidth]{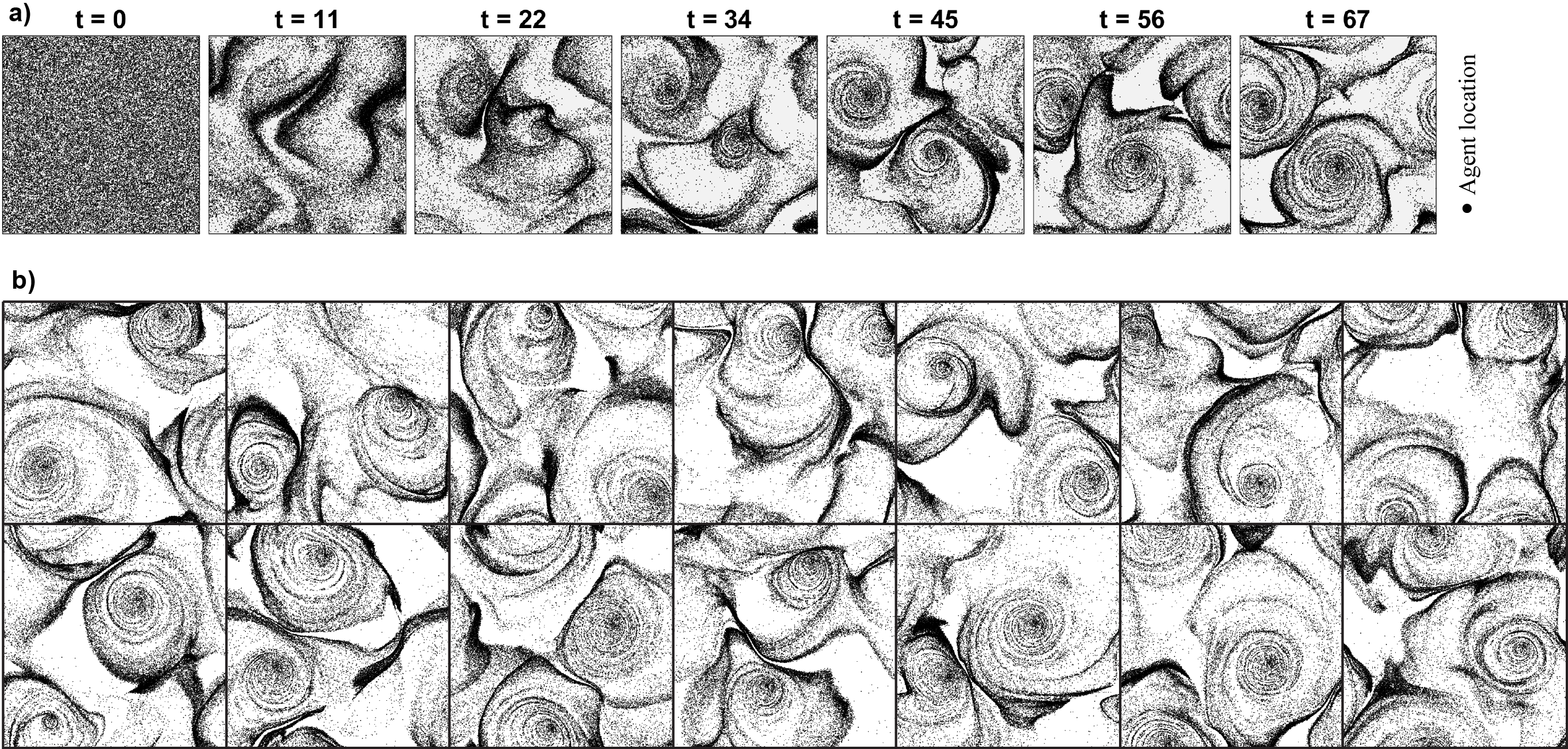}
    \caption{\textbf{Panel a)} shows seven snapshots of a simulation run, with $t$ renormalized by the number of interaction timescales \cite{ivarsen_onsager_2025-1}. \textbf{Panel b)} shows the end-states of 14 simulations, demonstrating that the space-filling twin-cell, or dipole, structure forms during every simulation that features sufficiently wide frustration $\Delta\omega$.}
    \label{fig:example}
\end{figure*}

%\tableofcontents

\section{Introduction}

\vspace{9pt}

%\color{black}
Active turbulence describes the complex, often unpredictable patterns of motion that emerge in driven, self-propelled matter, observed across a diverse spectrum of living and synthetic systems; such as bacterial swarms \cite{wensink_meso-scale_2012,aranson_bacterial_2022}, subcellular cytoskeletal extracts \cite{alert_active_2022}, and synthetic colloidal flocks \cite{bourgoin_kolmogorovian_2020,qi_emergence_2022}. Unusually for a turbulent phenomenon, active turbulence is driven-dissipative, in particular \textit{overdamped} ($\mathcal{R}e \approx 0$): chaotic transport is then sustained by continuous, non-conservative injection of energy at the microscale \cite{saintillan_active_2013,bhattacharjee_activity_2022}, instead of being the result of an inertial cascade. Understanding the hydrodynamic conditions under which the observed behaviour emerges remains a fundamental challenge to statistical physics, and existing approaches typically modify hydrodynamic field theories phenomenologically \cite{markovich_odd_2021,cavagna_flocking_2015,yang_hydrodynamics_2015}.

Recent experimental and computational work has begun mapping out the phenomenology of polar chiral active matter turbulence in particular. Ref.~\cite{meckeSimultaneousEmergenceActive2023} reported the simultaneous emergence of active turbulence and odd viscosity in a quasi-2D monolayer of vertical, spinning rod-shaped colloids, and Ref.~\cite{meckeChiralActiveSystems2024} demonstrated how substrate friction sets a damping length that bounds vortex sizes. Ref.~\cite{dewitPatternFormationTurbulent2024} demonstrated, in simulation and theory, that the wavelength of cascade-induced patterns in chiral active fluids is fixed by the magnitude of the odd viscosity. Ref.~\cite{chen_phases_2024} examined spinning particles at intermediate Reynolds numbers, observing localised \textit{vortlets} that flock and condense into a chiral active phase, extending chiral active phenomenology beyond the strictly overdamped regime.

The field is in need of microscopic theoretical building blocks for such observations. Using a minimalist, two-dimensional Kuramoto--Sakaguchi-type model of polar chiral agents as self-propelled phase oscillators, Ref.~\cite{ivarsen_onsager_2025-1} found that such a phase-slaved motile oscillator ensemble can drive a conservative inertial flow of phase rigidity that structurally resembles shallow-water Euler turbulence. The present article supplies the agent-level dynamics behind that observation. Specifically, we derive a formal isomorphism between the model and a motile, disordered, resistively shunted Josephson array; we lift the dynamics from $S^1$ to $S^2$ to identify the polar alignment torque (the Kuramoto term) with the Gilbert-damping term of the Landau--Lifshitz--Gilbert equation; and we close the resulting bath--fluid cycle through a kinetic Turing instability derived from the underlying Vlasov--Fokker--Planck equation. Thus, the agent inertia engineered into existing hydrodynamic models of chiral flocks \cite{toner_flocks_1998,yang_hydrodynamics_2015,cavagna_flocking_2015} may find a microscopic origin in the synchronization stiffness of driven agent ensembles.

%\color{black} 

\subsection*{The model}

The model in question is characterized by three simple rules, namely that its agents are \textit{(1)} self-propelled, \textit{(2)} polar, or flocking, favoring alignment, and \textit{(3)} chiral, with intrinsic rotation \cite{ivarsen_onsager_2025-1}. The model is unique in that it collapses these three properties into a single degree of freedom, phase, $\phi_i$,
\begin{equation} \label{eq:collapsed}
    \dot{\mathbf{r}}_i = v_0 \mathbf{\hat{n}}(\phi_i),
\end{equation}
where $\mathbf{r}_i$ is an agent's position, $v_0$ its constant swim speed, and $\mathbf{\hat{n}}(\phi_i)=(\cos\phi_i,\;\sin\phi_i)$ is the polar unit vector for internal phase $\phi_i\in [0, 2\pi)$. The time evolution in phase is given by, 
\begin{equation} \label{eq:kuramoto}
    \dot{\phi}_i = \omega_i + a_0 R \sin(\Psi - \phi_i) + \eta_i(t),
\end{equation}
where $\omega_i$ is an agent's intrinsic chirality, or \textit{frustration}, $\eta_i(t)$ is a Gaussian white noise term, and $a_0 R \sin(\Psi - \phi_i)$ is a localized Kuramoto-Sakaguchi-coupling \cite{acebron_kuramoto_2005,de_smet_partial_2007}. The local order parameter magnitude $R(\mathbf{x},t)\in[0,1]$ and mean phase $\Psi(\mathbf{x},t)$ are defined explicitly in Eq.~(\ref{eq:kernel}) below; throughout, $R$ measures local synchronization strength and $\Psi$ the local mean phase of the polar field. With this in hand, we note that
\begin{multline} \label{eq:kernel}
    Z(\mathbf{x}, t) = R(\mathbf{x}, t) e^{i\Psi(\mathbf{x}, t)} = \\ = \int_{\mathcal{D}} G(|\mathbf{x} - \mathbf{x}'|) \left[ \sum_{j=1}^N \delta(\mathbf{x}' - \mathbf{x}_j(t)) e^{i\phi_j(t)} \right] d\mathbf{x}',
\end{multline}
is the local complex order parameter field, expressed as the convolution of the microscopic agent distribution with a finite-range interaction kernel $G(|\mathbf{x}-\mathbf{x}'|)$, representing attractive forces between agents that align. By defining a continuum field theory for Eqs.~(\ref{eq:kuramoto}) and (\ref{eq:kernel}),  and in Appendix~A, we establish that the ensemble of active agents is subject to a kinetic Turing instability, triggered by the interplay between $Z$ (activation) and $\omega_i$ (inhibition).

Figure~\ref{fig:example}a) exhibits an example numerical simulation of our agent ensemble (see Section~III for numerical information), showing individual oscillators are black dots, or point-clouds, in seven temporal snapshots of the simulation box. Figure~\ref{fig:example}b) shows 14 end-state configurations of the simulation, demonstrating the persistent emergence of a large, space-filling dipole structure of agents with opposite chirality.

Of some importance, Ref.~\cite{ivarsen_onsager_2025-1}  explicitly considered a renormalized fluid element in the model, by the implementation of a low-pass filter scaled by the interaction correlation timescale. Ref.~\cite{ivarsen_onsager_2025-1} motivated this step in terms of shallow water theory, demonstrating that Eqs.~(\ref{eq:kuramoto}) and (\ref{eq:kernel}) thereby lead to vortex clustering and a dipole-like configuration (see Figure~\ref{fig:example}). We refer to Appendix~B, where we recap the derivation of a hydrodynamic limit for our model. The end state of the renormalized 2D flow is a pair of oppositely-signed vortices of topological charge, produced by consecutive \textit{shock-mergers}: locally ordered patches collide, fragment into phase slips (releasing entropy into the bath), and are then re-collected into the bulk inertial flow by the kinetic Turing instability. %The remainder of this article shows that the foregoing cycle is the depinning dynamics of a motile disordered Josephson array.

However, an important question remains. Our active agent ensemble is demonstrably overdamped and dissipative, yet the model exhibits an inverse, conservative energy cascade. How does a system that should fundamentally decay instead pump energy towards the largest scales? The resolution lies in considering the discrete distribution of agents as a spatially extended, disordered Josephson junction array. The depinning transition between `running' and `trapped' states then generates a macroscopic synchronization stiffness, or supercurrents of phase gradient information. By extending the model's equations to three dimensions (3D), we eventually derive a theoretical foundation indicating that the phase rigidity of the agent ensemble supports \textit{Goldstone mode spin waves} that supplements the two-dimensional (2D) model with the azimuthal precession around the mean-field (order parameter) of the 3D active chiral agents. We shall contrast these spin waves, which are ferromagnetic magnons, against the `second sound' waves of the Toner-Tu field theory, delineating the prospective universality of the Landau-Lifshitz-Gilbert (LLG) equation.

%\vspace{-8pt}
\section{Theoretical Foundation}
%\vspace{-8pt}

In chiral flocks, motile agents that possess polar symmetry and seek to align, phenotypic heterogeneity \cite{ackermann_functional_2015} produces an intrinsic frequency dispersion among the flock constituents, in much the same way that semiconductor fabrication introduces a dispersion in the superconductor chips called Josephson junctions \cite{josephson_possible_1962,tinkham_introduction_2004}. Those microelectrical chips are essentially coupled ensembles of quantum oscillators with a non-zero dispersion $\Delta\omega$ in their natural frequencies $\omega_i$, that synchronize (phase-lock) via a collective (global) coupling $K$ if the coupling strength is able to overcome the ensemble's fabrication heterogeneity, or dispersion, $\Delta\omega$. Crucially, the coupling $K$ is naturally described with the Kuramoto-model
\cite{wiesenfeld_synchronization_1996}, readily providing a theoretical foundation to understand the thermodynamics of our minimal model, which features a long-range, locally mediated Kuramoto interaction (Eqs.~\ref{eq:kuramoto}, \ref{eq:kernel}).

In this article, we shall consider Eq.~(\ref{eq:kuramoto}) as the overdamped Langevin equation, following Ref.~\cite{ambegaokar_voltage_1969}, which leads to the tilted washboard potential \cite{tinkham_introduction_2004}. Given that our model's agents are essentially massless, we shall derive a formal mapping to the \textit{resistively shunted junction} model in the overdamped limit \cite{stewart_current-voltage_1968,mccumber_effect_1968}.

We begin by writing Eq.~(\ref{eq:kuramoto}) as,
\begin{equation} \label{eq:kuramoto2}
    \dot{\phi}_i = -\frac{\partial V_\text{eff}}{\partial\phi_i}+\eta_i(t),
\end{equation}
where,
\begin{equation} \label{eq:veff}
    V_\text{eff}(\phi_i) = -a_0R\cos(\Psi-\phi_i) - \omega_i\phi_i,
\end{equation}
is the classic tilted washboard potential characterized by two competing energy scales, the potential wells $a_0R$ (which synchronize agents), and the tilt $\omega_i$ (which de-synchronize agents). %From Eq.~(\ref{eq:veff}) we observe that agents can be `trapped' when $a_0R>\omega_i$, `running' when $a_0R<\omega_i$, and `marginal' when $a_0R\approx\omega_i$.

In this dynamic, the stochastic noise $\eta_i(t)$ and the intrinsic frequency dispersion in $\omega_i$ function as sub-gap leakage \cite{danner_injection_2021,fang_subgap_2023} driving the irreducible background entropy production described by Ref.~\cite{marov_self-organization_2013}. This continuous dissipation ensures that the emergent inertial fluid retains a finite `thermal viscosity' that prevents the system from freezing into a glass-state, even deep within the phase-locked regime \cite{ivarsen_onsager_2025-1}. This mechanism establishes a direct connection to Ref.~\cite{benz_coherent_1991}, which demonstrated that the 2D topology of an inherently disordered Josephson array allows for coherent microwave emissions despite significant microscopic inhomogeneity (see also Ref.~\cite{kurtscheid_thermodynamics_2025}).

\subsection{The Adler equation}

To characterize the thermodynamics of Ref.~\cite{ivarsen_onsager_2025-1}'s  fluid-theoretically derived ``shock cycle'' between the active, dissipative bath and an ordered, inertial state, we shall derive an agent's phase velocity in the co-moving Lagrangian frame of the local order parameter. We start by decomposing an agent's phase velocity,
\begin{equation} \label{eq:phicomp}
    \phi_i(t) = \Psi_i(t) + \delta\phi_i(t),
\end{equation}
where $\Psi_i(t)=\Psi(\mathbf{r}_i(t),t)$ is the local phase at agent $i$'s location, representing the collective phase evolution, and $\delta\phi_i(t)$ represents an agent's relative phase phase evolution. Taking the time derivative of Eq.~(\ref{eq:phicomp}) yields,
\begin{equation} \label{eq:phirel}
    \dot{\phi_i} = \dot{\Psi}+\dot{\delta\phi},
\end{equation}
from which we define,
\begin{equation} \label{eq:vslip}
    v_\text{slip} \equiv  |\dot{\delta\phi_i}| = |\dot{\phi_i}-\dot{\Psi_i}|,
\end{equation}
as an agent's phase slip velocity. Here, we identify $\dot{\phi_i}$ as an agent's deterministic phase velocity (Eqs.~\ref{eq:kuramoto}, \ref{eq:veff}) and $\dot{\Psi_i}$ as the phase velocity of the potential well seen by that agent, meaning that $v_\text{slip}$ represents an agent's phase speed relative to a potential well. By quantifying $v_\text{slip}$ (Eq.~\ref{eq:vslip}) we can therefore measure whether an agent is being thermalized into the active bath, or whether it is contributing to an inertial Euler flow in $R^2$ \cite{ivarsen_onsager_2025-1}.

This realization allows us to pinpoint the mathematical isomorphism that explains why the system bifurcates into a hydrodynamic (trapped) state and an active bath (running) state. We use Eq.~(\ref{eq:kuramoto}) deterministically to rewrite Eq.~(\ref{eq:phirel}) in form of,
\begin{equation} \label{eq:adler}
    \dot{\delta\phi_i} = \omega_i-\dot{\Psi_i} - a_0R\sin\delta\phi,
\end{equation}
which is, \textit{in exact terms,} the Adler equation \cite{bhansali_gen-adler_2009,gandhi_dynamics_2015} for a dirty, or disordered, Josephson array \cite{wiesenfeld_synchronization_1996}, with individual junctions driven by an effective current $I_\text{eff}$ \cite{danner_injection_2021}. The Gaussian white noise $\eta_i(t)$ from Eq.~(\ref{eq:kuramoto}) is omitted from Eq.~(\ref{eq:adler}) for the deterministic analysis that follows; it is reinstated statistically below (Eq.~\ref{eq:vslipana}) as a fluctuating tilt that broadens the effective-current distribution $P(I_\text{eff})$ and drives the system stochastically across the depinning threshold. This treatment, standard in the resistively shunted junction literature \cite{ambegaokar_voltage_1969,danner_injection_2021}, is justified provided $\eta_i$ is fast compared to the bulk evolution of $\Psi$, as is the case here.
\begin{equation} \label{eq:ieff}
    I_\text{eff}=\omega_i-\dot{\Psi_i}.
\end{equation}
Here, we identify $\Delta\omega$,
\begin{equation} \label{eq:deltaomega}
    \langle\omega_i-\dot{\Psi_i}\rangle = \Delta\omega,
\end{equation}
the dispersion in the agents' intrinsic frequency, corresponding to inertial fluid's \textit{topography} in the shallow water isomorphism derived in Ref.~\cite{ivarsen_onsager_2025-1}. Consistent with this inference, the effective current in Eq.~(\ref{eq:ieff}) is a driving force that tries to break local synchronization. Setting Eq.~(\ref{eq:adler}) to zero and solving for $\delta\phi$ yields the critical current,
\begin{equation} \label{eq:icrit}
    I_c=a_0R,
\end{equation}
which likewise follows from considering the potential well depths in the tilted washboard potential (Eq.~\ref{eq:veff}). When $|I_\text{eff}|<I_c$, the local coupling force $a_0R$ is strong enough to trap the agent, which subsequently locks onto a stable fixed angle and swims with the local group. Conversely, when $|I_\text{eff}|>I_c$, the agent's intrinsic frustration exceeds the local maximum coupling force. The agent runs or slips relative to the group, with the phase continuously rotating.

\begin{figure*}
    \centering
    \includegraphics[width=\textwidth]{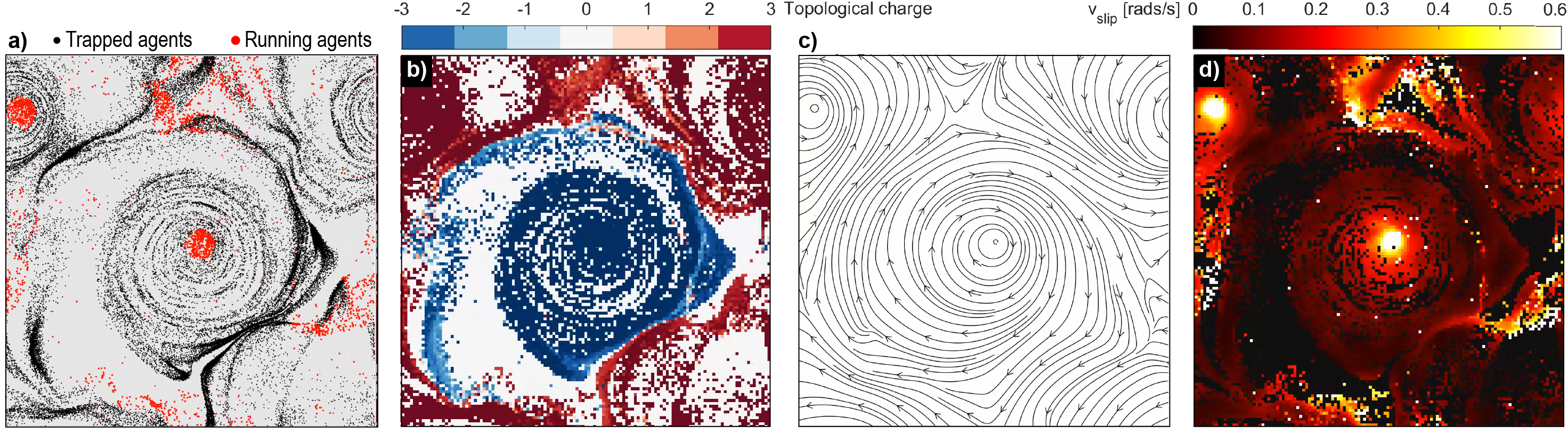}
    \caption{\textbf{A simulated ensemble of polar chiral active matter agents.} Panels a--d) show end-state snapshots of the entire simulation space, with agent point-cloud \textbf{(panel a)}, topological charge \textbf{(panel b)}, flow-lines of circulation in order $R^2$ \textit{(panel c)}, and agent slip velocity $v_\text{slip}$ \textbf{(panel d)}, with respective colorscales where applicable. In panel a), trapped agents are shown in black while running agents are shown in red.}
    \label{fig:art0}
\end{figure*}

\subsection{A motile Josephson junction array}

In equivalent terms, for Josephson junctions, when $I_\text{eff}<I_c$ an oscillator famously locks onto a constant value. However, whereas the critical current of a Josephson junction depends on material qualities, $I_c=a_0R$ is in the present model \textit{entirely dynamic} and depends on local synchronization. Still, we observe complete isomorphism between Eq.~(\ref{eq:adler}) and the Adler equation for a Josephson junction with in-series resistance \cite{danner_injection_2021}. When $I_\text{eff}>I_c$ the Josephson junction cannot hold onto the phase, acquires a voltage, and subsequently produces entropy through dissipative currents. Equivalently, in our model, when an agent runs or slips relative to the group, it acquires the voltage,
\begin{equation}
    V_\text{eff} = |\dot{\phi}_i-\dot{\Psi}_i| = v_\text{slip}.
\end{equation}
The resistive motion of the agent then thermalizes energy into the active bath.

We can thus formalize Ref.~\cite{ivarsen_onsager_2025-1}'s description of the model's central thermodynamic process; when agents start to slip, $R$ drops, lowering the local critical current $I_c$. This causes neighbours to slip, forcing a cascade in energy, resulting in a locally disorganized state ($R\approx0$). However, this state is linearly unstable to the kinetic Turing instability (Appendix~A). Once agents have thermalized, they re-organize. In the tilted washboard formalism, as $R$ grows, the potential wells deepen, drawing more agents in. Their intrinsic rotation $\omega_i$ is now synchronized by spatial segregation, producing a distinct \textit{topography} for the inertial flow \cite{ivarsen_onsager_2025-1}.

Following Ref.~\cite{danner_injection_2021}'s demonstration of how individual Josephson oscillators respond to external forcing, we calculate the time required for a $2\pi$ phase slip, 
\begin{equation} 
T_\text{slip} = \int_0^{2\pi}\;\frac{d\delta\phi}{\dot{\delta\phi}}=\frac{2\pi}{\sqrt{I_\text{eff}^2-I_c^2}},
\end{equation}
where we inserted the Adler equation into the integral (Eq.~\ref{eq:adler}). The integral assumes that $\dot\Psi_i$ varies slowly on the timescale $T_\text{slip}$ of an individual $2\pi$ slip, i.e.\ that the bulk hydrodynamic evolution of the order-parameter field is slow compared to the local Adler dynamics. This is the standard adiabatic approximation underpinning Adler's analysis of injection locking \cite{danner_injection_2021,bhansali_gen-adler_2009}, and is justified post hoc in the present model by the empirical agreement, in the high-disorder limit, between the deterministic prediction (Eq.~\ref{eq:vslip2}) and the simulated $\langle v_\text{slip}\rangle$ shown in Figure~\ref{fig:iv}. Consequently, the instantaneous slip velocity is 
\begin{equation} \label{eq:vslip2} 
v_\text{slip} = \frac{2\pi}{T_\text{slip}} = \sqrt{I_\text{eff}^2-I_c^2}, 
\end{equation} 
where we note that whereas Eq.~(\ref{eq:vslip}) measures an agent's realized slip velocity, Eq.~(\ref{eq:vslip2}) provides a theoretical prediction for the slip velocity based on the established response function of an ideal Josephson junction; the convergence of the two at high disorder will provide empirical weight behind our central thesis.

Examining the asymptotic limits of Eq.~(\ref{eq:vslip2}) reveals two distinct dynamical regimes in the deterministic limit (see also Ref.~\cite{likharev_dynamics_1986}): 
\begin{equation} \label{eq:simp} v_{\text{det}} = \begin{cases} \sqrt{I_{\text{eff}}^2 - I_c^2} \sim \sqrt{I_{\text{eff}} - I_c} & \text{ for } I_{\text{eff}} \ge I_c \\\\ 0 & \text{ for } I_{\text{eff}} < I_c \end{cases} \end{equation}
In equivalent terms, we follow Ref.~\cite{danner_injection_2021}'s formalism for single junctions, adapting it for the ensemble average, $\langle v_\text{slip}\rangle$, obtained by integrating Eq.~(\ref{eq:vslip2}) over the probability distribution $P(I_\text{eff})$ (see Eq.~\ref{eq:ieff} below). This accounts for noise as a fluctuating tilt that statistically drives the system across the critical threshold: 
\begin{equation} \label{eq:vslipana} 
\langle v_\text{slip} \rangle = \int_{|I_\text{eff}| > I_c} P(I_\text{eff}) \sqrt{I_\text{eff}^2 - I_c^2} \, dI_\text{eff}. 
\end{equation}

Eq.~(\ref{eq:vslipana}) is expected to conform to the specific 'soft knee' geometry identified in Ref.~\cite{danner_injection_2021} and stands in direct correspondence with Ref.~\cite{wiesenfeld_synchronization_1996}, which established that the macroscopic resistance of a disordered Josephson array scales with the fraction of running junctions, demarcating %. The shared critical exponent of 0.5 highlights the universality of 
the phase transition from a ‘superconducting glass’ (pinned, or trapped, phases) to a resistive metal (running phases).

In what follows, we demonstrate that our minimal active matter model effectively realizes a motile Josephson array. By embedding the discrete oscillator dynamics of Ref.~\cite{danner_injection_2021} into intrinsically motile (self-propelled) agents, we introduce a transport mechanism absent in standard static arrays \cite{wiesenfeld_synchronization_1996}. In our model, dissipative excitations are physically advected by the running agents themselves. This active transport expels disorder from the ordered bulk and concentrates it into defect cores, segregating the system into a dual-phase fluid. The segregation mimics the thermodynamic state preparation observed in driven optical systems \cite{kurtscheid_thermodynamics_2025}, where the \textit{interplay of drive and dissipation stabilizes distinct macroscopic phases} (see also Refs.~\cite{benz_coherent_1991,newrock_two-dimensional_2000}). Specifically, we observe \textit{(1)} an  inviscid-like fluid (adhering to shallow water hydrodynamics) of \textit{trapped} agents whose effective inertial mass density was found by Ref.~\cite{ivarsen_onsager_2025-1} to scale as the squared order parameter, $\lambda \propto R^2$ -- a phase-stiffness or emergent inertia that is given a microscopic origin in Section~\ref{sec:llg} below -- and \textit{(2)} a dissipative resistive fluid of \textit{running} agents. This resistive phase acts as the thermodynamic sink that ultimately sustains the cycle, permitting the regeneration of order via a kinetic Turing instability (derived in Appendix~A).

%\vspace{-8pt}
\section{Empirical Results}
%\vspace{-8pt}

The model (Eqs.~\ref{eq:kuramoto}, \ref{eq:kernel}) is implemented in a bi-periodic square domain of dimensions $2L \times 2L$ (with $L$ the physical half-width of the domain) containing $N = 50{,}000$ agents; $L=20$~grid units, $v_0 = 0.5$~grid units per second, $a_0=0.2$ (see also Ref.~\cite{ivarsen_onsager_2025-1}). The agents evolve in continuous space via a standard Euler integration scheme; macroscopic fields --- in particular the complex order parameter $Z(\mathbf{r},t)$ and the local coherence $R(\mathbf{r},t)$ --- are evaluated on a fixed $128 \times 128$ Eulerian grid by binning the discrete agent distribution and convolving with the Gaussian kernel $G(r) = (2\pi\sigma^2)^{-1}\exp(-r^2/2\sigma^2)$, where $\sigma=3$~grid units, via two-dimensional Fast Fourier Transforms. The convolution is performed at every integration step $\Delta t = 0.02$~s, providing continuous mean-field feedback to the agents. Each simulation is initialised in an ``active soup'' state of random uniform phases $\phi_i \in [0, 2\pi)$ and uniformly distributed positions, and evolved for $T_\mathrm{max} = 400$~s ($20{,}000$ steps, or $67$~interaction lengths $\tau=\sigma/v_0$). Topological charge is computed as the discrete circulation of the unit phase vector $\hat{\mathbf{n}}(\phi_i)$ around closed plaquettes of the same $128 \times 128$ grid (positive integers for vortices, negative for antivortices).

Figure~\ref{fig:art0} presents snapshots of the fully formed dipole structure, with panels a--c) showing agent point cloud, topological charge, and inertial flowlines, respectively. Panel d) shows the slip velocity $v_{slip}$ (Eq.~\ref{eq:vslip}), for the entire simulation space.

\begin{figure}
    \centering
    \includegraphics[width=.5\textwidth]{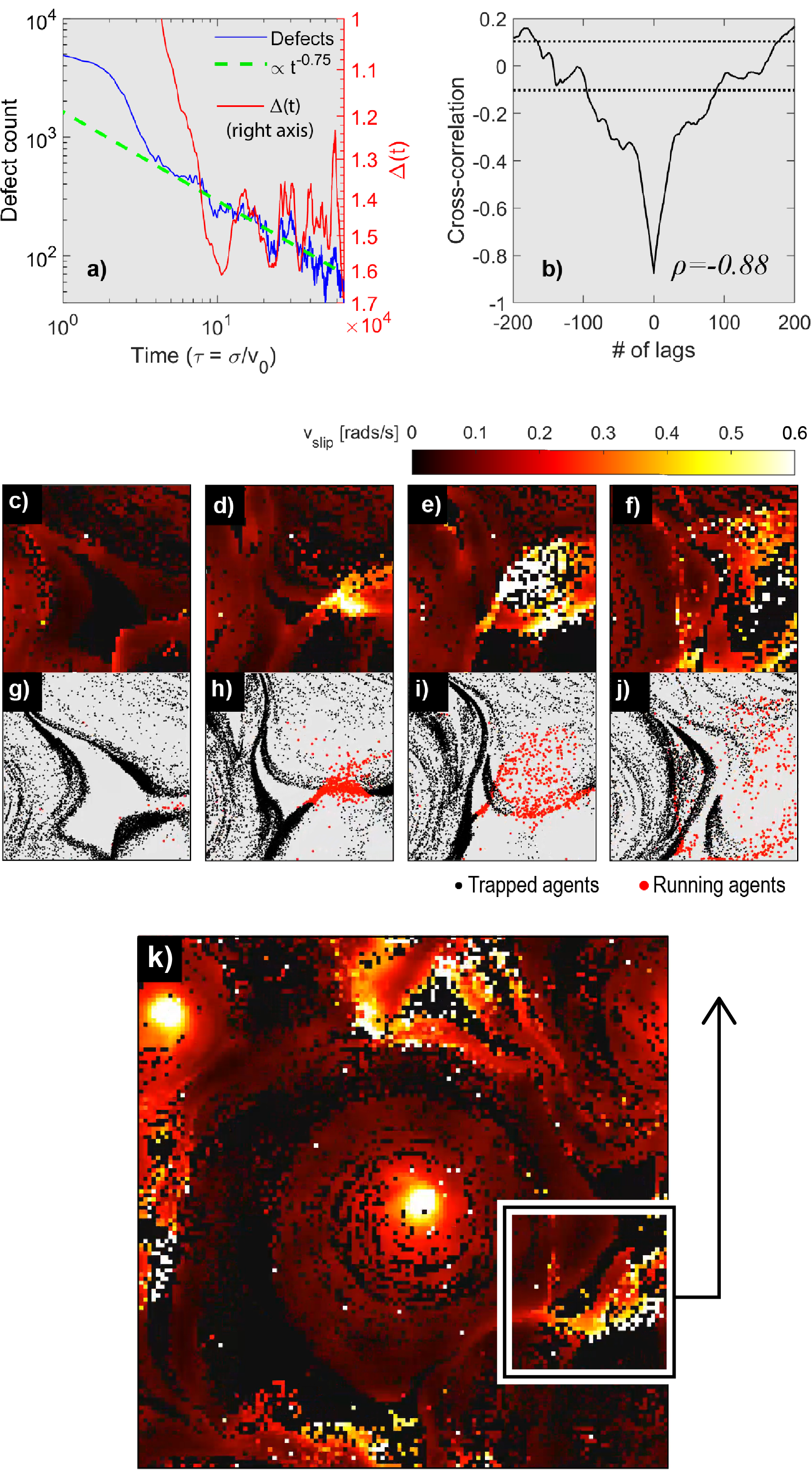}
    \caption{\textbf{Panel a)} shows defect count as a function of normalized simulation time (left $y$-axis) with a $t^{-0.75}$ powerlaw fit shown with a green, dashed line, and the average potential barrier height $\Delta(t)$ (Eq.~\ref{eq:barrier}) in red (right $y$-axis, inverted). Time has been renormalized by the model's interaction length scale $\tau= \sigma/v_0$, where $\sigma$ is the kernel size and $v_0$ the constant swim speed. \textbf{Panel b)} shows a cross-correlation analysis (using Pearson correlation at various lags) of the latter two timeseries, for normalized times $>5$. Estimated confidence intervals for the cross-correlation are shown with dashed, black lines. \textbf{Panels c--j)} show inset $v_\text{slip}$ and trapped/running agent pointclouds, data from a representative subregion of Figure~\ref{fig:art0}d (location indicated in \textbf{panel k}), in rapid succession-snapshots, highlighting ongoing shock merger, agent thermalization, and subsequent re-organization. See Video~S1 in the Supplementary Materials for a video of this simulation run.}
    \label{fig:art}
\end{figure}

Figure~\ref{fig:art}c--j), showing four prior $v_{slip}$-snapshots of a limited portion of the simulation space, offer an instructional explanation of the model's core thermodynamic mechanism. An ordered ($R\approx1$) region shatters into phase slips (active turbulence), generating entropy; this is the shock merger mechanism identified by Ref.~\cite{ivarsen_onsager_2025-1}, which we derive in Appendix~B. The agents inside this now-disordered patch are then quickly \textit{reorganized} into the inertial fluid by the kinetic Turing instability, triggered by the reaction-diffusion dynamic produced by the competition between intrinsic chirality, or frustration, and Kuramoto synchronization.

The reorganization step that closes the bath--fluid cycle is a kinetic Turing instability of the spatially extended Kuramoto--Sakaguchi system, derived in Appendix~A by linearising the Vlasov--Fokker--Planck equation around the incoherent homogeneous state. The central result of that Appendix is that the critical coupling $a_\mathrm{crit}(k)$ at which a perturbation of wavenumber $k$ first goes unstable is non-monotonic in $k$: short-wavelength perturbations are suppressed by the finite-range alignment kernel $\hat{G}(k)$, while long-wavelength perturbations are suppressed by the kinetic susceptibility $\chi(k)$, which depends on the dispersion $\Delta\omega$ in intrinsic frequencies --- the same temperature-like control parameter that dictates the I--V curve in Fig.~\ref{fig:iv} below. The minimum of $a_\mathrm{crit}(k)$ lies at a finite wavenumber where the alignment kernel still has reach and the kinetic dephasing is not yet decisive: the chiral-active-matter analog of the breakdown of Landau damping in plasma physics, where the collective drive overcomes the kinetic phase-mixing of the constituent oscillators [see, e.g., Refs.~\cite{penrose_electrostatic_1960,strogatz_coupled_1992}]. The instability re-collects the running, dissipative bath into ordered patches that re-enter the trapped phase. %Combined with the depinning analysis of \S,II--IV and the LLG dispersion of \S,V, the kinetic Turing wavelength selection closes a depinning--propagation--reordering cycle. 
The shallow minimum of $a_\mathrm{crit}(k)$ identified in Figure~\ref{fig:acrit} in Appendix~A implies that a band of wavenumbers, rather than a single sharp scale, is amplified --- consistent with the multi-vortex morphology of the agent ensemble in Fig.~\ref{fig:art0}.

Figure~\ref{fig:art}a, b) demonstrate the foregoing dynamic in rigorous terms. First, we define the gap excess $\Delta(t)$, a metric derived from the potential barrier height of the overdamped Adler equation \cite{danner_injection_2021}. That is, we sum the potential heights for the trapped population,
\begin{equation} \label{eq:barrier}
    \Delta(t) = \sum_\text{trapped}\left( a_0R - |\omega_i-\dot{\Psi}_i|\right),
\end{equation}
where the argument $t$ indicates that the quantity is a timeseries. $\Delta(t)$ thus measures the sum total depths of the local potential wells \cite{ambegaokar_voltage_1969} ($\mathcal{U}_B$, termed potential barrier height in Ref.~\cite{danner_injection_2021}) that confine each trapped agent's phase.

Figure~\ref{fig:art}a) shows topological defect count (blue line) going down with a characteristic scaling of $t^{-0.75}$ \cite{carnevale_evolution_1991,larichev_weakly_1991}, which, in the case of our model, was identified by Ref.~\cite{ivarsen_onsager_2025-1} to be caused by shallow water shock dynamics. At the same time, and with a strong negative correlation coefficient of $-0.88$ (Figure~\ref{fig:art}b), the gap excess $\Delta(t)$, or locking strength of the trapped agents \cite{danner_injection_2021}, increase.

The $t^{-0.75}$ scaling in Fig.~\ref{fig:art}a) is the Carnevale--McWilliams scaling for the late-time decay of coherent vortices in 2D turbulent flows~\cite{carnevale_evolution_1991,larichev_weakly_1991}. In a system of long-range-interacting point vortices in two dimensions, pairs that approach within a few core radii either annihilate (opposite signs) or merge (same signs), in both cases reducing the vortex count while conserving total topological charge; energy decreases through the filamentation that accompanies each event. A self-similar scaling argument based on these conservation laws yields a power-law decay $N(t)\propto t^{-\xi}$ with $\xi$ in the range $0.7$--$0.8$ depending on the precise dynamical assumptions, and $\xi \approx 0.75$ in the original Carnevale--McWilliams numerical experiment~\cite{carnevale_evolution_1991}. The exponent is a signature of pairwise merger/annihilation rather than viscous diffusion of the vortex cores, and in the present model, the merger events are realised through a shock-merger process (see Appendix~B): phase-locked regions fragment at impact, and their topological defects coalesce or pair-annihilate.

The recovery of the canonical Carnevale--McWilliams exponent in Fig.~\ref{fig:art}a (and in Ref.~\cite{ivarsen_onsager_2025-1}) is therefore independent empirical evidence that, despite the specific microscopic mechanism, the topological defects of the agent ensemble are well described as a 2D vortex gas in the renormalized, or coarse-grained, flow.

We note that the associated rise in $\Delta(t)$ during shock mergers (Figure~\ref{fig:art}a) resembles the outcome of a topological heat pump: the annihilation of resistive defect cores ($\mathcal{U}_B\approx 0$) instantaneously throws agents into deep superconducting traps ($\mathcal{U}_B\gg0$), actively lowering the system's effective phase temperature (see Appendix~B for a hydrodynamic description of this process).

\begin{figure}
    \centering
   \includegraphics[width=0.5\textwidth]{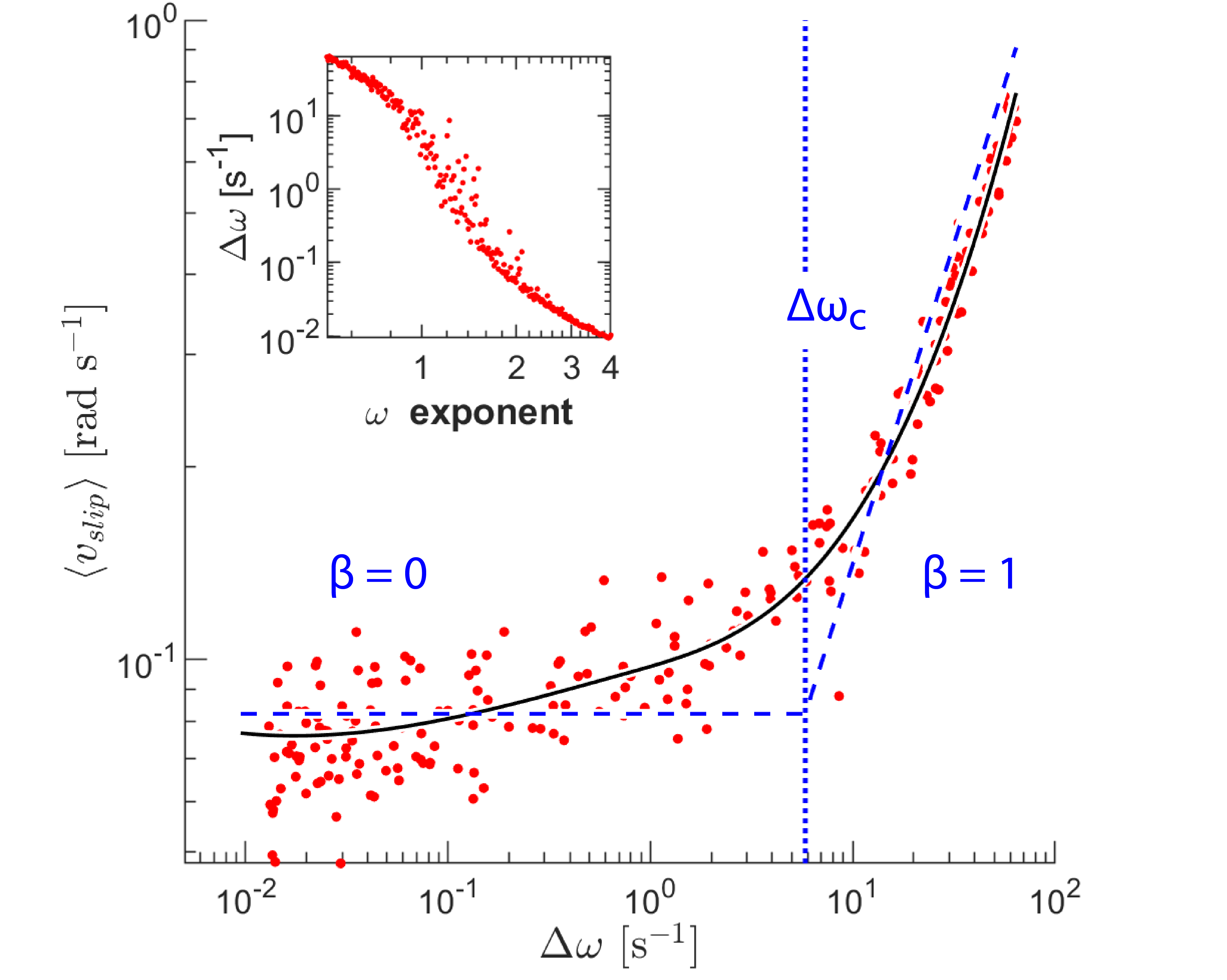}
    \caption{The result of 256 simulations systematically varying the exponent in the $\omega_i$ distribution, yielding a systematically varied disorder strength $\Delta\omega$ (keeping the peak of the $\omega_i$ unchanged at $0.2$~s$^{-1}$, $\Delta\omega$ is calculated using the standard deviation), shown in the inset panel. The main plot shows observed ensemble average of the slip velocity $\langle v_\text{slip}\rangle$ (voltage) against dispersion $\Delta\omega$ (current), in a log-log ``I-V'' plot. $\langle v_\text{slip} \rangle$ is calculated as the mean value of $v_\text{slip}$ for all $N$ agents, collected during four snapshots of the second half of each simulation. A solid black line shows a $4^\text{th}$ order Hermite polynomial fit \cite{derrico_slm-shape_2009}, while a dashed-blue line shows a $2^\text{nd}$ order (piece-wise linear) Hermite polynomial fit with slopes (exponent $\beta$) of 0 and 1. A vertical blue-dotted line indicates the empirically-determined critical threshold dispersion $\Delta\omega_c \approx 6,\text{s}^{-1}$, defined as the slope-0/slope-1 breakpoint of the piecewise-linear (2$^\text{nd}$ order) Hermite polynomial fit (dashed blue line) under nonlinear RMS minimisation. The observable onset of slip velocity occurs significantly earlier (around $\Delta \omega \approx 0.5 \, \text{s}^{-1}$), indicating substantial sub-gap leakage current driven by the intrinsic disorder of the array.}
    \label{fig:iv}
    %\vspace{-8pt}
\end{figure}

%\vspace{-8pt}
\subsection*{The I-V Curve} 

Next, we provide empirical support for our thesis by systematically varying the total agent disorder $\Delta\omega$, termed dispersion in Refs.~\cite{ivarsen_onsager_2025-1}, where $\Delta\omega \equiv \langle I_\text{eff}\rangle$ (Eq.~\ref{eq:deltaomega}). In each simulation the intrinsic chiralities $\{\omega_i\}$ are drawn from a truncated power-law distribution $P(\omega)\propto \omega^{-n}$ on $\omega\in[\omega_0, \omega_\text{max}]$ with $\omega_0 = 0.2~\text{s}^{-1}$ held fixed, so that the ensemble standard deviation $\Delta\omega$ -- our control parameter -- is determined entirely by the exponent $n$. We sweep $n$ from $0.5$ to $4$ over $256$ independent runs, yielding a Monte-Carlo realization of the current--voltage (I--V) curve for the model. Figure~\ref{fig:iv} shows the result as $\langle v_\text{slip}\rangle$, computed as the mean of $v_\text{slip}$ across all $N$ agents in four equally-spaced snapshots from the second half of each run, plotted against $\Delta\omega$. 

The threshold $\Delta\omega_c \approx 6~\text{s}^{-1}$ is extracted from the data: we overlay a piecewise-linear Hermite polynomial with slopes $\beta = 0$ (noise-floor) and $\beta = 1$ (Adler--Ohmic) on the log-log fit of Figure~\ref{fig:iv}, and identify $\Delta\omega_c$ as the breakpoint of that piecewise fit, itself determined by nonlinear least-squares minimisation of the residual against the data, making $\Delta\omega_c \approx 6~\text{s}^{-1}$ an empirical crossover scale.

To characterize the macroscopic response function without imposing a single-junction scaling exponent \textit{a priori}, we fit $\langle v_{\text{slip}} \rangle$ with a shape-preserving $4^\text{th}$-order Hermite polynomial \cite{derrico_slm-shape_2009} (solid black line). The fit is purely phenomenological -- it does not impose any particular $\beta$ -- and is chosen so that the local logarithmic slope can be compared to the 0--1 piecewise linear fit at every value of $\Delta\omega$. This allows us to detect the disorder-induced rounding of the saddle-node depinning transition without forcing the data into a specific exponent \cite{danner_injection_2021}.

Crucially, the solid-black fit in Figure~\ref{fig:iv} reveals a continuous evolution of the local scaling exponent. At very low dispersion ($\Delta\omega \ll \Delta\omega_c$), the response is dominated by a noise floor ($\beta = 0$), gradually rising towards $\Delta\omega_c\approx 6$~s$^{-1}$. The gradual rise in $\langle v_\text{slip}\rangle$ towards 6~s$^{-1}$ is consistent with sub-gap leakage currents smoothing out the $\beta=0.5$ transition \cite{danner_injection_2021}. As the dispersion exceeds the critical threshold, $\Delta\omega>\Delta\omega_c$, the system quickly and strictly adheres to a linear regime, asymptotically recovering the Adler-Ohmic limit ($\beta = 1$).

The classification of the minimal model as a disordered Josephson array is empirically demanded by the system's macroscopic response function (Figure~\ref{fig:iv}), which traverses the distinct Adler-Ohmic crossover predicted by the resistively shunted junction formalism \cite{likharev_dynamics_1986,wiesenfeld_synchronization_1996} with the disorder-induced rounding identified in Ref.~\cite{danner_injection_2021}, a signature of the Adler equation's saddle-node bifurcation. This distinguishes the model's central mechanism from standard viscous fluid transitions and justifies the identification of a tilted washboard potential \cite{ambegaokar_voltage_1969}. The recovery of the linear Ohmic limit ($\beta=1$) at high disorder is consistent with the dual-fluid description developed in Ref.~\cite{ivarsen_onsager_2025-1}: in this regime the trapped (synchronized) population is a vanishing minority and the macroscopic response is dominated by the resistive, running phase, exactly as the disordered Josephson array predicts \cite{wiesenfeld_synchronization_1996}. 

We note that the linear regime in Figure~\ref{fig:iv} spans less than a decade of $\Delta\omega$ in the run set. This is not a sampling limitation but a consequence of the physical range of the power-law exponent: distributions $P(\omega)\propto\omega^{-n}$ with $n \lesssim 0.5$ have ill-defined variance, and those with $n \gtrsim 4$ have negligibly populated high-frequency tails, so the accessible disorder window is bounded above by physical considerations rather than by computation. Within that window, the Adler--Ohmic linear regime is recovered cleanly at the upper end, and the full soft-knee crossover from $\beta=0$ through the disorder-broadened shoulder to $\beta=1$ is observed.

\section{Reconciling the model with real, three-dimensional chiral flocks}

The foregoing established empirical support for the Adler mechanism in our minimal model. However, the derived supercurrents of state information are \textit{maintaining} order in the chiral flock. In phenomenological models of real flocks, `second sound' waves propagate through this ordered ensemble \cite{toner_flocks_1998}. Is our polar, chiral active matter model able to reproduce realistic flocking behaviour?

\begin{figure}
    \centering
    \includegraphics[width=0.5\textwidth]{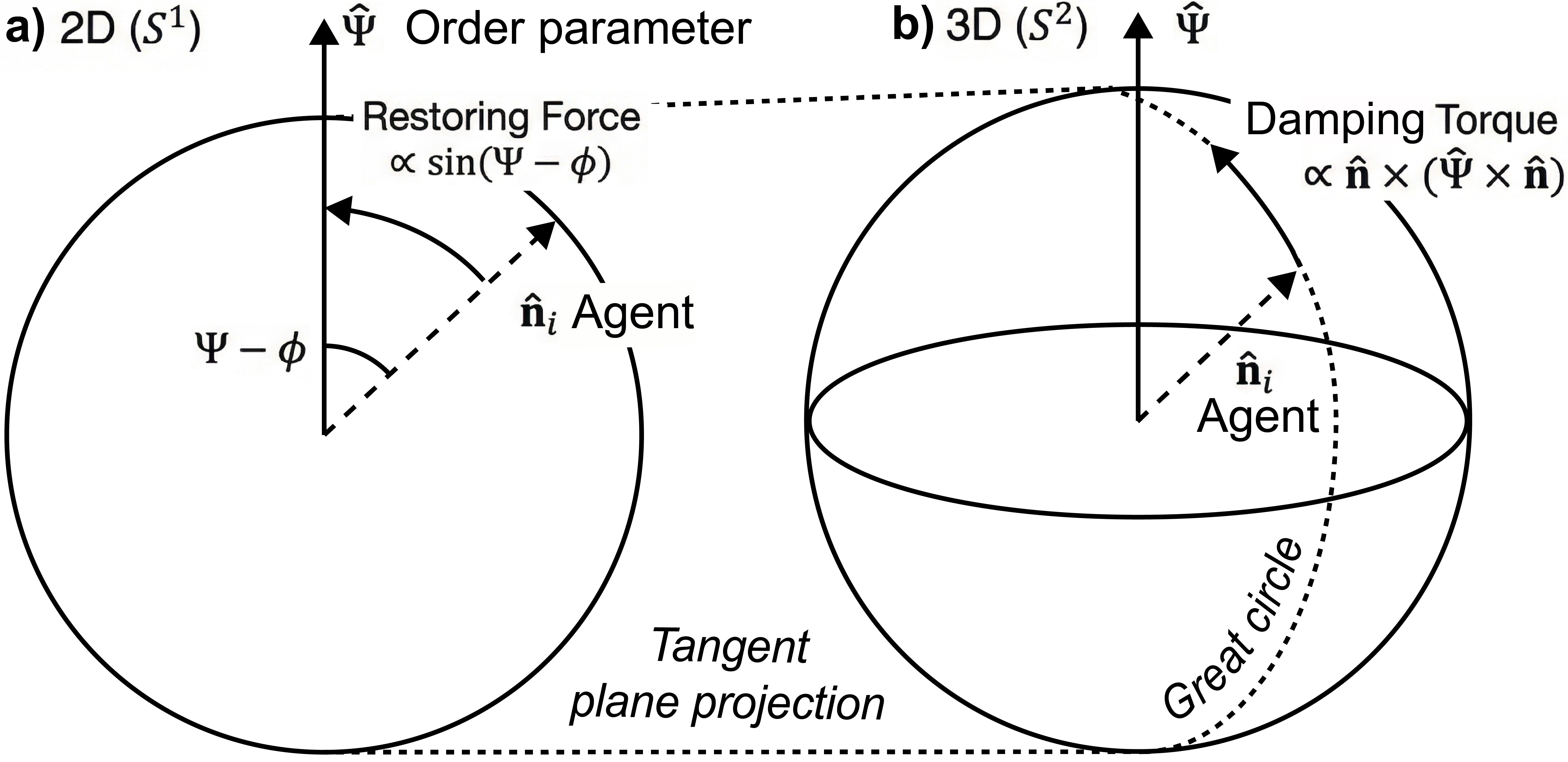}
    \caption{\textbf{Panel a):} the geometry of the 2D model, showing how the restoring force acts to align an agent's phase $\phi$ with the mean field $\Psi$. \textbf{Panel b):} the geometry of the 3D extension, where orientation is the unit vector on a spherical surface ($|\hat{\mathbf{n}}_i|=1$), and where the vector torque proportional to $(\hat{\mathbf{n}}_i \times \hat{\mathbf{\Psi}}) \times \hat{\mathbf{n}}_i$ (Eq.~\ref{eq:eom3d}) acts to align the agent with the mean field, and represents spin relaxing to align with a magnetic field \cite{lakshmanan_fascinating_2011}. This vector torque projects to the restoring force in panel a). }
    \label{fig:schematic}
\end{figure}

By increasing the degrees of freedom from 1 to 2, we can shift from the two-dimensional, minimal model to the real, three-dimensional dynamics of polar chiral flocks ($S^1 \to S^2$), illustrated in Figure~\ref{fig:schematic}. Eqs.~(\ref{eq:collapsed}, \ref{eq:kuramoto}) map a scalar phase $\phi_i$ to a two-dimensional $\mathbf{r}_i$. In three dimensions, the agent state becomes a unit phase vector $\hat{\mathbf{n}}_i$ on the unit sphere $S^2$. The agent's velocity is then $\dot{\mathbf{r}}_i = v_0 \hat{\mathbf{n}}_i$, with $|\hat{\mathbf{n}}_i| = 1$ (Figure~\ref{fig:schematic}b). The local order parameter $Z = R e^{i\Psi}$ (Eq.~\ref{eq:collapsed})  becomes a local director field vector $\mathbf{R} = R \hat{\mathbf{\Psi}}$, where $\hat{\mathbf{\Psi}}$ is the mean direction of the local neighborhood and $R$ is the coherence magnitude.

Next, to express the Adler dynamics (Eq.~\ref{eq:adler}), we note that $\omega_i$ (intrinsic rotation) becomes an intrinsic angular velocity vector $\boldsymbol{\omega}_i$. With this, the coupling term $a_0 R \sin(\Psi - \phi_i)$  represents a torque trying to align $\mathbf{n}_i$ with $\Psi$ (expressed as $\hat{\mathbf{n}}_i \times \hat{\mathbf{\Psi}}$). To keep the overdamped nature of the model, we then take a small-angle approximation of the angular velocity (near 0), yielding an inertia-less, linear expression (angular velocity $\propto$ torque). After this, we can express the equation of motion for an agent $i$ as,
\begin{equation} \label{eq:eom3d}
    \frac{d\hat{\mathbf{n}}_i}{dt} = (\boldsymbol{\omega}_i \times \hat{\mathbf{n}}_i) + a_0 R (\hat{\mathbf{n}}_i \times \hat{\mathbf{\Psi}}) \times \hat{\mathbf{n}}_i + \boldsymbol{\eta}_i(t),
\end{equation}
where we recognize the first term as intrinsic agent chirality, or frustration, and the second term as the localized Kuramoto-alignment term (Eq.~\ref{eq:kuramoto}). Here, we note that, in the spintronics of spin waves (ferromagnetic magnons), where the phase vector stays on the unit sphere (Landau-Lifshitz-Gilbert damping form \cite{johnson_interfacial_1985,lakshmanan_fascinating_2011}),  $\hat{\mathbf{n}}_i$ effectively relaxes towards $\hat{\mathbf{\Psi}}$ (Figure~\ref{fig:schematic}b).

\begin{figure}
    \centering
    \includegraphics[width=0.5\textwidth]{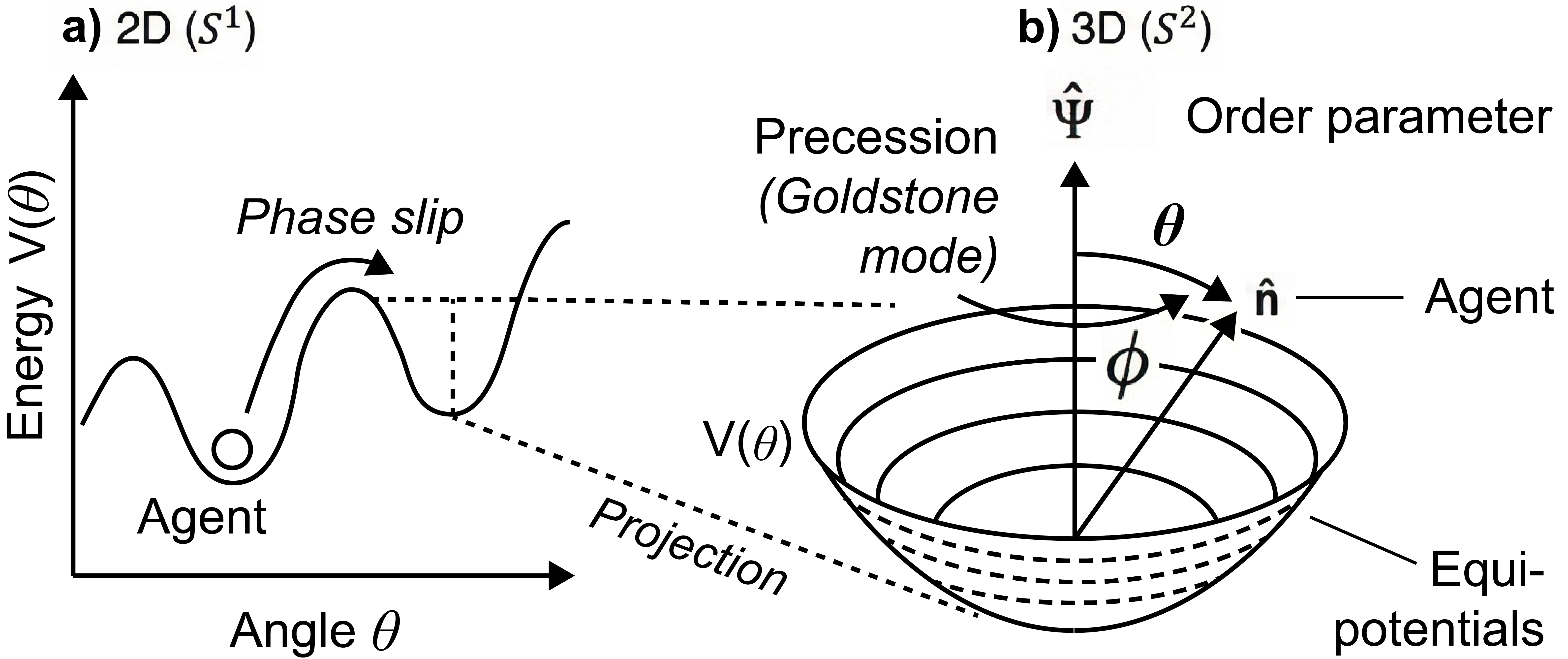}
    \caption{\textbf{The tilted washboard potential. Panel a):} the $S^1$ potential wells $V(\theta)$ of the tilted washboard potential, traversed by phase mismatch $\theta$. \textbf{Panel b):} the $S^2$ potential wells, where potential energy $V(\theta)\propto -\cos\theta$ is independent of the azimuthal angle $\phi$, which means that the potential projects onto $S^1$ in mismatch $\theta$. This enables effortless precession around the mean field and directly activates the Goldstone mode dynamics (Eq.~\ref{eq:dispersion}); a quasiparticle that advects phase gradients (turning stiffness), thereby mimicking inertial mass.  }
    \label{fig:schematic2}
\end{figure}

The angular velocity being strictly proportional to the applied torque allows for the gyroscopic damping by the energy of the dissipative system continuously decreasing, entirely equivalent with the washboard's potential wells in the two-dimensional description (Figure~\ref{fig:schematic2}), and entirely aligned with the damping term in  LLG spintronics \cite{lakshmanan_fascinating_2011}.

The decomposition of phase into a co-moving term with a perturbation (Eq.~\ref{eq:phicomp}), which led to the Adler equation in the two-dimensional case (Eq.~\ref{eq:adler}), now needs angular velocity $\boldsymbol{\Omega}_{\Psi}$, defined as,
\begin{equation}
    \frac{d\hat{\mathbf{\Psi}}}{dt} = \boldsymbol{\Omega}_{\Psi} \times \hat{\mathbf{\Psi}},
\end{equation}
after which the natural definition of a slip vector $\boldsymbol{\Omega}_{slip}$ becomes,
\begin{equation}
    \frac{d\hat{\mathbf{n}}_i}{dt} \bigg|_{slip} = \boldsymbol{\Omega}_{slip} \times \hat{\mathbf{n}}_i.
\end{equation}
Subtracting the frame rotation $\boldsymbol{\Omega}_{\Psi}$ from the agent's total motion yields the vector expression of Eq.~(\ref{eq:vslip2}),
\begin{equation}\label{eq:adlerrelation}
    \boldsymbol{\Omega}_{slip} = \underbrace{(\boldsymbol{\omega}_i - \boldsymbol{\Omega}_{\Psi})}_{\equiv\;\mathbf{I}_\text{eff}} - \underbrace{a_0 R (\hat{\mathbf{\Psi}} \times \hat{\mathbf{n}}_i)}_{\equiv\;\mathbf{I}_{c}}.
\end{equation}

That is, the minimal agent dynamics mapping a two-dimensional unit phase vector to a three-dimensional unit vector naturally leads to an Adler relation. Remarkably, this occurs without the explicit definition of a Kuramoto-term. The reason for this stem from the recognition of the LLG-equation from ferromagnetic theory, where a spin $\mathbf{S}$ naturally relaxes towards a magnetic field $\mathbf{H}$ (see Figure~\ref{fig:schematic}b). The three-dimensional synchronization in a chiral flock described in Eqs.~(\ref{eq:eom3d}) is therefore equivalent to magnetic damping.

Mathematically, the three-dimensional definitions preserved the symmetry group of the interaction while lifting the dimensionality, from which Kuramoto synchronization emerges without the ansatz.

Finally, we can derive a locking condition in three dimensions. The scalar slip velocity $v_{slip}$ defined in Eq.~(\ref{eq:vslip})  becomes the magnitude of the slip vector: $v_{slip}^{3D} \equiv || \boldsymbol{\Omega}_{slip} ||$, and setting this scalar equal to zero requires the effective driving current $\mathbf{I}_\text{eff}$ to be perfectly balanced by the restoring torque,
\begin{equation}
    \boldsymbol{\omega}_i - \boldsymbol{\Omega}_{\Psi} = a_0 R (\hat{\mathbf{\Psi}} \times \hat{\mathbf{n}}_i).
\end{equation}
Taking the magnitude of both sides yields,
\begin{equation}
    || \boldsymbol{\omega}_i - \boldsymbol{\Omega}_{\Psi} || = a_0 R |\sin(\theta)|,
\end{equation}
where $\theta$ is the inter-agent--flock phase mismatch ($\delta\phi$ in Eq.~\ref{eq:phicomp}). Since the maximum value of $|\sin(\theta)|$ is 1, we recover the exact Josephson inequality,
\begin{equation} \label{eq:adlerineq}
    || \mathbf{I}_\text{eff} || \le a_0 R.
\end{equation}

The derivation of the vector Adler relation (Eq.~\ref{eq:adlerrelation}) and the threshold critical current inequality (Eq.~\ref{eq:adlerineq}) from the $S^2$ constraint (Eq.~\ref{eq:eom3d}) ensures that the phase locking behaviour of the two-dimensional Josephson junctions remains intact. As we shall derive in the next section, the phase-locked 3D chiral flock precesses azimuthally, creating a Goldstone mode (topological quasiparticle).

\section{Overdamped spin waves as `second sound' in Toner-Tu theory}\label{sec:llg}

The previous section established a three-dimensional adherence to the Adler equations, where flock synchronization naturally arises from a Gilbert damping torque due by the geometry of the $S^2$ model (Eq.~\ref{eq:eom3d}). Having established the rigidity of the flock, the natural next step is to demonstrate that the predicted effects are observable in real 3D flocks. To do so, we must reconcile the 3D extension of our minimalist model with the standard model invoked to explain flocking in three dimensions: the Toner-Tu field theory \cite{toner_flocks_1998}.

We begin with the 3D equations of motion, Eq.~(\ref{eq:eom3d}), and we begin by examining the continuum limit $\hat{n}_i\to\hat{n}(\mathbf{r},t)$, approximating the local order parameter $\hat{\Psi}$ as the director field plus a Laplacian correction:
\begin{equation}
\hat{\Psi} \to \hat{n} + \xi^2 \nabla^2 \hat{n},
\end{equation}
where $\xi$ is the characteristic interaction scale. This approximation follows from the definition of $\hat{\Psi}$ as the average director within an interaction radius $\sigma$. Treating the agent distribution as a continuum, $\hat{\Psi}(\mathbf{r})$ is the convolution of $\hat{n}(\mathbf{r})$ with a local weighting kernel. A Taylor expansion of $\hat{n}(\mathbf{r} + \boldsymbol{\delta})$ averaged over an isotropic neighborhood eliminates first-order gradient terms ($\langle \boldsymbol{\delta} \rangle = 0$). The leading-order correction is therefore the quadratic term, proportional to $\nabla^2 \hat{n}$, which encodes the local field curvature. Consequently, $\xi^2$ scales with the square of the interaction radius ($\xi \sim \sigma^2$).

Substituting the continuum relations into the alignment term of Eq.~(\ref{eq:eom3d}) yields:
\begin{multline}
(\hat{n} \times \hat{\Psi}) \times \hat{n} \approx (\hat{n} \times (\hat{n} + \xi^2 \nabla^2 \hat{n})) \times \hat{n} = \\ = (\hat{n} \times \xi^2 \nabla^2 \hat{n}) \times \hat{n}.
\end{multline}
Using the identity $(\mathbf{A} \times \mathbf{B}) \times \mathbf{C} = (\mathbf{A} \cdot \mathbf{C})\mathbf{B} - (\mathbf{B} \cdot \mathbf{C})\mathbf{A}$, we obtain the continuum overdamped equation:
\begin{equation} \label{eq:stability2}
\frac{\partial \hat{n}}{\partial t} = - (a_0 R \xi^2) \hat{n} \times (\hat{n} \times \nabla^2 \hat{n}).
\end{equation}
This form is physically significant as it mirrors the dissipative damping term $\hat{n} \times (\hat{n} \times \mathbf{H})$ found in the LLG equation of spintronics \cite{lakshmanan_fascinating_2011}.

Eq.~(\ref{eq:stability2}) supplies the dissipative (Gilbert-damping) coefficient $\Gamma = a_0R\xi^2$. To obtain the conservative (precession) coefficient, we apply the same continuum substitution to the chirality, or frustration, term of Eq.~(\ref{eq:eom3d}). In the phase-locked regime, the intrinsic chirality of each agent aligns on average with the mean field,
\begin{equation} \label{eq:omega_align}
    \boldsymbol{\omega}_i \approx \omega_{0,i} \hat{\mathbf{\Psi}},
\end{equation}
so that the chirality torque becomes
\begin{equation} \label{eq:omega_prec}
    \boldsymbol{\omega}_i \times \hat{\mathbf{n}}_i \approx \omega_0 (\hat{\mathbf{\Psi}} \times \hat{\mathbf{n}}_i) \approx -(\omega_0 \xi^2)\, \hat{\mathbf{n}} \times \nabla^2 \hat{\mathbf{n}},
\end{equation}
where we have substituted the continuum approximation for $\hat{\mathbf{\Psi}}$ in the second step.

The chirality scale $\omega_0$ is not a free microscopic parameter; it is fixed by consistency of the resulting dynamics with the Landau--Lifshitz--Gilbert (LLG) form, in which the order-parameter magnitude $R$ acts akin to magnetization. Demanding that the dissipative and conservative coefficients combine to a Gilbert damping ratio $\alpha = \Gamma/\gamma$ that scales as the squared order parameter (the LLG hallmark of a coherent magnetization; see verification below) requires $\gamma = \Gamma/R^2$, equivalently
\begin{equation} \label{eq:gamma}
    \omega_0\xi^2 \;\equiv\; \gamma \;=\; \frac{a_0\xi^2}{R}.
\end{equation}
Physically, the inverse--$R$ scaling reflects the angular-momentum density of the unit phase vector on $S^2$: as the local order parameter (the magnetization analog) decreases, the precession rate of $\hat{\mathbf{n}}_i$ around $\hat{\mathbf{\Psi}}$ must increase to sustain a finite interaction torque. The same divergence signals the breakdown of the hydrodynamic description at defect cores ($R\to 0$), as we discuss below.

Combining Eqs.~(\ref{eq:stability2}) and (\ref{eq:gamma}), the continuum equation of motion takes the Landau--Lifshitz--Gilbert form
\begin{equation} \label{eq:dynamics}
\frac{\partial \hat{n}}{\partial t} = -\frac{a_0 \xi^2}{R} \hat{n} \times \nabla^2 \hat{n} - (a_0 R \xi^2)\, \hat{n} \times (\hat{n} \times \nabla^2 \hat{n}),
\end{equation}
which we recognize as the conservative precession term ($\propto \hat{n}\times\nabla^2\hat{n}$, the parity--breaking Goldstone mode of the broken rotational symmetry) and the dissipative Gilbert damping term ($\propto \hat{n}\times(\hat{n}\times\nabla^2\hat{n})$, a viscous restoring force). Identifying these with even (shear) and odd viscosities, the expressions,
\begin{equation} \label{eq:nuenuo}
    \nu_E = \Gamma =  a_0R\xi^2, \;\;\;\text{   and   }\;\;\; \nu_O= \gamma = \frac{a_0\xi^2}{R},
\end{equation}
follow directly from the Eq.~(\ref{eq:dynamics}) coefficients.

To verify the LLG mapping that motivated Eq.~(\ref{eq:gamma}), we compare Eq.~(\ref{eq:dynamics}) to the canonical Gilbert form,
\begin{equation}
    \frac{\partial\hat{n}}{\partial t} = -\frac{\gamma'}{1+\alpha^2} (\hat{n} \times \nabla^2\hat{n}) - \frac{\gamma' \alpha}{1+\alpha^2} \left[ \hat{n} \times (\hat{n} \times \nabla^2\hat{n}) \right],
\end{equation}
where $\alpha$ is the dimensionless Gilbert damping parameter. Term-by-term matching, using Eq.~(\ref{eq:nuenuo}), gives
\begin{equation} \label{eq:alpha}
    \alpha =  \frac{\Gamma}{\gamma} = \frac{a_0 R \xi^2}{\left( \frac{a_0 \xi^2}{R} \right)} = R^2,
\end{equation}
recovering, post-hoc, the Gilbert damping parameter as the squared order parameter, which is the result anticipated by Eq.~(\ref{eq:gamma}) and the physical justification for treating $R$ as the magnetization magnitude in the LLG mapping.

The foregoing allows us to express the viscosity terms as functions of the Gilbert damping term $\alpha$, 
\begin{equation}\label{eq:nuenuo2}
    \nu_E = \nu_0 \alpha^{1/2}, \;\;\;\text{   and   }\;\;\; \nu_O = \nu_0 \alpha^{-1/2},
\end{equation}
where $\nu_0 \equiv a_0 \xi^2$ is the bare interaction viscosity. Since $\alpha=R^2$, Eq.~(\ref{eq:nuenuo2}) represents the spintronic hydrodynamics of synchronization.

%\vspace{-8pt}
\subsection*{The dispersion relation}

We observe that as the local order parameter $R$ vanishes, the shear viscosity disappears, compelling the precession rate to diverge to sustain a finite interaction torque. This description remains valid only while the local timescale separation persists, breaking down precisely at the defect cores and at the boundaries ($R \rightarrow 0$) where the divergence signals the transition from a hydrodynamic superfluid to a strictly microscopic, overdamped gas. To describe this aspect of the model, we linearize the polarized agent (see Appendix~C), yielding the final total dispersion relation (Eq.~\ref{eq:dispersion_last}),
\begin{equation}\label{eq:dispersion}
\omega(k) = \pm \frac{a_0}{R} k^2 - i (a_0 R) k^2.
\end{equation}
The real part ($\pm a_0 R^{-1} k^2$) represents spin waves (ferromagnetic magnons), which diverge in frequency as the turning inertia vanishes ($R \to 0$). The imaginary part ($-i a_0 R k^2$) represents diffusive relaxation and vanishes in the disordered limit, confirming that the ordered flock is a rigid, wave-supporting medium \cite{toner_flocks_1998} surrounded by a chaotic, fast-relaxing bath.

The active, 3D, overdamped bath on the boundaries of a real chiral flock naturally leads to a conservative azimuthal turning inertia of a spintronic fluid \cite{lakshmanan_fascinating_2011}, by merit of the dimensional constraints of polar chiral alignment alone. The conservative fluid forms the characteristic flocking behaviour, and is strictly contingent on the agents individually adhering to the phase-locking condition from Josephson junction theory (Eq.~\ref{eq:adlerineq}). The conservation is the relaxation of the stress-energy of the sphere $S^2$, giving a geometric and thereby completely natural definition of synchronization in any system with polar alignment, by merit of the LLG equation.

\section{Discussion}

% 
%These findings compel the classification of active chiral matter as a high-temperature superconductor for information. While odd viscosity successfully captures the macroscopic fluidity of the state \cite{marmol_colloquium_2024,maitra_activity_2025}, our results demonstrate that its constitutive rigidity is fundamentally governed by Josephson critical currents $I_c$, sustained by the phase stiffness of the interaction network. Consequently, the Josephson inequality $|\omega-\dot{\Psi}|<I_c$ imposes a hard physical limit on the flock's maneuverability: there exists a maximum turning rate beyond which the effective `voltage drop' between agents exceeds the critical threshold, resulting in information loss via dissipative phase slippage.
%By bridging chiral active matter turbulence and superconductivity we have identified phase synchronization as a microscopic engine capable of creating macroscopic order, elucidating the core thermodynamic mechanism of new universality class for driven chiral flocks \cite{ivarsen_onsager_2025-1}.

\subsection{Superconductivity}

The recovery of strict disordered Josephson junction theory in the model equations follows from the alignment term $a_0R\sin\delta\phi$ in our model being mathematically indistinguishable from the Josphson energy $E_J\sin\psi$. This phase rigidity enables information supercurrents that are sustained by agents trapped in a tilted washboard potential, analogous to Cooper pair supercurrents \cite{ingold_cooper-pair_1994}. In Figure~\ref{fig:iv}, we provide  empirical support for the notion that the minimalist, two-dimensional model describes disordered, resistively shunted Josephson junctions \cite{wiesenfeld_synchronization_1996,danner_injection_2021}, capable of supporting supercurrents, which, as we shall defend with some rigour, facilitate a Goldstone mode-mediated transfer of phase gradients (rigidity) across an ensemble of polar chiral agents, carrying the topological protection of the bulk order.

However, in simpler terms, the supercurrents are the collective azimuthal rotation rate around the mean field of the spintronic fluid.

\subsection{Thermodynamic pump}

To see why the above holds, we identify the azimuthal angle of the $S^2$ potential as the scalar phase $\phi$ in the reduced $S^1$ limit. This degree of freedom is driven by the intrinsic chirality $\omega_i$ [and stochastic noise $\eta_i(t)$], providing the essential "frustration" that prevents the chiral flock from freezing into a trivial, fully organized state \cite{marov_self-organization_2013,ivarsen_onsager_2025-1}. Instead, the frustration fuels a continuous, natural precession within the washboard potential wells (Figure~\ref{fig:schematic2}b), generating the Goldstone mode quasiparticles (magnons) that sustain the fluid's inertia.

The two-dimensional model (Eqs.~\ref{eq:kuramoto}, \ref{eq:kernel}, Figure~\ref{fig:art0}) must therefore be taken to account for the tangent plane dynamics of the full 3D spintronic fluid. Crucially, this constraint identifies the physical origin of the system's inertia as the LLG equation \cite{lakshmanan_fascinating_2011}; just as a gyroscope resists reorientation due to angular momentum, the chiral flock resists flow curvature due to the conservation of this azimuthal precession.

Inside the two giant vortex cores, we observe in Figure~\ref{fig:art}a,d) that the agents are continuously phase-slipping.  In this limit, the critical current Eq.~(\ref{eq:icrit}) vanishes, reducing the Adler equation to its trivial kinematic limit ($\dot{\phi}_i\approx\omega_i$), and the spintronic fluid's capacity to support spin waves diverges, $\gamma\propto R^{-1}$ (Eq.~\ref{eq:gamma}). Consequently, the vortex cores are purely dissipative singularities where the capacity to sustain supercurrents has collapsed. This decoupling renders the cores unable to transmit phase stiffness or radiate coherent modes in the shallow water formalism, effectively silencing them within the inertial fluid \cite{ivarsen_onsager_2025-1}.

\subsection{Ferromagnetic magnons}

In the 3D chiral flock, agents precess around the mean-field, and the Goldstone mode yields spin waves in the form of ferromagnetic magnons. We derived the dispersion relation (Eqs.~\ref{eq:dispersion} and \ref{eq:dispersion_last}),
\begin{equation*}
\omega(k) = \pm \frac{a_0}{R} k^2 - i (a_0 R) k^2,
\end{equation*}
and we now make the following two key observations:

\textit{(1)} the real part of $\omega(k)$ corresponds to the conservative propagation of phase gradients (`trapped' supercurrents) that provide the Euler-like inertia, while the imaginary component captures the inevitable sub-gap leakage (diffusive relaxation and phase slips). The demonstrable stability of dipole structure thus express a delicate balance where the active chirality ($\omega_i$) pumps energy into the real, inertial modes, while the phase slips ($\beta=1$ regime in Figure~\ref{fig:iv}) provide the necessary dissipation to arrest the cascade. Equivalently, the divergence of the 3D spintronic fluid's precession inertia $\gamma\propto R^{-1}$ at the dissipative boundaries ($R\to 0$) creates a thermodynamic pressure. It is this dynamic process that acts as a thermodynamic state preparation cycle \cite{kurtscheid_thermodynamics_2025}, mimicking coherent microwave emissions in disordered Josephson arrays \cite{benz_coherent_1991}, effectively 'pumping' the system from the high-entropy running state back into the low-entropy trapped state, thereby sustaining the dipole structure.

\textit{(2)} the $k^2$ dependency in the dispersion relation corresponds to the Goldstone modes of the broken rotational symmetry (spin waves, $\omega(k)\propto k^2$), driving ferromagnetic magnons.

Here, we note that spin waves with $k^2$ dependency can be modeled with odd viscosity \cite{markovich_odd_2021}, and our model provides such odd viscosity (Eq.~\ref{eq:nuenuo2}), only that it is purely emergent from \textit{flock coherence} rather than a material quality, and it is only realized in the marginal synchronization regime. 

\begin{figure}
    \centering
    \includegraphics[width=0.5\textwidth]{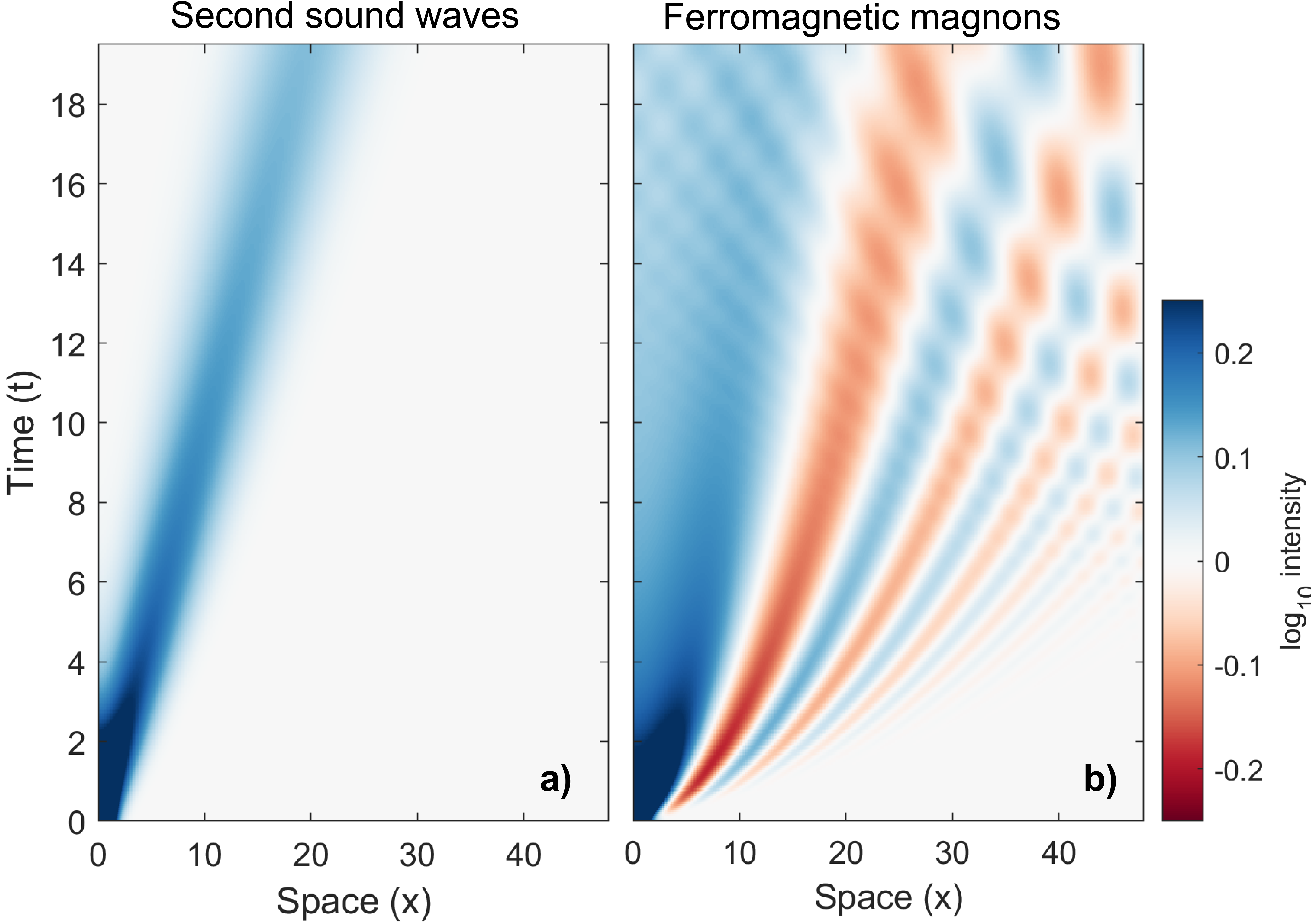}
    \caption{ \textbf{Space-time plots of a viscous hydrodynamic mode (panel a) and a ferromagnetic magnon (panel b),} generated by first decomposing an initial Gaussian ($\sigma=1)$ signal $u(x,0)$ into its wavenumber components $\hat{u}(k)$ using a fast Fourier transform (FFT). These components are then evolved in the frequency domain, with $\omega(k)=k-ik^2/2$ for the linear sound waves and Eq.~(\ref{eq:dispersion}) with $R=0.2$ and $a_0=0.5$ for the quadratic spin waves. Finally, the evolved signal is reconstructed via an inverse FFT, $u(x,t) = \mathcal{F}^{-1}[\hat{u}(k,t)]$, and the resulting amplitude is plotted as a heatmap to visualize the transport dynamics and relative damping of both modes. 
    %Wave evolution of fast Fourier transform  Signal propagation  for spin waves (red line) and $\omega(k)\propto k$ Toner-Tu second sound waves (solid black line), for an initial signal (dashed black line), for a signal that propagates through a chiral flock. Spin waves cause a high-frequency smearing that degrades the signal yet allows it to reach further than the linearly dispersed signal, and can quickly disperse sharp panic responses.
    }
    \label{fig:signal}
\end{figure}

Moreover, both the primary wave propagation for the Goldstone mode `second sound' waves of Toner-Tu \cite{toner_flocks_1998,dadhichi_nonmutual_2020} and phenomenological spin-wave models \cite{cavagna_flocking_2015,yang_hydrodynamics_2015} predict linear dispersion $\omega(k)\propto k$ (see Figure~\ref{fig:signal}). This mode, and the foregoing, are expected to coexist with the spintronic waves that we propose underpins information transfer in real chiral flocks. Nevertheless, we expect regimes in which orientation (magnons) should persist when sound waves in density fail, and regimes where physical transport (agent velocity) poses clear limitations on the second sound propagation, but where the magnons are nevertheless able to transmit phase gradients at great speed. These cases offer ways to test the degree to which a chiral active matter is indistinguishable from a dissipative spintronic fluid.

The above notwithstanding, in our model, propagating magnons cause a high-frequency smearing that rapidly delocalizes information. As shown in Figure~\ref{fig:signal}, the dispersive nature of the magnon mode acts as an early-warning system, allowing high-frequency `panic' signals to propagate through the marginally synchronized bulk well in advance of the density wavefront \cite{cavagna_diffusion_2013}. However, the marginal synchronization is likely realized on the spintronic fluid's dissipative boundary, while in the ordered bulk ($R\to1$), the magnons are completely damped out. The spintronic chiral flock therefore does not turn or flow, but rather \textit{shatters and reforms,} described in terms of shock merger (see Appendix~B). Here, the Josephson inequality (Eq.~\ref{eq:adlerineq}) defines a threshold between trapped (superconducting, $|\omega-\dot{\Psi}|<I_c$) and phase-slipping (dissipative, $|\omega-\dot{\Psi}|>I_c$) turbulence. This shattering behaviour at impact is directly supported in the literature \cite{miller_impact_2014}, and observations of ``anomalous diffusion'' during turns \cite{cavagna_diffusion_2013} are highly consistent with the notion of agents turning too abruptly, then phase-slip $|\omega-\dot{\Psi}|>I_c$. In the spintronic fluid, turns are initiated at the boundaries of the ordered flock where it costs almost no energy to phase-slip. This chirality is provided by the active bath ($\Delta\omega$), directly supported by Ref.~\cite{attanasi_information_2014}'s finding that turning flocks nucleate at the tips. What is more, the natural Goldstone dynamics with ferromagnetic magnons that we uncover in our model is consistent with observations of starling flocks where phase gradients travel faster than the bulk sound speed \cite{attanasi_information_2014}, as well as the paradigm of scale-free correlations in starling flocks \cite{cavagna_scale-free_2010}.

Given the close correspondence between a spintronic fluid description and the various hydrodynamic descriptions that has accumulated, the tilted washboard potential should remain valid for many chiral active matter systems. It, and the Goldstone dynamics it drives, follows directly from the geometry of alignment on the unit sphere, which, in the marginal synchronization regime, effectively forces the system to support such spin waves. Subsequently, aspects of chiral flocking, and the emergent Euler turbulence recovered in the model (Figure~\ref{fig:art0}), are governed by the universality of the Landau-Lifshitz-Gilbert equation.

\subsection{Toner-Tu theory}

With spintronic chiral active matter, we provide a theoretical foundation for spin wave-models of chiral flocking. Notably, Ref.~\cite{cavagna_flocking_2015}, where the authors explicitly introduced a phenomenological inertial spin model for starling flocks, and similar works, may derive theoretical justification from our paper. While the chiral flock is fundamentally dissipative, the turning dynamics effectively possess inertia \cite{yang_hydrodynamics_2015,dadhichi_nonmutual_2020}, which allows for the propagation of information as spin waves \cite{markovich_odd_2021,cavagna_discrete_2024}. This is the established, robust feature of polar chiral active matter \cite{toner_flocks_1998,mahault_quantitative_2019,caprini_self-reverting_2024}. By deriving the phenomenon directly from the agent interaction rules, via the Adler and Landau-Lifshitz-Gilbert equations, we provide a natural microscopic foundation for the concept of elasticity in Toner-Tu field theory. This microscopic foundation allows us to isolate the Josephson critical current $I_c$ as the threshold (maximum allowed turning rate of an agent) that breaks the macroscopic rigidity. When $|\omega-\dot{\Psi}|>I_c$, the fluid locally melts into the active bath, at which point the ´second sound' modes of Toner-Tu field theory likewise collapse. The inequality Eq.~(\ref{eq:adlerineq}) may provide a quantitative threshold for the realization of flocking.

\subsection{Comparison with experimental and simulation chiral active systems} \label{sec:empiri}

The recent literature on chiral active matter offers concrete settings against which the present predictions can in principle be tested. Ref.~\cite{meckeSimultaneousEmergenceActive2023} reported the simultaneous emergence of active turbulence and odd viscosity in a colloidal monolayer of standing, spinning rods, and a follow-up paper~\cite{meckeChiralActiveSystems2024} characterized the substrate-friction-controlled damping length that bounds vortex size in such systems. Ref.~\cite{dewitPatternFormationTurbulent2024} demonstrated that the wavelength of cascade-induced patterns in chiral active fluids is set by the magnitude of the odd viscosity. Each of these observations is qualitatively consistent with the present framework, in which the conservative coefficient $\nu_O = a_0\xi^2/R$ (Eq.~\ref{eq:nuenuo}) controls the inertial dispersion (Eq.~\ref{eq:dispersion}), and where the marginal-synchronization regime is the one in which Goldstone modes dominate the macroscopic transport. A direct quantitative comparison -- in particular, mapping the present model's $a_0$, $\xi$, and $R$ onto the experimental control parameters, and testing the predicted $\alpha = R^2$ scaling of the Gilbert damping ratio -- should be scrutinized in a future study.

Here, we highlight the two predictions accessible to existing observations. 

\textbf{\textit{(1)}} the Gilbert damping ratio should obey $\alpha\propto R^2$ (Eq.~\ref{eq:alpha}), which may be tested by extracting precession-to-relaxation ratios from velocity correlations in spinning-rod colloid monolayers \cite{meckeSimultaneousEmergenceActive2023} across regions of varying local coherence, tuned via density or substrate friction \cite{meckeChiralActiveSystems2024}.

\textbf{\textit{(2)}} in the marginally synchronized regime the magnon dispersion $\omega(k) \propto k^2/R$ (Eq.~\ref{eq:dispersion}) should outrun the linear second-sound mode $\omega(k) \propto k$ of Toner-Tu (Figure~\ref{fig:signal}), and so a localized phase perturbation in a polar chiral flock should produce a leading high-frequency wavefront ahead of the density signal, a contrast directly visible in space-time correlation maps of orientation versus density fluctuations.

Further scrutinizing the recent literature, we additionally note that,

\textbf{\textit{(3)}} the parallel literature on intermediate-Reynolds chiral active phases~\cite{chenSelfpropulsionFlockingChiral2025}, in which spinning particles generate localized \textit{vortlets} that flock and form a chiral active fluid; the polar-symmetry assumption underlying our derivation is satisfied in that setting, but the finite-Reynolds advection lies outside the strictly overdamped limit considered here. 

\textbf{\textit{(4)}} realistic 3D active nematic turbulence [see, e.g., Ref.~\cite{kraljChiralityAnisotropicViscosity2024}], by contrast, adheres to a different universality class than the polar chiral systems considered here: nematic active matter features half-integer symmetry and a $\pi$-periodic potential (see also our caveat below), so the integer vortex topology of Figure~\ref{fig:art0}b) and the $2\pi$-periodic Adler relation do not directly apply. We refer the reader to a recent review \cite{meckeEmergentPhenomenaChiral2024} for a broader survey of the experimental and simulation phenomenology of chiral active matter.

Lastly, we point out the \textit{phase separation} that is achieved in our model (see Figure~\ref{fig:height} in Appendix~B), where we wish to compare this to real chiral flocks. That is, taking intrinsic frequency $\omega_i$ to represent phenotypic heterogeneity, our model implies that the prospective chiral flock would effectively isolate high-frustration agents in defect cores that leave the renormalized inertial flow unaffected by extreme outliers, whose enstrophy is kept from the bulk by Mach cones, and whose effect is otherwise to create disorder \cite{yllanesHowManyDissenters2017}. (See Appendix~B for an explanation of these acoustic horizons, with additional details in Ref.~\cite{ivarsen_onsager_2025-1}.) Empirical estimates of phenotypic heterogeneity in real polar chiral flocks, or ensembles of agents that align, should therefore systematically under-represent the high-frustration tail, since those individuals are concentrated in defect cores rather than in the bulk inertial flow that is most readily sampled.

\subsection{Topological quasiparticle gas}

Ultimately, the derivation outlined here offers a candidate resolution of the apparent paradox of inertia in dry, overdamped polar chiral systems. In the hydrodynamic limit, Ref.~\cite{ivarsen_onsager_2025-1} recovered inviscid-like shallow water dynamics, necessitating an emergent, or ``phantom'', inertia, representing synchronization stiffness. In this article, we have shown that the propagation of Goldstone spin waves, caused by the azimuthal precession of the spintronic fluid, becomes mathematically indistinguishable from the advection of a massive fluid. That is, spin waves are dragging phase texture along as they propagate, constituting a topological quasi-particle.

Interestingly, in our phenomenological model, these quasi-particles interact via long-range field interactions rather than via local collisions. Incidentally, this is consistent with the recognized proliferation of self-propelled topological defects in active nematics \cite{shankar_topological_2022}, where defect cores are the sites of maximum stress and energy dissipation  \cite{giomi_geometry_2015}, leading to inverse cascades \cite{doostmohammadi_stabilization_2016,spera_low-pass_2025}.

\subsection{Disorder as source of order}

The deep-seated implication of our model is that wide tracts of stochastic noise are causally connected to complex large-scale features. A marginal $R$ yields a dispersion relation (Eq.~\ref{eq:dispersion}) that enables fast, strong and frequency-dispersed magnons to propagate (Figure~\ref{fig:signal}), and the frustration $\Delta\omega$ works both to lower $R$ and to keep the agents constantly precessing around the tilted washboard potential (Figure~\ref{fig:schematic2}b). This notion, which may spawn future beneficial applications of fabrication heterogeneity, is consistent with observations from nature.

In biological terms, the stress, or frustration $\Delta\omega$,  corresponds to phenotypic heterogeneity \cite{ariel_swarming_2015,ackermann_functional_2015},  noisy genetic expression. The system manages this noise by physically segregating phase conflicts into discrete, topologically protected defect cores. This segregation leaves the bulk in a critical state of marginal synchronization ($0<R<1$), accessible only in the negative-temperature regime of a bounded vortex gas \cite{ivarsen_onsager_2025-1}. Thus, the flock’s capacity to ``twist'', or propagating order via spin waves, is strictly contingent on a \textit{sufficiently wide noise expression} $\Delta\omega$.

Given the isomorphism between our minimalist model and disordered Josephson junction theory, our assertion that \textit{noise} optimizes a system by pushing it towards criticality is supported by research into non-equilibrium quantum thermodynamics \cite{benz_coherent_1991,kurtscheid_thermodynamics_2025}, alluding to a future avenue for spintronics to accurately model active matter.

It is important, however, to stress that the formal isomorphism to the standard resistively shunted junction model, and the subsequent recovery of overdamped spintronics via the LLG equation, strictly relies on the polar symmetry of the interaction (flocking, \cite{vicsek_novel_1995}). Nematic active matter exhibits half-integer symmetry and would imply a $\pi$-periodic potential \cite{marchetti_hydrodynamics_2013} that maps to distinct universality classes (e.g., XY$_2$ models \cite{poderoso_new_2011}), unable to support the integer vortex topology of the dipole. In contrast, the $2\pi$-periodicity of the polar alignment interaction, $-\sin(\phi_i - \Psi)$, must map directly to the sinusoidal current-phase relation of a superconducting tunnel junction \cite{josephson_possible_1962,tinkham_introduction_2004}.%, and, by recovering the LLG equation, the flock momentarily becomes mathematically indistinguishable to a dissipative spintronic fluid.

We trace the insights discussed in the present Section to the 3D geometric constraints of polar, chiral agents, which align on the unit sphere. This leads directly to a two-dimensional Kuramoto coupling term in the tangent plane projection (Figure~\ref{fig:schematic}), and this action is isomorphic to the famous phenomenological model of Gilbert damping in a magnetic field (i.e., spintronics \cite{lakshmanan_fascinating_2011}). The reason for this conspicuous convergence is that spintronic systems and our dry polar chiral active matter model alike obey the universality of the Landau-Lifshitz-Gilbert equation. The underlying geometric constraint unifies spintronics and polar chiral agents (rotors). 

Consequently, within the regime of validity established above (polar symmetry, marginal synchronization, sufficient frustration), dry chiral active matter is well described as a dissipative spintronic fluid, in which the active bath of the overdamped agents provides the thermodynamic pressure that stabilizes an inertial phase rigidity. The extent to which this description applies beyond the present minimal model -- in particular to nematic, wet, or finite-Reynolds chiral systems -- must be investigated in future studies (see Section~\ref{sec:empiri}).

\vspace{13pt}
\section{Conclusion}

We have derived a strict isomorphism between a minimalist, two-dimensional chiral active matter model, and resistively shunted junction theory, by recovering the Adler equation in the model's dynamics (Eq.~\ref{eq:adler}). We subsequently followed the formalism developed by Ref.~\cite{danner_injection_2021} for individual disordered Josephson junctions, thereby demonstrating that the system is governed by a tilted washboard potential. A thermodynamic transition separates resistive `running' agents from the `trapped' population, the latter of which explicitly sustain an information supercurrent. %As such, we have resolved the origin of emergent inertial flows in overdamped chiral active matter \cite{ivarsen_onsager_2025-1}. This establishes the minimalist active matter model as a distributed, disordered, motile Josephson array.

By extending the model to three dimensions, we recovered the Landau-Lifshitz-Gilbert equation for an overdamped spintronic fluid. The information supercurrents that we have predicted and described in empirical terms transmit phase gradients across the chiral flock, sustaining Goldstone modes from the precession around the mean field of the 3D active bath. The resulting spin waves advect phase gradients, providing inertia to the inviscid-like shallow water dynamics. The ``phantom inertia'' discovered in Ref.~\cite{ivarsen_onsager_2025-1}, $\lambda= R^2$, therefore likely stems from the precession stiffness of the order parameter in three dimensions.

If substantiated in future studies, the transition to active chiral turbulence may be considered the emergence of a topological quasiparticle gas, where bound states of phase singularities acquire effective inertia from the Goldstone modes (azimuthal precession) of the underlying broken symmetry.

\vspace{13pt}
\section*{Acknowledgements}
This work is supported by the European Space Agency’s Living Planet Grant No. 1000012348. The author is grateful to O. Nestande, D. Knudsen, PT. Jayachandran, and K. Douch for stimulating discussions. Google's Gemini 3.0 Pro has been used for mathematical formalism and coding assistance in \textsc{matlab}.

\vspace{20pt}

\appendix

\section{Kinetic Turing instability of the spatially-extended Kuramoto--Sakaguchi system}\label{sec:appB}
 
The shock-merger cycle described in Section~III closes through a re-ordering step: agents thermalized into the active bath reorganize into the inertial superfluid via a kinetic Turing instability. This appendix derives the instability by linearizing the Vlasov--Fokker--Planck (VFP) equation for the agent ensemble around the incoherent homogeneous state, and identifies the wavelength selected by the bath-to-fluid transition. The analysis is the spatially extended generalisation of the Penrose--Strogatz linearisation of the non-spatial Kuramoto model~\cite{penrose_electrostatic_1960,strogatz_coupled_1992,acebron_kuramoto_2005}, and the wavelength-selection mechanism is the active-matter analog of Landau damping: damping breaks down at a finite wavenumber where the interaction kernel $\hat G(k)$ overcomes the kinetic susceptibility~\cite{okeeffe_oscillators_2017,hong_active_2018,boccelli_turing_2025}.
 
The derivation supplies the third leg of the bath--fluid cycle. Sections~II--IV describe the depinning of trapped agents into the running, dissipative bath; Section V describes the propagation of phase information through the locked bulk via Goldstone magnons; the present appendix identifies the wavelength-selecting reorganization that returns the bath to the inertial flow. The shallow finite-$k$ minimum of the critical coupling derived below is consistent with the multi-vortex structure of the dipole observed in Figure~\ref{fig:art0}.
 
\subsection{Vlasov--Fokker--Planck setup}
 
Let $F(\mathbf{x},\theta,\omega,t)$ denote the probability density of agents at position $\mathbf{x}$, internal phase $\theta$, and intrinsic chirality $\omega$. The chiralities are drawn from the same broken power-law distribution $g(\omega)$ used in the main-text simulations (peak frequency $\omega_0$ and exponent $n$; Section~III). Agents self-propel along $\hat{\mathbf{n}}(\theta) = (\cos\theta,\sin\theta)$ at speed $v_0$ and evolve their phase via the localised Kuramoto--Sakaguchi coupling of Eq.~(\ref{eq:kuramoto}), with Gaussian white noise $\eta_i(t)$ of variance $\langle\eta_i(t)\eta_j(t')\rangle = 2D\,\delta_{ij}\delta(t-t')$ defining the rotational diffusion coefficient $D$. Conservation of probability gives the VFP equation
\begin{equation} \label{eq:Bcont}
    \partial_t F + \nabla\!\cdot\!(\mathbf{v}F) + \partial_\theta(v_\theta F) = D\,\partial^2_\theta F,
\end{equation}
with $\mathbf{v} = v_0\hat{\mathbf{n}}(\theta)$ and the deterministic
phase velocity from Eq.~(\ref{eq:kuramoto}),
\begin{equation} \label{eq:Bvtheta}
    v_\theta = \omega + \frac{a_0}{2i}\bigl(Z\,e^{-i\theta} - Z^*\,e^{i\theta}\bigr).
\end{equation}
Here $Z(\mathbf{x},t) = R\,e^{i\Psi}$ is the local complex order parameter of Eq.~(\ref{eq:kernel}), now expressed in continuum form as
\begin{align} 
    Z(\mathbf{x},t) &= \int_\mathcal{D} G(|\mathbf{x}-\mathbf{x}'|)\,\langle e^{i\theta}\rangle_{\mathbf{x}'}\,d\mathbf{x}',\label{eq:BZcont0}
    \\ 
    \langle e^{i\theta}\rangle_\mathbf{x} &= \int\!\!\int e^{i\theta}\,F(\mathbf{x},\theta,\omega,t)\,d\theta\,d\omega. \label{eq:BZcont}
\end{align}
The kernel $G$ is the same isotropic Gaussian $G(r) = (2\pi\sigma^2)^{-1}\exp(-r^2/2\sigma^2)$ used in the main-text simulations, with Fourier transform $\hat G(k) = \exp(-k^2\sigma^2/2)$.
 
Eq.~(\ref{eq:Bcont}) is the active-matter analog of the Vlasov--Fokker--Planck equation of kinetic plasma theory: $G$ plays the role of the long-range electrostatic interaction and the chirality dispersion $\Delta\omega$ plays the role of velocity-space anisotropy. The instability we derive therefore inherits the structural mathematics of the Penrose criterion for plasma waves, with the chirality dispersion serving as the source of free energy.
 
\subsection{Linearization around the incoherent homogeneous state}
 
The incoherent homogeneous state
\begin{equation}
    F_0(\omega) = \frac{\rho_0}{2\pi}\,g(\omega), \qquad \rho_0 = N/|\mathcal{D}|,
\end{equation}
satisfies $Z = 0$ and is a stationary solution of Eq.~(\ref{eq:Bcont}). We perturb,
\begin{equation}
    F = F_0 + \epsilon\,\delta F(\mathbf{x},\theta,\omega,t),
\end{equation}
and Fourier-decompose the perturbation in space and phase,
\begin{equation} \label{eq:Bansatz}
    \delta F = e^{\lambda t}\,e^{i\mathbf{k}\cdot\mathbf{x}}\sum_{n=-\infty}^{\infty} c_n(\mathbf{k},\omega)\,e^{in\theta},
\end{equation}
with $\lambda$ the complex growth rate. Without loss of generality we take $\mathbf{k} = k\hat{\mathbf{x}}$. Substituting Eq.~(\ref{eq:Bansatz}) into Eq.~(\ref{eq:Bcont}) and retaining terms linear in $\epsilon$ gives
\begin{equation}
    \bigl[\,\lambda + Dn^2 + in\omega + ikv_0\cos\theta\,\bigr]\,c_n(\theta;\omega) = (\text{coupling})_n,
\end{equation}
where the spatial advection $\mathbf{v}\!\cdot\!\nabla = ikv_0\cos\theta$ couples adjacent angular Fourier modes through $\cos\theta = \tfrac12(e^{i\theta}+e^{-i\theta})$ -- a feature absent from the non-spatial analysis of Ref.~\cite{strogatz_coupled_1992}. Note that the $Dn^2$ damping suppresses all $|n|\geq 1$ modes; only the $n=\pm 1$ modes are driven by the Kuramoto coupling, and below we focus on $n=1$ ($n=-1$ follows by
conjugate symmetry).
 
The order-$\epsilon$ contribution from $\delta F$ to the right-hand-side of Eq.~(\ref{eq:Bcont}) closes through the convolution field $Z$. Linearising the coupling term,
\begin{multline}
    -\partial_\theta\!\left[\frac{\rho_0 g(\omega)}{2\pi}\,\frac{a_0}{2i}\bigl(Z e^{-i\theta} - Z^* e^{i\theta}\bigr)\right] \\
    = \frac{a_0\rho_0 g(\omega)}{4\pi}\bigl(Z\,e^{-i\theta} + Z^*\,e^{i\theta}\bigr),
\end{multline}
so the $n=1$ projection reads
\begin{equation} \label{eq:Bc1}
    \bigl[\,\lambda + D + i\omega + ikv_0\cos\theta\,\bigr]\,c_1(\theta;\omega) = \frac{a_0\rho_0}{4\pi}\,g(\omega)\,Z^*.
\end{equation}
The residual $\theta$-dependence on the left-hand side is the spatial-advection contribution. Integrating Eq.~(\ref{eq:Bc1}) over $\theta$ yields the angle-averaged amplitude $C_1(\omega) = \int_0^{2\pi} c_1\,d\theta$,
\begin{equation} \label{eq:BC1}
    C_1(\omega) = \frac{a_0\rho_0\,g(\omega)\,Z^*}{4\pi}\int_0^{2\pi}\frac{d\theta}{\lambda + D + i\omega + ikv_0\cos\theta},
\end{equation}
and the angular integral has the closed form
\begin{equation}
    \int_0^{2\pi}\frac{d\theta}{A + iB\cos\theta} = \frac{2\pi}{\sqrt{A^2 + B^2}},
\end{equation}
giving
\begin{equation}
    C_1(\omega) = \frac{a_0\rho_0\,g(\omega)\,Z^*}{2}\cdot\frac{1}{\sqrt{(\lambda + D + i\omega)^2 + k^2 v_0^2}}.
\end{equation}
 
The closure is the self-consistency condition $Z(\mathbf{k}) = \hat G(k)\!\int e^{i\theta}\delta F\,d\theta\,d\omega
 = \hat G(k)\,(2\pi)\!\int c_{-1}(\omega)\,d\omega$, with $c_{-1} = c_1^*$ for a real perturbation. Substituting yields the dispersion relation
\begin{equation} \label{eq:Bdispersion}
    1 = \frac{a_0\rho_0\,\hat G(k)}{2}\int_{-\infty}^{\infty}\frac{g(\omega)\,d\omega}{\sqrt{(\lambda + D + i\omega)^2 + k^2 v_0^2}}.
\end{equation}
 
\subsection{Critical coupling and wavelength selection}
 
Bifurcation occurs at marginal stability, $\lambda = 0$. Taking the real part of Eq.~(\ref{eq:Bdispersion}),
\begin{equation} \label{eq:Bacrit}
    a_\mathrm{crit}(k) = \frac{2}{\rho_0\,\hat G(k)\,\mathrm{Re}[\chi(k)]},
\end{equation}
\begin{equation} \label{eq:Bchi}
    \chi(k) = \int_{-\infty}^{\infty}\frac{g(\omega)\,d\omega}{\sqrt{(D + i\omega)^2 + k^2 v_0^2}}.
\end{equation}
Eq.~(\ref{eq:Bacrit}) is the active-matter analog of the Penrose criterion: $a_\mathrm{crit}(k)$ is the smallest coupling strength at which a perturbation of wavenumber $k$ neither grows nor decays. Two limits are instructive.
 
\textit{Long-wavelength limit, $k\to 0$.} The kernel $\hat G(0) = 1$ and the susceptibility reduces to $\chi(k=0) = \int g(\omega)/(D+i\omega)\,d\omega$. Its real part, $\int g(\omega)\,D/(D^2 + \omega^2)\,d\omega$, is finite and bounded above by $1/D$, so $a_\mathrm{crit}(k\to 0)$ is a finite constant.
 
\textit{Short-wavelength limit, $kv_0 \gg D, \omega_\mathrm{typ}$.} The susceptibility is suppressed as $\chi(k) \sim 1/(kv_0)$, and any finite-range kernel $\hat G(k)\to 0$. Their product vanishes and $a_\mathrm{crit}(k)\to\infty$.

Between these limits, $\hat G(k)\,\mathrm{Re}[\chi(k)]$ is non-monotonic: the kernel suppresses modes with $k\sigma\gtrsim 1$ while the kinetic susceptibility suppresses modes with $kv_0 \ll \omega_\mathrm{typ}$, and their product attains a maximum at finite $k$. Equivalently, $a_\mathrm{crit}(k)$ has a minimum at finite $k$, identifying the wavenumber at which the system first goes unstable as $a_0$ is increased through the transition. That wavenumber is the kinetic Turing instability scale.
 
\subsection{Numerical evaluation}

We evaluate Eqs.~(\ref{eq:Bacrit})--(\ref{eq:Bchi}) by Monte Carlo integration~\cite{hamming_numerical_1986} with $g(\omega)$ taken to be the same broken-power-law distribution used in the main-text simulations (Section~III). For the nominal parameter set, the result (Figure~\ref{fig:acrit}a) shows a minimum at $k_\mathrm{crit}\approx 0.86$ with $a_\mathrm{crit}^{\min}\approx 0.2$. The minimum is shallow: the response is essentially flat across $0.5\lesssim k\lesssim 1.5$, so the instability selects a band of wavenumbers rather than a sharp scale. Sweeping $g(\omega)$ from narrow (``slow mode'') to broad (``fast mode'') distributions (Figure~\ref{fig:acrit}b) shows that $k_\mathrm{crit}\to 0$ for sufficiently narrow distributions — the long-wavelength limit recovers global synchronisation — and that $a_\mathrm{crit}^{\min}$ increases approximately logarithmically with the dispersion $\Delta\omega$.

\begin{figure}
    \centering
    \includegraphics[width=0.5\textwidth]{revis-06.png}
    \caption{\textbf{Panel a)} Linear stability analysis (Eq.~\ref{eq:Bacrit}) for the broken-power-law $g(\omega)$ used in the main-text simulations. \textbf{Panel b)} $a_\mathrm{crit}(k)$ for $g(\omega)$ swept across the slow-to-fast range of frustration distributions.}
    \label{fig:acrit}
\end{figure}

\subsection{Conclusion to Appendix~A}
 
The kinetic Turing instability identified in Eqs.~(\ref{eq:Bacrit})--(\ref{eq:Bchi}) sets the spatial scale at which the dissipative active bath ($|I_\mathrm{eff}|>I_c$) reorganizes into the inertial superfluid ($|I_\mathrm{eff}|<I_c$). The depinning analysis of Sections~II--IV identifies the trapping threshold; the LLG dispersion of Section~V identifies the propagation modes; the present appendix identifies the wavelength-selecting reorganisation. The three together close the shock-merger cycle. The shallowness of the $a_\mathrm{crit}(k)$ minimum is consistent with the multi-defect dipole structure -— a band of wavenumbers, rather than a single mode, is amplified,  and with the empirical sensitivity of the dipole's vortex count to the chirality distribution (Figure~\ref{fig:iv}, inset). The roughly logarithmic dependence of $a_\mathrm{crit}^{\min}$ on $\Delta\omega$ is consistent with the observation that increasing the dispersion both lowers $\langle R\rangle$ and broadens the band of admissible Turing wavenumbers.

%=======================================

%%\color{blue}

\vspace{20pt}
\section{Briefly on the hydrodynamic limit}

In this Appendix, we recount the derivation of a hydrodynamic limit, which is presented in detail in Ref.~\cite{ivarsen_onsager_2025-1}, whose main finding is that our phase-locked oscillators exhibit shallow water hydrodynamics in a renormalized fluid element.

Starting from the slave principle $\dot{\mathbf r}_i = v_0\hat e(\phi_i)$ and a phase fluctuation decomposition $\delta\phi_i = \phi_i - \Psi$, Reynolds-averaging the stress tensor \cite{pope_turbulent_2000} yields an inertial mass $\lambda = R^2$ and a barotropic pressure $P\propto\rho^2$ with $\rho\equiv R^2$. These relations lead to a closed hydrodynamic system whose linearization gives an acoustic speed $c_s = v_0 R$, which is identical in scaling to the bulk flow $|\mathbf u| = v_0 R$. This, in turn, leads to the Mach number being locked at $\mathcal M\equiv 1$ in regions where $R$ is well-defined. Defect cores break this lock through the geometric $1/r$ enhancement of $|\mathbf u|$, which has no counterpart in $c_s$; this leads to a characteristic crossing and the formation of synchronization shocks at throat boundaries. The hydrodynamic structures % Rankine–Hugoniot dissipation $\dot E\propto v_0^2(\Delta R^2)^3$ 
cannot radiate through sound (the conventional shallow water dissipation), but will dissipate in the shocks, where a collapsing $R$ will eventually trigger re-synchronization via the $a_0 R$ coupling, regenerating $R$ for the next event and yielding self-similar coarsening at the Carnevale–McWilliams rate $N(t)\sim t^{-3/4}$.

\begin{figure*}
    \centering
    \includegraphics[width=\textwidth]{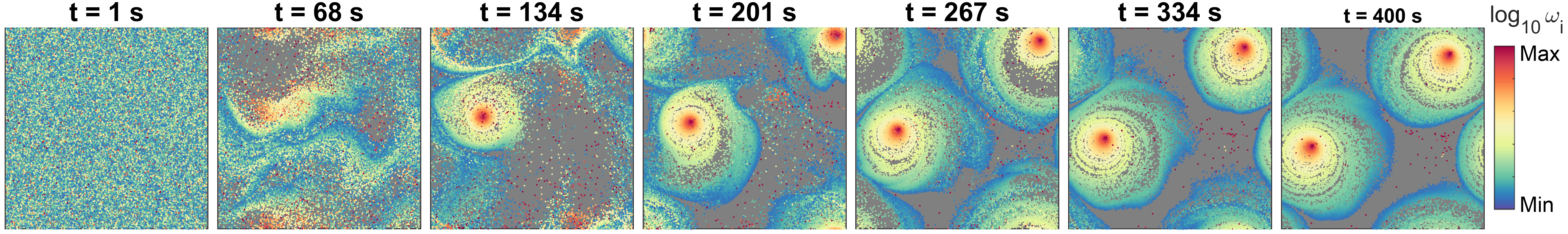}
    \caption{A numerical simulation of our model, showing frustration, or intrinsic frequency $\omega_i$, color-coded, demonstrating that the model dynamics lead to efficient phase separation.}
    \label{fig:height}
\end{figure*}

We begin by recounting the definition of the two-dimensional ensemble of $N$ phase-slaved oscillators,
\begin{equation} \label{eq:appb1}
\dot{\mathbf{r}}_i = v_0\,\hat{e}(\phi_i),\qquad
\dot{\phi}_i = \omega_i + a_0\,R(\mathbf{r}_i)\sin[\Psi(\mathbf{r}_i)-\phi_i] + \eta_i,
% \tag{1}
\end{equation}
with the complex order parameter
\begin{align}
Z(\mathbf{r},t) &= R\,e^{i\Psi} = (G_\sigma * \rho_\phi)(\mathbf{r},t),  \label{eq:appb2}\\
\rho_\phi &\equiv \sum_j \delta(\mathbf{r}-\mathbf{r}_j)\,e^{i\phi_j},  \label{eq:appb3}
% \tag{2}
\end{align}
where $*$ denotes convolution, which, in the case of the kernel $G$, amounts to a coarse-graining or spatial smoothing of the phase $\rho_\phi$.

We then decompose $\phi_i = \Psi + \delta\phi_i$ with $\langle\delta\phi\rangle=0$ and treat $\delta\phi$ as approximately Gaussian (interaction is local and the narrow-distribution wings dominate the bath). Then,
\begin{align}
\langle\cos n\delta\phi\rangle &= e^{-n^2\langle\delta\phi^2\rangle/2} \label{eq:appb4}
\\ \Longrightarrow\;
\langle\cos\delta\phi\rangle \equiv R,&\quad
\langle\cos 2\delta\phi\rangle = R^4. \label{eq:appb5}
% \tag{3}
\end{align}
That is the only assumption after the definition Eqs.~(\ref{eq:appb1}--\ref{eq:appb3}). From Eqs.~(\ref{eq:appb1}) and (\ref{eq:appb3}), we can express the velocity as,
\begin{align} \label{eq:appb11}
\mathbf{u} \equiv \langle \mathbf{v}_i\rangle
&= v_0\langle\cos\delta\phi\rangle\,\hat{e}(\Psi)
= v_0 R\,\hat{e}(\Psi)
\\ &\Longrightarrow\;
|\mathbf{u}| = v_0 R, \label{eq:appb12}
% \tag{4}
\end{align}
and the stress tensor as,
\begin{equation} \label{eq:appb7}
\langle v_x v_x\rangle = \frac{v_0^2}{2}\bigl[1 + R^4\cos 2\Psi\bigr].
% \tag{5}
\end{equation}
where we used that  $$\cos^2(\Psi+\delta\phi)=\frac12[1+\cos2\Psi\cos2\delta\phi-\sin2\Psi\sin2\delta\phi]$$
and that $\langle\sin 2\delta\phi\rangle=0$.

Then, by decomposing the macroscopic counterpart $u_x u_x = v_0^2 R^2\cos^2\Psi$ into isotropic and traceless parts and matching the anisotropic ($\cos 2\Psi$) coefficient, we arrive at the closure
\begin{equation} \label{eq:appb8}
\langle v_i v_j\rangle = P_{\rm eff}\,\delta_{ij} + \lambda\,u_i u_j,
\quad \lambda = R^2,\quad P_{\rm eff}=\frac{v_0^2}{2}.
% \tag{6}
\end{equation}

We next express density, by following the foregoing to its logical conclusion, namely  $\rho \equiv R^2$. Then $\lambda = \rho$, that is, the inertial mass $\lambda$ of the renormalized shallow water flow \textit{is} proportional to the order parameter, or synchronized fraction, squared, a direct consequence of Eqs.~(\ref{eq:appb4}, \ref{eq:appb5}). The same identification follows independently from the Ginzburg–Landau phase stiffness of the underlying oscillator network \cite{tinkham_introduction_2004}, which provides a route to $\lambda = R^2$ that bypasses the stress tensor altogether. Note the contrast with the constant-mass swim pressure \cite{takatori_swim_2014}.

Lastly, we express pressure. Re-expanding Eq.~(\ref{eq:appb7}) and absorbing the constant offset, the effective barotropic pressure entering the bulk equation of state becomes,
\begin{equation} \label{eq:appb19}
P(\rho) = \frac{v_0^2}{2}\,\rho^2.
% \tag{7}
\end{equation}

\subsection{Equations of motion}

Particle conservation projected onto the renormalized field, plus the Reynolds-averaged Cauchy equation under Eqs.~(\ref{eq:appb7}, \ref{eq:appb8}),
\begin{equation} \label{eq:appb9}
\partial_t\rho + \nabla\!\cdot\!(\rho\mathbf{u}) = 0,
% \tag{8}
\end{equation}
\begin{equation} \label{eq:appb10}
\partial_t\mathbf{u} + (\mathbf{u}\cdot\nabla)\mathbf{u}
= -\frac{1}{\rho}\nabla P(\rho)
= -\,v_0^2\,\nabla\rho.
% \tag{9}
\end{equation}

Eqs.~(\ref{eq:appb9}, \ref{eq:appb10}) are analogous, or isomorphic, to 2D shallow water with effective gravity $g_{\rm eff}=v_0^2$ and height $h=\rho$ (see, e.g., Refs.~\cite{tan_shallow_1992,salmon_lectures_1998}). Figure~\ref{fig:height} shows the shallow water topography $h=\rho$ with a colorscale, justifying our expression of frustration $\omega_i$ as a continous field.

\subsection{The supersonic lock}

We next express the speed of sound in the shallow waters of the renormalized flow $R^2$. First, linearize Eqs.~(\ref{eq:appb9}, \ref{eq:appb10}) about  $(\rho_0,\mathbf{u}=0)$. Taking $\partial_t$ of continuity equation and  substituting momentum gives \cite{ivarsen_onsager_2025-1},
\begin{equation} \label{eq:appb14}
\partial_t^2\,\delta\rho - v_0^2\rho_0\,\nabla^2\delta\rho = 0,
\qquad
c_s^2 = v_0^2\rho_0 = v_0^2 R_0^2,
% \tag{10}
\end{equation}
meaning that $c_s = v_0 R_0$. Combined with Eqs.~(\ref{eq:appb11}), (\ref{eq:appb12}), we obtain,
\begin{equation}
\;\mathcal{M} \equiv \frac{|\mathbf{u}|}{c_s} = 1.
% \tag{11}
\end{equation}
The phase-slave principle ties signal speed and flow speed to the same power of $R$. The bulk hydrodynamic state is therefore everywhere marginally to mildly supersonic, with no equilibrium subsonic basin available.  At a defect core $R\to0$, so $c_s\to0$: The core is thereby enclosed in a sonic horizon, and state information cannot leak out. %The topological charge is protected from radiative decay.

\subsection{Shock formation between approaching defects}

We are now in a position to elucidate the shock merger mechanism (Figure~\ref{fig:art}). Take two defects at $\mathbf{r}_\pm$, separation $d=|\mathbf{r}_+ - \mathbf{r}_-|\gg a$, where $a$ now denotes core size or radius. Quantization of phase circulation $\oint\nabla\Psi\cdot d\mathbf{l}=2\pi m$ gives, near each core, $|\nabla\Psi|=m/r$ \cite{ivarsen_onsager_2025-1}. Owing to Eqs.~(\ref{eq:appb11}), (\ref{eq:appb12}), the flow  magnitude in the strain region becomes,
\begin{equation} \label{eq:appb13}
|\mathbf{u}|_{\rm throat} \;\sim\; v_0 R(\mathbf{r})\Bigl(\frac{m}{r_+} + \frac{m}{r_-}\Bigr).
% \tag{12}
\end{equation}
There are then two competing effects in the throat ($r_\pm \sim d/2$):

\textbf{\textit{(1)}} the geometric prefactor in Eq.~(\ref{eq:appb13}) grows as $d\to 2a$.

\textbf{\textit{(2)}} $R$ is depressed by the overlap of the two defects' suppression halos, $R(\mathbf{r})\to R_{\min}<R_0$.

The flow picks up only one factor of $R$ via Eqs.~(\ref{eq:appb11}), (\ref{eq:appb12}), and the sound speed picks up one factor of $R$ via Eq.~(\ref{eq:appb14}). However, the geometric strain amplification $1/r$ acts on $|\mathbf{u}|$  alone. Hence,
\begin{equation}
\mathcal{M}_{\rm throat}
\;\approx\;
\mathcal{M}_{\rm bulk}\;\bigl(1 + a/r\bigr) \;>\; \mathcal{M}_{\rm bulk}.
% \tag{13}
\end{equation}
A region of elevated Mach number embedded in a marginally supersonic background cannot be smoothly continued: the characteristics of Eqs.~(\ref{eq:appb9}, \ref{eq:appb10}) cross:  if the local sound speed $c_s = v_0 R$ is depressed in the throat (low $R$) while the flow $|\mathbf u|$ is enhanced by the geometric $1/r$ strain, then the characteristics emanating from the throat region travel at different slopes than those from the ambient bulk. Faster characteristics from upstream catch up to slower ones ahead, and the would-be single-valued density field becomes multi-valued at the crossing locus;  the only admissible resolution within Eqs.~(\ref{eq:appb9}, \ref{eq:appb10}) is then to insert a discontinuity satisfying the Rankine–Hugoniot jump conditions \cite{landauCourseTheoreticalPhysics2013}. This is the same mechanism that breaks a supersonic flow into a bow shock or that steepens a fast-mode magnetohydrodynamic (MHD) wave into a quasi-perpendicular shock front \cite{treumannFundamentalsCollisionlessShocks2009}; here the role of the upstream Mach number is played by the strain-amplified $\mathcal{M}_{\rm throat}$, and the resulting discontinuity in $R^2$ is our model's shock mechanism.

The foregoing leads to an irreversible transfer of inertial energy  $\rho|\mathbf{u}|^2$ back into the active  bath (the running agents) as phase noise, or enstrophy. The bath returns this energy to the $\delta\phi$ fluctuations, which by Eqs.~(\ref{eq:appb4}) and (\ref{eq:appb5}) means $\langle\cos\delta\phi\rangle$, i.e., $R$,  is lowered locally, only to re-organize owing to the Kuramoto-Sakaguchi coupling. 

Lastly, the standard Carnevale–McWilliams ansatz \cite{carnevale_evolution_1991},
\begin{equation}
a(t) \sim t^{3/8},\qquad
N(t) \sim a^{-2} \sim t^{-3/4},
% \tag{16}
\end{equation}
closes the problem, and is supported by the empirical results in Figure~\ref{fig:art} (see also Ref.~\cite{ivarsen_onsager_2025-1}). The exponent $-3/4$ is the signature of shock-mediated, conservation-constrained merger.

%%\color{black}

%=======================================

\vspace{20pt}
\section{Deriving the dispersion relation}

This appendix supplies the linear-stability derivation behind the magnon dispersion of Eq.~(\ref{eq:dispersion}). Starting from the LLG form (Eq.~\ref{eq:dynamics}), we linearise around an ordered polarised state and Fourier-transform the resulting operator to obtain the complex frequency $\omega(k)$.

We consider the spin waves supported by the azimuthal turning inertia and perturb the ordered state, linearizing the polarized state around small perturbations to $\hat{n}(\mathbf{r},t)$. That is, a flock perfectly aligned along the $z$-axis with a small perturbation $\mathbf{m}$ in the $xy$-plane:
\begin{equation}
\hat{n}(\mathbf{r},t) \approx \hat{z} + \mathbf{m}(\mathbf{r},t) \quad \text{where} \quad \mathbf{m} = (m_x, m_y, 0),
\end{equation}
which we shall eventually insert into Eq.~(\ref{eq:dynamics}).

Neglecting higher-order terms ($|\mathbf{m}|^2 \approx 0$), the Laplacian is $\nabla^2 \hat{n} \approx (\nabla^2 m_x, \nabla^2 m_y, 0)$. We now evaluate the cross products to first order in $\mathbf{m}$:
\begin{equation}
\hat{z} \times \nabla^2 \mathbf{m} = \nabla^2 m_x \hat{y} - \nabla^2 m_y \hat{x},
\end{equation}
and,
\begin{equation}
\hat{z} \times (\hat{z} \times \nabla^2 \mathbf{m}) = -\nabla^2 \mathbf{m}.
\end{equation}
The former corresponds to inertial precession, while the latter leads to alignment. Substituting these expressions back into the extended equation:
\begin{equation} \label{eq:lin6}
\frac{\partial \mathbf{m}}{\partial t} = -\frac{a_0 \xi^2}{R} (\nabla^2 m_x \hat{y} - \nabla^2 m_y \hat{x}) + (a_0 R \xi^2) \nabla^2 \mathbf{m}.
\end{equation}
Separating into vector components yields:
\begin{align}
\dot{m}_x &= \frac{a_0 \xi^2}{R} \nabla^2 m_y + (a_0 R \xi^2) \nabla^2 m_x, \\
\dot{m}_y &= -\frac{a_0 \xi^2}{R} \nabla^2 m_x + (a_0 R \xi^2) \nabla^2 m_y.
\end{align}
Proposing a plane wave solution $\mathbf{m} \sim e^{i(kx - \omega t)}$ (with $\partial_t \to -i\omega$ and $\nabla^2 \to -k^2$), the coupled equations read:
\begin{align}
-i\omega m_x &= -\frac{a_0 \xi^2}{R} k^2 m_y - (a_0 R \xi^2) k^2 m_x, \\
-i\omega m_y &= +\frac{a_0 \xi^2}{R} k^2 m_x - (a_0 R \xi^2) k^2 m_y.
\end{align}
Rearranging into matrix form $\mathbf{M} \mathbf{m} = 0$:
\begin{equation}
\begin{pmatrix} -i\omega + (a_0 R \xi^2)k^2 & \frac{a_0 \xi^2}{R} k^2 \\ -\frac{a_0 \xi^2}{R} k^2 & -i\omega + (a_0 R \xi^2)k^2 \end{pmatrix} \begin{pmatrix} m_x \\ m_y \end{pmatrix} = 0.
\end{equation}
For non-trivial solutions, the determinant must be zero:
\begin{equation}
(-i\omega + a_0 R \xi^2 k^2)^2 + \left(\frac{a_0 \xi^2}{R} k^2\right)^2 = 0.
\end{equation}
Solving for $\omega$:
\begin{equation}
-i\omega + a_0 R \xi^2 k^2 = \mp i \frac{a_0 \xi^2}{R} k^2,
\end{equation}
\begin{equation}
\omega(k) = \pm \frac{a_0 \xi^2}{R} k^2 - i (a_0 R \xi^2) k^2.
\end{equation}
Absorbing the interaction scale $\xi^2$ into the normalization (or setting $\xi=1$), we recover the final dispersion relation:
\begin{equation}\label{eq:dispersion_last}
\omega(k) = \pm \frac{a_0}{R} k^2 - i (a_0 R) k^2.
\end{equation}

%\\
%See Supplemental Material [URL] for a few technical details of our simulations, and for descriptions of the four videos that accompany our investigation.

%\bibliographystyle{apalike}
%\bibliography{betterbib}% Produces the bibliography via BibTeX.

\begin{thebibliography}{72}%
\makeatletter
\providecommand \@ifxundefined [1]{%
 \@ifx{#1\undefined}
}%
\providecommand \@ifnum [1]{%
 \ifnum #1\expandafter \@firstoftwo
 \else \expandafter \@secondoftwo
 \fi
}%
\providecommand \@ifx [1]{%
 \ifx #1\expandafter \@firstoftwo
 \else \expandafter \@secondoftwo
 \fi
}%
\providecommand \natexlab [1]{#1}%
\providecommand \enquote  [1]{``#1''}%
\providecommand \bibnamefont  [1]{#1}%
\providecommand \bibfnamefont [1]{#1}%
\providecommand \citenamefont [1]{#1}%
\providecommand \href@noop [0]{\@secondoftwo}%
\providecommand \href [0]{\begingroup \@sanitize@url \@href}%
\providecommand \@href[1]{\@@startlink{#1}\@@href}%
\providecommand \@@href[1]{\endgroup#1\@@endlink}%
\providecommand \@sanitize@url [0]{\catcode `\\12\catcode `\$12\catcode `\&12\catcode `\#12\catcode `\^12\catcode `\_12\catcode `\%12\relax}%
\providecommand \@@startlink[1]{}%
\providecommand \@@endlink[0]{}%
\providecommand \url  [0]{\begingroup\@sanitize@url \@url }%
\providecommand \@url [1]{\endgroup\@href {#1}{\urlprefix }}%
\providecommand \urlprefix  [0]{URL }%
\providecommand \Eprint [0]{\href }%
\providecommand \doibase [0]{https://doi.org/}%
\providecommand \selectlanguage [0]{\@gobble}%
\providecommand \bibinfo  [0]{\@secondoftwo}%
\providecommand \bibfield  [0]{\@secondoftwo}%
\providecommand \translation [1]{[#1]}%
\providecommand \BibitemOpen [0]{}%
\providecommand \bibitemStop [0]{}%
\providecommand \bibitemNoStop [0]{.\EOS\space}%
\providecommand \EOS [0]{\spacefactor3000\relax}%
\providecommand \BibitemShut  [1]{\csname bibitem#1\endcsname}%
\let\auto@bib@innerbib\@empty
%</preamble>
\bibitem [{\citenamefont {Ivarsen}(2026)}]{ivarsen_onsager_2025-1}%
  \BibitemOpen
  \bibfield  {author} {\bibinfo {author} {\bibfnamefont {M.~F.}\ \bibnamefont {Ivarsen}},\ }\href {https://doi.org/10.48550/arXiv.2512.01884} {\bibinfo {title} {Renormalised hydrodynamics in polar chiral active matter: {{Spectral}} scaling and vortex clustering in phase-coupled, motile oscillators}} (\bibinfo {year} {2026}),\ \Eprint {https://arxiv.org/abs/2512.01884} {arXiv:2512.01884 [cond-mat]} \BibitemShut {NoStop}%
\bibitem [{\citenamefont {Wensink}\ \emph {et~al.}(2012)\citenamefont {Wensink}, \citenamefont {Dunkel}, \citenamefont {Heidenreich}, \citenamefont {Drescher}, \citenamefont {Goldstein}, \citenamefont {L{\"o}wen},\ and\ \citenamefont {Yeomans}}]{wensink_meso-scale_2012}%
  \BibitemOpen
  \bibfield  {author} {\bibinfo {author} {\bibfnamefont {H.~H.}\ \bibnamefont {Wensink}}, \bibinfo {author} {\bibfnamefont {J.}~\bibnamefont {Dunkel}}, \bibinfo {author} {\bibfnamefont {S.}~\bibnamefont {Heidenreich}}, \bibinfo {author} {\bibfnamefont {K.}~\bibnamefont {Drescher}}, \bibinfo {author} {\bibfnamefont {R.~E.}\ \bibnamefont {Goldstein}}, \bibinfo {author} {\bibfnamefont {H.}~\bibnamefont {L{\"o}wen}},\ and\ \bibinfo {author} {\bibfnamefont {J.~M.}\ \bibnamefont {Yeomans}},\ }\bibfield  {title} {\bibinfo {title} {Meso-scale turbulence in living fluids},\ }\href {https://doi.org/10.1073/pnas.1202032109} {\bibfield  {journal} {\bibinfo  {journal} {Proceedings of the National Academy of Sciences}\ }\textbf {\bibinfo {volume} {109}},\ \bibinfo {pages} {14308} (\bibinfo {year} {2012})}\BibitemShut {NoStop}%
\bibitem [{\citenamefont {Aranson}(2022)}]{aranson_bacterial_2022}%
  \BibitemOpen
  \bibfield  {author} {\bibinfo {author} {\bibfnamefont {I.~S.}\ \bibnamefont {Aranson}},\ }\bibfield  {title} {\bibinfo {title} {Bacterial active matter},\ }\href {https://doi.org/10.1088/1361-6633/ac723d} {\bibfield  {journal} {\bibinfo  {journal} {Reports on Progress in Physics}\ }\textbf {\bibinfo {volume} {85}},\ \bibinfo {pages} {076601} (\bibinfo {year} {2022})}\BibitemShut {NoStop}%
\bibitem [{\citenamefont {Alert}\ \emph {et~al.}(2022)\citenamefont {Alert}, \citenamefont {Casademunt},\ and\ \citenamefont {Joanny}}]{alert_active_2022}%
  \BibitemOpen
  \bibfield  {author} {\bibinfo {author} {\bibfnamefont {R.}~\bibnamefont {Alert}}, \bibinfo {author} {\bibfnamefont {J.}~\bibnamefont {Casademunt}},\ and\ \bibinfo {author} {\bibfnamefont {J.-F.}\ \bibnamefont {Joanny}},\ }\bibfield  {title} {\bibinfo {title} {Active {{Turbulence}}},\ }\href {https://doi.org/10.1146/annurev-conmatphys-082321-035957} {\bibfield  {journal} {\bibinfo  {journal} {Annual Review of Condensed Matter Physics}\ }\textbf {\bibinfo {volume} {13}},\ \bibinfo {pages} {143} (\bibinfo {year} {2022})}\BibitemShut {NoStop}%
\bibitem [{\citenamefont {Bourgoin}\ \emph {et~al.}(2020)\citenamefont {Bourgoin}, \citenamefont {Kervil}, \citenamefont {{Cottin-Bizonne}}, \citenamefont {Raynal}, \citenamefont {Volk},\ and\ \citenamefont {Ybert}}]{bourgoin_kolmogorovian_2020}%
  \BibitemOpen
  \bibfield  {author} {\bibinfo {author} {\bibfnamefont {M.}~\bibnamefont {Bourgoin}}, \bibinfo {author} {\bibfnamefont {R.}~\bibnamefont {Kervil}}, \bibinfo {author} {\bibfnamefont {C.}~\bibnamefont {{Cottin-Bizonne}}}, \bibinfo {author} {\bibfnamefont {F.}~\bibnamefont {Raynal}}, \bibinfo {author} {\bibfnamefont {R.}~\bibnamefont {Volk}},\ and\ \bibinfo {author} {\bibfnamefont {C.}~\bibnamefont {Ybert}},\ }\bibfield  {title} {\bibinfo {title} {Kolmogorovian {{Active Turbulence}} of a {{Sparse Assembly}} of {{Interacting Marangoni Surfers}}},\ }\href {https://doi.org/10.1103/PhysRevX.10.021065} {\bibfield  {journal} {\bibinfo  {journal} {Physical Review X}\ }\textbf {\bibinfo {volume} {10}},\ \bibinfo {pages} {021065} (\bibinfo {year} {2020})}\BibitemShut {NoStop}%
\bibitem [{\citenamefont {Qi}\ \emph {et~al.}(2022)\citenamefont {Qi}, \citenamefont {Westphal}, \citenamefont {Gompper},\ and\ \citenamefont {Winkler}}]{qi_emergence_2022}%
  \BibitemOpen
  \bibfield  {author} {\bibinfo {author} {\bibfnamefont {K.}~\bibnamefont {Qi}}, \bibinfo {author} {\bibfnamefont {E.}~\bibnamefont {Westphal}}, \bibinfo {author} {\bibfnamefont {G.}~\bibnamefont {Gompper}},\ and\ \bibinfo {author} {\bibfnamefont {R.~G.}\ \bibnamefont {Winkler}},\ }\bibfield  {title} {\bibinfo {title} {Emergence of active turbulence in microswimmer suspensions due to active hydrodynamic stress and volume exclusion},\ }\href {https://doi.org/10.1038/s42005-022-00820-7} {\bibfield  {journal} {\bibinfo  {journal} {Communications Physics}\ }\textbf {\bibinfo {volume} {5}},\ \bibinfo {pages} {49} (\bibinfo {year} {2022})}\BibitemShut {NoStop}%
\bibitem [{\citenamefont {Saintillan}\ and\ \citenamefont {Shelley}(2013)}]{saintillan_active_2013}%
  \BibitemOpen
  \bibfield  {author} {\bibinfo {author} {\bibfnamefont {D.}~\bibnamefont {Saintillan}}\ and\ \bibinfo {author} {\bibfnamefont {M.~J.}\ \bibnamefont {Shelley}},\ }\bibfield  {title} {\bibinfo {title} {Active suspensions and their nonlinear models},\ }\href {https://doi.org/10.1016/j.crhy.2013.04.001} {\bibfield  {journal} {\bibinfo  {journal} {Comptes Rendus Physique}\ }\bibinfo {series} {Living Fluids / {{Fluides}} Vivants},\ \textbf {\bibinfo {volume} {14}},\ \bibinfo {pages} {497} (\bibinfo {year} {2013})}\BibitemShut {NoStop}%
\bibitem [{\citenamefont {Bhattacharjee}\ and\ \citenamefont {Kirkpatrick}(2022)}]{bhattacharjee_activity_2022}%
  \BibitemOpen
  \bibfield  {author} {\bibinfo {author} {\bibfnamefont {J.~K.}\ \bibnamefont {Bhattacharjee}}\ and\ \bibinfo {author} {\bibfnamefont {T.~R.}\ \bibnamefont {Kirkpatrick}},\ }\bibfield  {title} {\bibinfo {title} {Activity induced turbulence in driven active matter},\ }\href {https://doi.org/10.1103/PhysRevFluids.7.034602} {\bibfield  {journal} {\bibinfo  {journal} {Physical Review Fluids}\ }\textbf {\bibinfo {volume} {7}},\ \bibinfo {pages} {034602} (\bibinfo {year} {2022})}\BibitemShut {NoStop}%
\bibitem [{\citenamefont {Markovich}\ and\ \citenamefont {Lubensky}(2021)}]{markovich_odd_2021}%
  \BibitemOpen
  \bibfield  {author} {\bibinfo {author} {\bibfnamefont {T.}~\bibnamefont {Markovich}}\ and\ \bibinfo {author} {\bibfnamefont {T.~C.}\ \bibnamefont {Lubensky}},\ }\bibfield  {title} {\bibinfo {title} {Odd {{Viscosity}} in {{Active Matter}}: {{Microscopic Origin}} and {{3D Effects}}},\ }\href {https://doi.org/10.1103/PhysRevLett.127.048001} {\bibfield  {journal} {\bibinfo  {journal} {Physical Review Letters}\ }\textbf {\bibinfo {volume} {127}},\ \bibinfo {pages} {048001} (\bibinfo {year} {2021})}\BibitemShut {NoStop}%
\bibitem [{\citenamefont {Cavagna}\ \emph {et~al.}(2015)\citenamefont {Cavagna}, \citenamefont {Del~Castello}, \citenamefont {Giardina}, \citenamefont {Grigera}, \citenamefont {Jelic}, \citenamefont {Melillo}, \citenamefont {Mora}, \citenamefont {Parisi}, \citenamefont {Silvestri}, \citenamefont {Viale},\ and\ \citenamefont {Walczak}}]{cavagna_flocking_2015}%
  \BibitemOpen
  \bibfield  {author} {\bibinfo {author} {\bibfnamefont {A.}~\bibnamefont {Cavagna}}, \bibinfo {author} {\bibfnamefont {L.}~\bibnamefont {Del~Castello}}, \bibinfo {author} {\bibfnamefont {I.}~\bibnamefont {Giardina}}, \bibinfo {author} {\bibfnamefont {T.}~\bibnamefont {Grigera}}, \bibinfo {author} {\bibfnamefont {A.}~\bibnamefont {Jelic}}, \bibinfo {author} {\bibfnamefont {S.}~\bibnamefont {Melillo}}, \bibinfo {author} {\bibfnamefont {T.}~\bibnamefont {Mora}}, \bibinfo {author} {\bibfnamefont {L.}~\bibnamefont {Parisi}}, \bibinfo {author} {\bibfnamefont {E.}~\bibnamefont {Silvestri}}, \bibinfo {author} {\bibfnamefont {M.}~\bibnamefont {Viale}},\ and\ \bibinfo {author} {\bibfnamefont {A.~M.}\ \bibnamefont {Walczak}},\ }\bibfield  {title} {\bibinfo {title} {Flocking and {{Turning}}: A {{New Model}} for {{Self-organized Collective Motion}}},\ }\href {https://doi.org/10.1007/s10955-014-1119-3} {\bibfield  {journal} {\bibinfo  {journal} {Journal of Statistical Physics}\ }\textbf {\bibinfo {volume} {158}},\
  \bibinfo {pages} {601} (\bibinfo {year} {2015})}\BibitemShut {NoStop}%
\bibitem [{\citenamefont {Yang}\ and\ \citenamefont {Marchetti}(2015)}]{yang_hydrodynamics_2015}%
  \BibitemOpen
  \bibfield  {author} {\bibinfo {author} {\bibfnamefont {X.}~\bibnamefont {Yang}}\ and\ \bibinfo {author} {\bibfnamefont {M.~C.}\ \bibnamefont {Marchetti}},\ }\bibfield  {title} {\bibinfo {title} {Hydrodynamics of {{Turning Flocks}}},\ }\href {https://doi.org/10.1103/PhysRevLett.115.258101} {\bibfield  {journal} {\bibinfo  {journal} {Physical Review Letters}\ }\textbf {\bibinfo {volume} {115}},\ \bibinfo {pages} {258101} (\bibinfo {year} {2015})}\BibitemShut {NoStop}%
\bibitem [{\citenamefont {Mecke}\ \emph {et~al.}(2023)\citenamefont {Mecke}, \citenamefont {Gao}, \citenamefont {Ram{\'i}rez~Medina}, \citenamefont {Aarts}, \citenamefont {Gompper},\ and\ \citenamefont {Ripoll}}]{meckeSimultaneousEmergenceActive2023}%
  \BibitemOpen
  \bibfield  {author} {\bibinfo {author} {\bibfnamefont {J.}~\bibnamefont {Mecke}}, \bibinfo {author} {\bibfnamefont {Y.}~\bibnamefont {Gao}}, \bibinfo {author} {\bibfnamefont {C.~A.}\ \bibnamefont {Ram{\'i}rez~Medina}}, \bibinfo {author} {\bibfnamefont {D.~G. A.~L.}\ \bibnamefont {Aarts}}, \bibinfo {author} {\bibfnamefont {G.}~\bibnamefont {Gompper}},\ and\ \bibinfo {author} {\bibfnamefont {M.}~\bibnamefont {Ripoll}},\ }\bibfield  {title} {\bibinfo {title} {Simultaneous emergence of active turbulence and odd viscosity in a colloidal chiral active system},\ }\href {https://doi.org/10.1038/s42005-023-01442-3} {\bibfield  {journal} {\bibinfo  {journal} {Communications Physics}\ }\textbf {\bibinfo {volume} {6}},\ \bibinfo {pages} {324} (\bibinfo {year} {2023})}\BibitemShut {NoStop}%
\bibitem [{\citenamefont {Mecke}\ \emph {et~al.}(2024{\natexlab{a}})\citenamefont {Mecke}, \citenamefont {Gao}, \citenamefont {Gompper},\ and\ \citenamefont {Ripoll}}]{meckeChiralActiveSystems2024}%
  \BibitemOpen
  \bibfield  {author} {\bibinfo {author} {\bibfnamefont {J.}~\bibnamefont {Mecke}}, \bibinfo {author} {\bibfnamefont {Y.}~\bibnamefont {Gao}}, \bibinfo {author} {\bibfnamefont {G.}~\bibnamefont {Gompper}},\ and\ \bibinfo {author} {\bibfnamefont {M.}~\bibnamefont {Ripoll}},\ }\bibfield  {title} {\bibinfo {title} {Chiral active systems near a substrate: {{Emergent}} damping length controlled by fluid friction},\ }\href {https://doi.org/10.1038/s42005-024-01817-0} {\bibfield  {journal} {\bibinfo  {journal} {Communications Physics}\ }\textbf {\bibinfo {volume} {7}},\ \bibinfo {pages} {332} (\bibinfo {year} {2024}{\natexlab{a}})}\BibitemShut {NoStop}%
\bibitem [{\citenamefont {{de Wit}}\ \emph {et~al.}(2024)\citenamefont {{de Wit}}, \citenamefont {Fruchart}, \citenamefont {Khain}, \citenamefont {Toschi},\ and\ \citenamefont {Vitelli}}]{dewitPatternFormationTurbulent2024}%
  \BibitemOpen
  \bibfield  {author} {\bibinfo {author} {\bibfnamefont {X.~M.}\ \bibnamefont {{de Wit}}}, \bibinfo {author} {\bibfnamefont {M.}~\bibnamefont {Fruchart}}, \bibinfo {author} {\bibfnamefont {T.}~\bibnamefont {Khain}}, \bibinfo {author} {\bibfnamefont {F.}~\bibnamefont {Toschi}},\ and\ \bibinfo {author} {\bibfnamefont {V.}~\bibnamefont {Vitelli}},\ }\bibfield  {title} {\bibinfo {title} {Pattern formation by turbulent cascades},\ }\href {https://doi.org/10.1038/s41586-024-07074-z} {\bibfield  {journal} {\bibinfo  {journal} {Nature}\ }\textbf {\bibinfo {volume} {627}},\ \bibinfo {pages} {515} (\bibinfo {year} {2024})}\BibitemShut {NoStop}%
\bibitem [{\citenamefont {Chen}\ \emph {et~al.}(2024)\citenamefont {Chen}, \citenamefont {Zhang}, \citenamefont {Wang}, \citenamefont {Sun},\ and\ \citenamefont {Meng}}]{chen_phases_2024}%
  \BibitemOpen
  \bibfield  {author} {\bibinfo {author} {\bibfnamefont {B.-B.}\ \bibnamefont {Chen}}, \bibinfo {author} {\bibfnamefont {X.}~\bibnamefont {Zhang}}, \bibinfo {author} {\bibfnamefont {Y.}~\bibnamefont {Wang}}, \bibinfo {author} {\bibfnamefont {K.}~\bibnamefont {Sun}},\ and\ \bibinfo {author} {\bibfnamefont {Z.~Y.}\ \bibnamefont {Meng}},\ }\bibfield  {title} {\bibinfo {title} {Phases of \$(2+1)\textbackslash mathrm\textbraceleft{{D}}\textbraceright\$ {{SO}}(5) {{Nonlinear Sigma Model}} with a {{Topological Term}} on a {{Sphere}}: {{Multicritical Point}} and {{Disorder Phase}}},\ }\href {https://doi.org/10.1103/PhysRevLett.132.246503} {\bibfield  {journal} {\bibinfo  {journal} {Physical Review Letters}\ }\textbf {\bibinfo {volume} {132}},\ \bibinfo {pages} {246503} (\bibinfo {year} {2024})}\BibitemShut {NoStop}%
\bibitem [{\citenamefont {Toner}\ and\ \citenamefont {Tu}(1998)}]{toner_flocks_1998}%
  \BibitemOpen
  \bibfield  {author} {\bibinfo {author} {\bibfnamefont {J.}~\bibnamefont {Toner}}\ and\ \bibinfo {author} {\bibfnamefont {Y.}~\bibnamefont {Tu}},\ }\bibfield  {title} {\bibinfo {title} {Flocks, herds, and schools: {{A}} quantitative theory of flocking},\ }\href {https://doi.org/10.1103/PhysRevE.58.4828} {\bibfield  {journal} {\bibinfo  {journal} {Physical Review E}\ }\textbf {\bibinfo {volume} {58}},\ \bibinfo {pages} {4828} (\bibinfo {year} {1998})}\BibitemShut {NoStop}%
\bibitem [{\citenamefont {Acebr{\'o}n}\ \emph {et~al.}(2005)\citenamefont {Acebr{\'o}n}, \citenamefont {Bonilla}, \citenamefont {P{\'e}rez~Vicente}, \citenamefont {Ritort},\ and\ \citenamefont {Spigler}}]{acebron_kuramoto_2005}%
  \BibitemOpen
  \bibfield  {author} {\bibinfo {author} {\bibfnamefont {J.~A.}\ \bibnamefont {Acebr{\'o}n}}, \bibinfo {author} {\bibfnamefont {L.~L.}\ \bibnamefont {Bonilla}}, \bibinfo {author} {\bibfnamefont {C.~J.}\ \bibnamefont {P{\'e}rez~Vicente}}, \bibinfo {author} {\bibfnamefont {F.}~\bibnamefont {Ritort}},\ and\ \bibinfo {author} {\bibfnamefont {R.}~\bibnamefont {Spigler}},\ }\bibfield  {title} {\bibinfo {title} {The {{Kuramoto}} model: {{A}} simple paradigm for synchronization phenomena},\ }\href {https://doi.org/10.1103/RevModPhys.77.137} {\bibfield  {journal} {\bibinfo  {journal} {Reviews of Modern Physics}\ }\textbf {\bibinfo {volume} {77}},\ \bibinfo {pages} {137} (\bibinfo {year} {2005})}\BibitemShut {NoStop}%
\bibitem [{\citenamefont {De~Smet}\ and\ \citenamefont {Aeyels}(2007)}]{de_smet_partial_2007}%
  \BibitemOpen
  \bibfield  {author} {\bibinfo {author} {\bibfnamefont {F.}~\bibnamefont {De~Smet}}\ and\ \bibinfo {author} {\bibfnamefont {D.}~\bibnamefont {Aeyels}},\ }\bibfield  {title} {\bibinfo {title} {Partial entrainment in the finite {{Kuramoto}}--{{Sakaguchi}} model},\ }\href {https://doi.org/10.1016/j.physd.2007.06.025} {\bibfield  {journal} {\bibinfo  {journal} {Physica D: Nonlinear Phenomena}\ }\textbf {\bibinfo {volume} {234}},\ \bibinfo {pages} {81} (\bibinfo {year} {2007})}\BibitemShut {NoStop}%
\bibitem [{\citenamefont {Ackermann}(2015)}]{ackermann_functional_2015}%
  \BibitemOpen
  \bibfield  {author} {\bibinfo {author} {\bibfnamefont {M.}~\bibnamefont {Ackermann}},\ }\bibfield  {title} {\bibinfo {title} {A functional perspective on phenotypic heterogeneity in microorganisms},\ }\href {https://doi.org/10.1038/nrmicro3491} {\bibfield  {journal} {\bibinfo  {journal} {Nature Reviews Microbiology}\ }\textbf {\bibinfo {volume} {13}},\ \bibinfo {pages} {497} (\bibinfo {year} {2015})}\BibitemShut {NoStop}%
\bibitem [{\citenamefont {Josephson}(1962)}]{josephson_possible_1962}%
  \BibitemOpen
  \bibfield  {author} {\bibinfo {author} {\bibfnamefont {B.~D.}\ \bibnamefont {Josephson}},\ }\bibfield  {title} {\bibinfo {title} {Possible new effects in superconductive tunnelling},\ }\href {https://doi.org/10.1016/0031-9163(62)91369-0} {\bibfield  {journal} {\bibinfo  {journal} {Physics Letters}\ }\textbf {\bibinfo {volume} {1}},\ \bibinfo {pages} {251} (\bibinfo {year} {1962})}\BibitemShut {NoStop}%
\bibitem [{\citenamefont {Tinkham}(2004)}]{tinkham_introduction_2004}%
  \BibitemOpen
  \bibfield  {author} {\bibinfo {author} {\bibfnamefont {M.}~\bibnamefont {Tinkham}},\ }\href@noop {} {\emph {\bibinfo {title} {Introduction to {{Superconductivity}}}}}\ (\bibinfo  {publisher} {Courier Corporation},\ \bibinfo {year} {2004})\BibitemShut {NoStop}%
\bibitem [{\citenamefont {Wiesenfeld}\ \emph {et~al.}(1996)\citenamefont {Wiesenfeld}, \citenamefont {Colet},\ and\ \citenamefont {Strogatz}}]{wiesenfeld_synchronization_1996}%
  \BibitemOpen
  \bibfield  {author} {\bibinfo {author} {\bibfnamefont {K.}~\bibnamefont {Wiesenfeld}}, \bibinfo {author} {\bibfnamefont {P.}~\bibnamefont {Colet}},\ and\ \bibinfo {author} {\bibfnamefont {S.~H.}\ \bibnamefont {Strogatz}},\ }\bibfield  {title} {\bibinfo {title} {Synchronization {{Transitions}} in a {{Disordered Josephson Series Array}}},\ }\href {https://doi.org/10.1103/PhysRevLett.76.404} {\bibfield  {journal} {\bibinfo  {journal} {Physical Review Letters}\ }\textbf {\bibinfo {volume} {76}},\ \bibinfo {pages} {404} (\bibinfo {year} {1996})}\BibitemShut {NoStop}%
\bibitem [{\citenamefont {Ambegaokar}\ and\ \citenamefont {Halperin}(1969)}]{ambegaokar_voltage_1969}%
  \BibitemOpen
  \bibfield  {author} {\bibinfo {author} {\bibfnamefont {V.}~\bibnamefont {Ambegaokar}}\ and\ \bibinfo {author} {\bibfnamefont {B.~I.}\ \bibnamefont {Halperin}},\ }\bibfield  {title} {\bibinfo {title} {Voltage {{Due}} to {{Thermal Noise}} in the dc {{Josephson Effect}}},\ }\href {https://doi.org/10.1103/PhysRevLett.22.1364} {\bibfield  {journal} {\bibinfo  {journal} {Physical Review Letters}\ }\textbf {\bibinfo {volume} {22}},\ \bibinfo {pages} {1364} (\bibinfo {year} {1969})}\BibitemShut {NoStop}%
\bibitem [{\citenamefont {Stewart}(1968)}]{stewart_current-voltage_1968}%
  \BibitemOpen
  \bibfield  {author} {\bibinfo {author} {\bibfnamefont {W.~C.}\ \bibnamefont {Stewart}},\ }\bibfield  {title} {\bibinfo {title} {Current-{{Voltage Characteristics}} of {{Josephson Junctions}}},\ }\href {https://doi.org/10.1063/1.1651991} {\bibfield  {journal} {\bibinfo  {journal} {Applied Physics Letters}\ }\textbf {\bibinfo {volume} {12}},\ \bibinfo {pages} {277} (\bibinfo {year} {1968})}\BibitemShut {NoStop}%
\bibitem [{\citenamefont {McCumber}(1968)}]{mccumber_effect_1968}%
  \BibitemOpen
  \bibfield  {author} {\bibinfo {author} {\bibfnamefont {D.~E.}\ \bibnamefont {McCumber}},\ }\bibfield  {title} {\bibinfo {title} {Effect of ac {{Impedance}} on dc {{Voltage}}-{{Current Characteristics}} of {{Superconductor Weak}}-{{Link Junctions}}},\ }\href {https://doi.org/10.1063/1.1656743} {\bibfield  {journal} {\bibinfo  {journal} {Journal of Applied Physics}\ }\textbf {\bibinfo {volume} {39}},\ \bibinfo {pages} {3113} (\bibinfo {year} {1968})}\BibitemShut {NoStop}%
\bibitem [{\citenamefont {Danner}\ \emph {et~al.}(2021)\citenamefont {Danner}, \citenamefont {Padurariu}, \citenamefont {Ankerhold},\ and\ \citenamefont {Kubala}}]{danner_injection_2021}%
  \BibitemOpen
  \bibfield  {author} {\bibinfo {author} {\bibfnamefont {L.}~\bibnamefont {Danner}}, \bibinfo {author} {\bibfnamefont {C.}~\bibnamefont {Padurariu}}, \bibinfo {author} {\bibfnamefont {J.}~\bibnamefont {Ankerhold}},\ and\ \bibinfo {author} {\bibfnamefont {B.}~\bibnamefont {Kubala}},\ }\bibfield  {title} {\bibinfo {title} {Injection locking and synchronization in {{Josephson}} photonics devices},\ }\href {https://doi.org/10.1103/PhysRevB.104.054517} {\bibfield  {journal} {\bibinfo  {journal} {Physical Review B}\ }\textbf {\bibinfo {volume} {104}},\ \bibinfo {pages} {054517} (\bibinfo {year} {2021})}\BibitemShut {NoStop}%
\bibitem [{\citenamefont {Fang}\ \emph {et~al.}(2023)\citenamefont {Fang}, \citenamefont {Han}, \citenamefont {Chesi},\ and\ \citenamefont {Choi}}]{fang_subgap_2023}%
  \BibitemOpen
  \bibfield  {author} {\bibinfo {author} {\bibfnamefont {Y.}~\bibnamefont {Fang}}, \bibinfo {author} {\bibfnamefont {S.}~\bibnamefont {Han}}, \bibinfo {author} {\bibfnamefont {S.}~\bibnamefont {Chesi}},\ and\ \bibinfo {author} {\bibfnamefont {M.-S.}\ \bibnamefont {Choi}},\ }\bibfield  {title} {\bibinfo {title} {Subgap modes in two-dimensional magnetic {{Josephson}} junctions},\ }\href {https://doi.org/10.1103/PhysRevB.107.115114} {\bibfield  {journal} {\bibinfo  {journal} {Physical Review B}\ }\textbf {\bibinfo {volume} {107}},\ \bibinfo {pages} {115114} (\bibinfo {year} {2023})}\BibitemShut {NoStop}%
\bibitem [{\citenamefont {Marov}\ and\ \citenamefont {Kolesnichenko}(2013)}]{marov_self-organization_2013}%
  \BibitemOpen
  \bibfield  {author} {\bibinfo {author} {\bibfnamefont {M.~Y.}\ \bibnamefont {Marov}}\ and\ \bibinfo {author} {\bibfnamefont {A.~V.}\ \bibnamefont {Kolesnichenko}},\ }\bibfield  {title} {\bibinfo {title} {Self-{{Organization}} of {{Developed Turbulence}} and {{Formation Mechanisms}} of {{Coherent Structures}}},\ }in\ \href {https://doi.org/10.1007/978-1-4614-5155-6_6} {\emph {\bibinfo {booktitle} {Turbulence and {{Self-Organization}}: {{Modeling Astrophysical Objects}}}}},\ \bibinfo {editor} {edited by\ \bibinfo {editor} {\bibfnamefont {M.~Y.}\ \bibnamefont {Marov}}\ and\ \bibinfo {editor} {\bibfnamefont {A.~V.}\ \bibnamefont {Kolesnichenko}}}\ (\bibinfo  {publisher} {Springer},\ \bibinfo {address} {New York, NY},\ \bibinfo {year} {2013})\ pp.\ \bibinfo {pages} {373--423}\BibitemShut {NoStop}%
\bibitem [{\citenamefont {Benz}\ and\ \citenamefont {Burroughs}(1991)}]{benz_coherent_1991}%
  \BibitemOpen
  \bibfield  {author} {\bibinfo {author} {\bibfnamefont {S.~P.}\ \bibnamefont {Benz}}\ and\ \bibinfo {author} {\bibfnamefont {C.~J.}\ \bibnamefont {Burroughs}},\ }\bibfield  {title} {\bibinfo {title} {Coherent emission from two-dimensional {{Josephson}} junction arrays},\ }\href {https://doi.org/10.1063/1.104993} {\bibfield  {journal} {\bibinfo  {journal} {Applied Physics Letters}\ }\textbf {\bibinfo {volume} {58}},\ \bibinfo {pages} {2162} (\bibinfo {year} {1991})}\BibitemShut {NoStop}%
\bibitem [{\citenamefont {Kurtscheid}\ \emph {et~al.}(2025)\citenamefont {Kurtscheid}, \citenamefont {Redmann}, \citenamefont {Vewinger}, \citenamefont {Schmitt},\ and\ \citenamefont {Weitz}}]{kurtscheid_thermodynamics_2025}%
  \BibitemOpen
  \bibfield  {author} {\bibinfo {author} {\bibfnamefont {C.}~\bibnamefont {Kurtscheid}}, \bibinfo {author} {\bibfnamefont {A.}~\bibnamefont {Redmann}}, \bibinfo {author} {\bibfnamefont {F.}~\bibnamefont {Vewinger}}, \bibinfo {author} {\bibfnamefont {J.}~\bibnamefont {Schmitt}},\ and\ \bibinfo {author} {\bibfnamefont {M.}~\bibnamefont {Weitz}},\ }\bibfield  {title} {\bibinfo {title} {Thermodynamics and {{State Preparation}} in a {{Two-State System}} of {{Light}}},\ }\href {https://doi.org/10.1103/kynj-l87s} {\bibfield  {journal} {\bibinfo  {journal} {Physical Review Letters}\ }\textbf {\bibinfo {volume} {135}},\ \bibinfo {pages} {160406} (\bibinfo {year} {2025})}\BibitemShut {NoStop}%
\bibitem [{\citenamefont {Bhansali}\ and\ \citenamefont {Roychowdhury}(2009)}]{bhansali_gen-adler_2009}%
  \BibitemOpen
  \bibfield  {author} {\bibinfo {author} {\bibfnamefont {P.}~\bibnamefont {Bhansali}}\ and\ \bibinfo {author} {\bibfnamefont {J.}~\bibnamefont {Roychowdhury}},\ }\bibfield  {title} {\bibinfo {title} {Gen-{{Adler}}: {{The}} generalized {{Adler}}'s equation for injection locking analysis in oscillators},\ }in\ \href {https://doi.org/10.1109/ASPDAC.2009.4796533} {\emph {\bibinfo {booktitle} {2009 {{Asia}} and {{South Pacific Design Automation Conference}}}}}\ (\bibinfo {year} {2009})\ pp.\ \bibinfo {pages} {522--527}\BibitemShut {NoStop}%
\bibitem [{\citenamefont {Gandhi}\ \emph {et~al.}(2015)\citenamefont {Gandhi}, \citenamefont {Knobloch},\ and\ \citenamefont {Beaume}}]{gandhi_dynamics_2015}%
  \BibitemOpen
  \bibfield  {author} {\bibinfo {author} {\bibfnamefont {P.}~\bibnamefont {Gandhi}}, \bibinfo {author} {\bibfnamefont {E.}~\bibnamefont {Knobloch}},\ and\ \bibinfo {author} {\bibfnamefont {C.}~\bibnamefont {Beaume}},\ }\bibfield  {title} {\bibinfo {title} {Dynamics of phase slips in systems with time-periodic modulation},\ }\href {https://doi.org/10.1103/PhysRevE.92.062914} {\bibfield  {journal} {\bibinfo  {journal} {Physical Review E}\ }\textbf {\bibinfo {volume} {92}},\ \bibinfo {pages} {062914} (\bibinfo {year} {2015})}\BibitemShut {NoStop}%
\bibitem [{\citenamefont {Likharev}(1986)}]{likharev_dynamics_1986}%
  \BibitemOpen
  \bibfield  {author} {\bibinfo {author} {\bibfnamefont {K.~K.}\ \bibnamefont {Likharev}},\ }\href@noop {} {\emph {\bibinfo {title} {Dynamics of {{Josephson Junctions}} and {{Circuits}}}}}\ (\bibinfo  {publisher} {Routledge},\ \bibinfo {address} {London},\ \bibinfo {year} {1986})\BibitemShut {NoStop}%
\bibitem [{\citenamefont {Newrock}\ \emph {et~al.}(2000)\citenamefont {Newrock}, \citenamefont {Lobb}, \citenamefont {Geigenm{\"u}ller},\ and\ \citenamefont {Octavio}}]{newrock_two-dimensional_2000}%
  \BibitemOpen
  \bibfield  {author} {\bibinfo {author} {\bibfnamefont {R.~S.}\ \bibnamefont {Newrock}}, \bibinfo {author} {\bibfnamefont {C.~J.}\ \bibnamefont {Lobb}}, \bibinfo {author} {\bibfnamefont {U.}~\bibnamefont {Geigenm{\"u}ller}},\ and\ \bibinfo {author} {\bibfnamefont {M.}~\bibnamefont {Octavio}},\ }\bibfield  {title} {\bibinfo {title} {The two-dimensional physics of {{Josephson}} junction arrays},\ }in\ \href@noop {} {\emph {\bibinfo {booktitle} {Solid {{State Physics}}}}},\ Vol.~\bibinfo {volume} {54}\ (\bibinfo  {publisher} {Elsevier},\ \bibinfo {year} {2000})\ pp.\ \bibinfo {pages} {263--512}\BibitemShut {NoStop}%
\bibitem [{\citenamefont {Penrose}(1960)}]{penrose_electrostatic_1960}%
  \BibitemOpen
  \bibfield  {author} {\bibinfo {author} {\bibfnamefont {O.}~\bibnamefont {Penrose}},\ }\bibfield  {title} {\bibinfo {title} {Electrostatic {{Instabilities}} of a {{Uniform Non}}-{{Maxwellian Plasma}}},\ }\href {https://doi.org/10.1063/1.1706024} {\bibfield  {journal} {\bibinfo  {journal} {The Physics of Fluids}\ }\textbf {\bibinfo {volume} {3}},\ \bibinfo {pages} {258} (\bibinfo {year} {1960})}\BibitemShut {NoStop}%
\bibitem [{\citenamefont {Strogatz}\ \emph {et~al.}(1992)\citenamefont {Strogatz}, \citenamefont {Mirollo},\ and\ \citenamefont {Matthews}}]{strogatz_coupled_1992}%
  \BibitemOpen
  \bibfield  {author} {\bibinfo {author} {\bibfnamefont {S.~H.}\ \bibnamefont {Strogatz}}, \bibinfo {author} {\bibfnamefont {R.~E.}\ \bibnamefont {Mirollo}},\ and\ \bibinfo {author} {\bibfnamefont {P.~C.}\ \bibnamefont {Matthews}},\ }\bibfield  {title} {\bibinfo {title} {Coupled nonlinear oscillators below the synchronization threshold: {{Relaxation}} by generalized {{Landau}} damping},\ }\href {https://doi.org/10.1103/PhysRevLett.68.2730} {\bibfield  {journal} {\bibinfo  {journal} {Physical Review Letters}\ }\textbf {\bibinfo {volume} {68}},\ \bibinfo {pages} {2730} (\bibinfo {year} {1992})}\BibitemShut {NoStop}%
\bibitem [{\citenamefont {Carnevale}\ \emph {et~al.}(1991)\citenamefont {Carnevale}, \citenamefont {McWilliams}, \citenamefont {Pomeau}, \citenamefont {Weiss},\ and\ \citenamefont {Young}}]{carnevale_evolution_1991}%
  \BibitemOpen
  \bibfield  {author} {\bibinfo {author} {\bibfnamefont {G.~F.}\ \bibnamefont {Carnevale}}, \bibinfo {author} {\bibfnamefont {J.~C.}\ \bibnamefont {McWilliams}}, \bibinfo {author} {\bibfnamefont {Y.}~\bibnamefont {Pomeau}}, \bibinfo {author} {\bibfnamefont {J.~B.}\ \bibnamefont {Weiss}},\ and\ \bibinfo {author} {\bibfnamefont {W.~R.}\ \bibnamefont {Young}},\ }\bibfield  {title} {\bibinfo {title} {Evolution of vortex statistics in two-dimensional turbulence},\ }\href {https://doi.org/10.1103/PhysRevLett.66.2735} {\bibfield  {journal} {\bibinfo  {journal} {Physical Review Letters}\ }\textbf {\bibinfo {volume} {66}},\ \bibinfo {pages} {2735} (\bibinfo {year} {1991})}\BibitemShut {NoStop}%
\bibitem [{\citenamefont {Larichev}\ and\ \citenamefont {McWilliams}(1991)}]{larichev_weakly_1991}%
  \BibitemOpen
  \bibfield  {author} {\bibinfo {author} {\bibfnamefont {V.~D.}\ \bibnamefont {Larichev}}\ and\ \bibinfo {author} {\bibfnamefont {J.~C.}\ \bibnamefont {McWilliams}},\ }\bibfield  {title} {\bibinfo {title} {Weakly decaying turbulence in an equivalent-barotropic fluid},\ }\href {https://doi.org/10.1063/1.857970} {\bibfield  {journal} {\bibinfo  {journal} {Physics of Fluids A: Fluid Dynamics}\ }\textbf {\bibinfo {volume} {3}},\ \bibinfo {pages} {938} (\bibinfo {year} {1991})}\BibitemShut {NoStop}%
\bibitem [{\citenamefont {D'Errico}(2009)}]{derrico_slm-shape_2009}%
  \BibitemOpen
  \bibfield  {author} {\bibinfo {author} {\bibfnamefont {J.}~\bibnamefont {D'Errico}},\ }\bibfield  {title} {\bibinfo {title} {{{SLM-shape}} language modeling},\ }\href@noop {} {\bibfield  {journal} {\bibinfo  {journal} {SLM-Shape Language Modeling.. http://www. mathworks. com/matlabcentral/fileexchange/24443-slm-shape-language-modeling: Mathworks}\ } (\bibinfo {year} {2009})}\BibitemShut {NoStop}%
\bibitem [{\citenamefont {Lakshmanan}(2011)}]{lakshmanan_fascinating_2011}%
  \BibitemOpen
  \bibfield  {author} {\bibinfo {author} {\bibfnamefont {M.}~\bibnamefont {Lakshmanan}},\ }\bibfield  {title} {\bibinfo {title} {The fascinating world of the {{Landau}}--{{Lifshitz}}--{{Gilbert}} equation: An overview},\ }\href {https://doi.org/10.1098/rsta.2010.0319} {\bibfield  {journal} {\bibinfo  {journal} {Philosophical Transactions of the Royal Society A: Mathematical, Physical and Engineering Sciences}\ }\textbf {\bibinfo {volume} {369}},\ \bibinfo {pages} {1280} (\bibinfo {year} {2011})}\BibitemShut {NoStop}%
\bibitem [{\citenamefont {Johnson}\ and\ \citenamefont {Silsbee}(1985)}]{johnson_interfacial_1985}%
  \BibitemOpen
  \bibfield  {author} {\bibinfo {author} {\bibfnamefont {M.}~\bibnamefont {Johnson}}\ and\ \bibinfo {author} {\bibfnamefont {R.~H.}\ \bibnamefont {Silsbee}},\ }\bibfield  {title} {\bibinfo {title} {Interfacial charge-spin coupling: {{Injection}} and detection of spin magnetization in metals},\ }\href {https://doi.org/10.1103/PhysRevLett.55.1790} {\bibfield  {journal} {\bibinfo  {journal} {Physical Review Letters}\ }\textbf {\bibinfo {volume} {55}},\ \bibinfo {pages} {1790} (\bibinfo {year} {1985})}\BibitemShut {NoStop}%
\bibitem [{\citenamefont {Ingold}\ \emph {et~al.}(1994)\citenamefont {Ingold}, \citenamefont {Grabert},\ and\ \citenamefont {Eberhardt}}]{ingold_cooper-pair_1994}%
  \BibitemOpen
  \bibfield  {author} {\bibinfo {author} {\bibfnamefont {G.-L.}\ \bibnamefont {Ingold}}, \bibinfo {author} {\bibfnamefont {H.}~\bibnamefont {Grabert}},\ and\ \bibinfo {author} {\bibfnamefont {U.}~\bibnamefont {Eberhardt}},\ }\bibfield  {title} {\bibinfo {title} {Cooper-pair current through ultrasmall {{Josephson}} junctions},\ }\href {https://doi.org/10.1103/PhysRevB.50.395} {\bibfield  {journal} {\bibinfo  {journal} {Physical Review B}\ }\textbf {\bibinfo {volume} {50}},\ \bibinfo {pages} {395} (\bibinfo {year} {1994})}\BibitemShut {NoStop}%
\bibitem [{\citenamefont {Dadhichi}\ \emph {et~al.}(2020)\citenamefont {Dadhichi}, \citenamefont {Kethapelli}, \citenamefont {Chajwa}, \citenamefont {Ramaswamy},\ and\ \citenamefont {Maitra}}]{dadhichi_nonmutual_2020}%
  \BibitemOpen
  \bibfield  {author} {\bibinfo {author} {\bibfnamefont {L.~P.}\ \bibnamefont {Dadhichi}}, \bibinfo {author} {\bibfnamefont {J.}~\bibnamefont {Kethapelli}}, \bibinfo {author} {\bibfnamefont {R.}~\bibnamefont {Chajwa}}, \bibinfo {author} {\bibfnamefont {S.}~\bibnamefont {Ramaswamy}},\ and\ \bibinfo {author} {\bibfnamefont {A.}~\bibnamefont {Maitra}},\ }\bibfield  {title} {\bibinfo {title} {Nonmutual torques and the unimportance of motility for long-range order in two-dimensional flocks},\ }\href {https://doi.org/10.1103/PhysRevE.101.052601} {\bibfield  {journal} {\bibinfo  {journal} {Physical Review E}\ }\textbf {\bibinfo {volume} {101}},\ \bibinfo {pages} {052601} (\bibinfo {year} {2020})}\BibitemShut {NoStop}%
\bibitem [{\citenamefont {Cavagna}\ \emph {et~al.}(2013)\citenamefont {Cavagna}, \citenamefont {Queir{\'o}s}, \citenamefont {Giardina}, \citenamefont {Stefanini},\ and\ \citenamefont {Viale}}]{cavagna_diffusion_2013}%
  \BibitemOpen
  \bibfield  {author} {\bibinfo {author} {\bibfnamefont {A.}~\bibnamefont {Cavagna}}, \bibinfo {author} {\bibfnamefont {S.~M.~D.}\ \bibnamefont {Queir{\'o}s}}, \bibinfo {author} {\bibfnamefont {I.}~\bibnamefont {Giardina}}, \bibinfo {author} {\bibfnamefont {F.}~\bibnamefont {Stefanini}},\ and\ \bibinfo {author} {\bibfnamefont {M.}~\bibnamefont {Viale}},\ }\bibfield  {title} {\bibinfo {title} {Diffusion of individual birds in starling flocks},\ }\href {https://doi.org/10.1098/rspb.2012.2484} {\bibfield  {journal} {\bibinfo  {journal} {Proceedings of the Royal Society B: Biological Sciences}\ }\textbf {\bibinfo {volume} {280}},\ \bibinfo {pages} {20122484} (\bibinfo {year} {2013})}\BibitemShut {NoStop}%
\bibitem [{\citenamefont {Miller}\ and\ \citenamefont {Ouellette}(2014)}]{miller_impact_2014}%
  \BibitemOpen
  \bibfield  {author} {\bibinfo {author} {\bibfnamefont {P.~W.}\ \bibnamefont {Miller}}\ and\ \bibinfo {author} {\bibfnamefont {N.~T.}\ \bibnamefont {Ouellette}},\ }\bibfield  {title} {\bibinfo {title} {Impact fragmentation of model flocks},\ }\href {https://doi.org/10.1103/PhysRevE.89.042806} {\bibfield  {journal} {\bibinfo  {journal} {Physical Review E}\ }\textbf {\bibinfo {volume} {89}},\ \bibinfo {pages} {042806} (\bibinfo {year} {2014})}\BibitemShut {NoStop}%
\bibitem [{\citenamefont {Attanasi}\ \emph {et~al.}(2014)\citenamefont {Attanasi}, \citenamefont {Cavagna}, \citenamefont {Del~Castello}, \citenamefont {Giardina}, \citenamefont {Grigera}, \citenamefont {Jeli{\'c}}, \citenamefont {Melillo}, \citenamefont {Parisi}, \citenamefont {Pohl}, \citenamefont {Shen},\ and\ \citenamefont {Viale}}]{attanasi_information_2014}%
  \BibitemOpen
  \bibfield  {author} {\bibinfo {author} {\bibfnamefont {A.}~\bibnamefont {Attanasi}}, \bibinfo {author} {\bibfnamefont {A.}~\bibnamefont {Cavagna}}, \bibinfo {author} {\bibfnamefont {L.}~\bibnamefont {Del~Castello}}, \bibinfo {author} {\bibfnamefont {I.}~\bibnamefont {Giardina}}, \bibinfo {author} {\bibfnamefont {T.~S.}\ \bibnamefont {Grigera}}, \bibinfo {author} {\bibfnamefont {A.}~\bibnamefont {Jeli{\'c}}}, \bibinfo {author} {\bibfnamefont {S.}~\bibnamefont {Melillo}}, \bibinfo {author} {\bibfnamefont {L.}~\bibnamefont {Parisi}}, \bibinfo {author} {\bibfnamefont {O.}~\bibnamefont {Pohl}}, \bibinfo {author} {\bibfnamefont {E.}~\bibnamefont {Shen}},\ and\ \bibinfo {author} {\bibfnamefont {M.}~\bibnamefont {Viale}},\ }\bibfield  {title} {\bibinfo {title} {Information transfer and behavioural inertia in starling flocks},\ }\href {https://doi.org/10.1038/nphys3035} {\bibfield  {journal} {\bibinfo  {journal} {Nature Physics}\ }\textbf {\bibinfo {volume} {10}},\ \bibinfo {pages} {691} (\bibinfo {year}
  {2014})}\BibitemShut {NoStop}%
\bibitem [{\citenamefont {Cavagna}\ \emph {et~al.}(2010)\citenamefont {Cavagna}, \citenamefont {Cimarelli}, \citenamefont {Giardina}, \citenamefont {Parisi}, \citenamefont {Santagati}, \citenamefont {Stefanini},\ and\ \citenamefont {Viale}}]{cavagna_scale-free_2010}%
  \BibitemOpen
  \bibfield  {author} {\bibinfo {author} {\bibfnamefont {A.}~\bibnamefont {Cavagna}}, \bibinfo {author} {\bibfnamefont {A.}~\bibnamefont {Cimarelli}}, \bibinfo {author} {\bibfnamefont {I.}~\bibnamefont {Giardina}}, \bibinfo {author} {\bibfnamefont {G.}~\bibnamefont {Parisi}}, \bibinfo {author} {\bibfnamefont {R.}~\bibnamefont {Santagati}}, \bibinfo {author} {\bibfnamefont {F.}~\bibnamefont {Stefanini}},\ and\ \bibinfo {author} {\bibfnamefont {M.}~\bibnamefont {Viale}},\ }\bibfield  {title} {\bibinfo {title} {Scale-free correlations in starling flocks},\ }\href {https://doi.org/10.1073/pnas.1005766107} {\bibfield  {journal} {\bibinfo  {journal} {Proceedings of the National Academy of Sciences}\ }\textbf {\bibinfo {volume} {107}},\ \bibinfo {pages} {11865} (\bibinfo {year} {2010})}\BibitemShut {NoStop}%
\bibitem [{\citenamefont {Cavagna}\ \emph {et~al.}(2024)\citenamefont {Cavagna}, \citenamefont {Crist{\'i}n}, \citenamefont {Giardina}, \citenamefont {Grigera},\ and\ \citenamefont {Veca}}]{cavagna_discrete_2024}%
  \BibitemOpen
  \bibfield  {author} {\bibinfo {author} {\bibfnamefont {A.}~\bibnamefont {Cavagna}}, \bibinfo {author} {\bibfnamefont {J.}~\bibnamefont {Crist{\'i}n}}, \bibinfo {author} {\bibfnamefont {I.}~\bibnamefont {Giardina}}, \bibinfo {author} {\bibfnamefont {T.~S.}\ \bibnamefont {Grigera}},\ and\ \bibinfo {author} {\bibfnamefont {M.}~\bibnamefont {Veca}},\ }\bibfield  {title} {\bibinfo {title} {Discrete {{Laplacian}} thermostat for flocks and swarms: The fully conserved {{Inertial Spin Model}}},\ }\href {https://doi.org/10.1088/1751-8121/ad7ca0} {\bibfield  {journal} {\bibinfo  {journal} {Journal of Physics A: Mathematical and Theoretical}\ }\textbf {\bibinfo {volume} {57}},\ \bibinfo {pages} {415002} (\bibinfo {year} {2024})}\BibitemShut {NoStop}%
\bibitem [{\citenamefont {Mahault}\ \emph {et~al.}(2019)\citenamefont {Mahault}, \citenamefont {Ginelli},\ and\ \citenamefont {Chat{\'e}}}]{mahault_quantitative_2019}%
  \BibitemOpen
  \bibfield  {author} {\bibinfo {author} {\bibfnamefont {B.}~\bibnamefont {Mahault}}, \bibinfo {author} {\bibfnamefont {F.}~\bibnamefont {Ginelli}},\ and\ \bibinfo {author} {\bibfnamefont {H.}~\bibnamefont {Chat{\'e}}},\ }\bibfield  {title} {\bibinfo {title} {Quantitative {{Assessment}} of the {{Toner}} and {{Tu Theory}} of {{Polar Flocks}}},\ }\href {https://doi.org/10.1103/PhysRevLett.123.218001} {\bibfield  {journal} {\bibinfo  {journal} {Physical Review Letters}\ }\textbf {\bibinfo {volume} {123}},\ \bibinfo {pages} {218001} (\bibinfo {year} {2019})}\BibitemShut {NoStop}%
\bibitem [{\citenamefont {Caprini}\ \emph {et~al.}(2024)\citenamefont {Caprini}, \citenamefont {Liebchen},\ and\ \citenamefont {L{\"o}wen}}]{caprini_self-reverting_2024}%
  \BibitemOpen
  \bibfield  {author} {\bibinfo {author} {\bibfnamefont {L.}~\bibnamefont {Caprini}}, \bibinfo {author} {\bibfnamefont {B.}~\bibnamefont {Liebchen}},\ and\ \bibinfo {author} {\bibfnamefont {H.}~\bibnamefont {L{\"o}wen}},\ }\bibfield  {title} {\bibinfo {title} {Self-reverting vortices in chiral active matter},\ }\href {https://doi.org/10.1038/s42005-024-01637-2} {\bibfield  {journal} {\bibinfo  {journal} {Communications Physics}\ }\textbf {\bibinfo {volume} {7}},\ \bibinfo {pages} {153} (\bibinfo {year} {2024})}\BibitemShut {NoStop}%
\bibitem [{\citenamefont {Chen}\ \emph {et~al.}(2025)\citenamefont {Chen}, \citenamefont {Weady}, \citenamefont {Atis}, \citenamefont {Matsuzawa}, \citenamefont {Shelley},\ and\ \citenamefont {Irvine}}]{chenSelfpropulsionFlockingChiral2025}%
  \BibitemOpen
  \bibfield  {author} {\bibinfo {author} {\bibfnamefont {P.}~\bibnamefont {Chen}}, \bibinfo {author} {\bibfnamefont {S.}~\bibnamefont {Weady}}, \bibinfo {author} {\bibfnamefont {S.}~\bibnamefont {Atis}}, \bibinfo {author} {\bibfnamefont {T.}~\bibnamefont {Matsuzawa}}, \bibinfo {author} {\bibfnamefont {M.~J.}\ \bibnamefont {Shelley}},\ and\ \bibinfo {author} {\bibfnamefont {W.~T.~M.}\ \bibnamefont {Irvine}},\ }\bibfield  {title} {\bibinfo {title} {Self-propulsion, flocking and chiral active phases from particles spinning at intermediate {{Reynolds}} numbers},\ }\href {https://doi.org/10.1038/s41567-024-02651-5} {\bibfield  {journal} {\bibinfo  {journal} {Nature Physics}\ }\textbf {\bibinfo {volume} {21}},\ \bibinfo {pages} {146} (\bibinfo {year} {2025})}\BibitemShut {NoStop}%
\bibitem [{\citenamefont {Kralj}\ \emph {et~al.}(2024)\citenamefont {Kralj}, \citenamefont {Ravnik},\ and\ \citenamefont {Kos}}]{kraljChiralityAnisotropicViscosity2024}%
  \BibitemOpen
  \bibfield  {author} {\bibinfo {author} {\bibfnamefont {N.}~\bibnamefont {Kralj}}, \bibinfo {author} {\bibfnamefont {M.}~\bibnamefont {Ravnik}},\ and\ \bibinfo {author} {\bibfnamefont {{\v Z}.}~\bibnamefont {Kos}},\ }\bibfield  {title} {\bibinfo {title} {Chirality, anisotropic viscosity and elastic anisotropy in three-dimensional active nematic turbulence},\ }\href {https://doi.org/10.1038/s42005-024-01720-8} {\bibfield  {journal} {\bibinfo  {journal} {Communications Physics}\ }\textbf {\bibinfo {volume} {7}},\ \bibinfo {pages} {222} (\bibinfo {year} {2024})}\BibitemShut {NoStop}%
\bibitem [{\citenamefont {Mecke}\ \emph {et~al.}(2024{\natexlab{b}})\citenamefont {Mecke}, \citenamefont {Nketsiah}, \citenamefont {Li},\ and\ \citenamefont {Gao}}]{meckeEmergentPhenomenaChiral2024}%
  \BibitemOpen
  \bibfield  {author} {\bibinfo {author} {\bibfnamefont {J.}~\bibnamefont {Mecke}}, \bibinfo {author} {\bibfnamefont {J.~O.}\ \bibnamefont {Nketsiah}}, \bibinfo {author} {\bibfnamefont {R.}~\bibnamefont {Li}},\ and\ \bibinfo {author} {\bibfnamefont {Y.}~\bibnamefont {Gao}},\ }\bibfield  {title} {\bibinfo {title} {Emergent phenomena in chiral active matter},\ }\href {https://doi.org/10.1360/nso/20230086} {\bibfield  {journal} {\bibinfo  {journal} {National Science Open}\ }\textbf {\bibinfo {volume} {3}},\ \bibinfo {pages} {20230086} (\bibinfo {year} {2024}{\natexlab{b}})}\BibitemShut {NoStop}%
\bibitem [{\citenamefont {Yllanes}\ \emph {et~al.}(2017)\citenamefont {Yllanes}, \citenamefont {Leoni},\ and\ \citenamefont {Marchetti}}]{yllanesHowManyDissenters2017}%
  \BibitemOpen
  \bibfield  {author} {\bibinfo {author} {\bibfnamefont {D.}~\bibnamefont {Yllanes}}, \bibinfo {author} {\bibfnamefont {M.}~\bibnamefont {Leoni}},\ and\ \bibinfo {author} {\bibfnamefont {M.~C.}\ \bibnamefont {Marchetti}},\ }\bibfield  {title} {\bibinfo {title} {How many dissenters does it take to disorder a flock?},\ }\href {https://doi.org/10.1088/1367-2630/aa8ed7} {\bibfield  {journal} {\bibinfo  {journal} {New Journal of Physics}\ }\textbf {\bibinfo {volume} {19}},\ \bibinfo {pages} {103026} (\bibinfo {year} {2017})}\BibitemShut {NoStop}%
\bibitem [{\citenamefont {Shankar}\ \emph {et~al.}(2022)\citenamefont {Shankar}, \citenamefont {Souslov}, \citenamefont {Bowick}, \citenamefont {Marchetti},\ and\ \citenamefont {Vitelli}}]{shankar_topological_2022}%
  \BibitemOpen
  \bibfield  {author} {\bibinfo {author} {\bibfnamefont {S.}~\bibnamefont {Shankar}}, \bibinfo {author} {\bibfnamefont {A.}~\bibnamefont {Souslov}}, \bibinfo {author} {\bibfnamefont {M.~J.}\ \bibnamefont {Bowick}}, \bibinfo {author} {\bibfnamefont {M.~C.}\ \bibnamefont {Marchetti}},\ and\ \bibinfo {author} {\bibfnamefont {V.}~\bibnamefont {Vitelli}},\ }\bibfield  {title} {\bibinfo {title} {Topological active matter},\ }\href {https://doi.org/10.1038/s42254-022-00445-3} {\bibfield  {journal} {\bibinfo  {journal} {Nature Reviews Physics}\ }\textbf {\bibinfo {volume} {4}},\ \bibinfo {pages} {380} (\bibinfo {year} {2022})}\BibitemShut {NoStop}%
\bibitem [{\citenamefont {Giomi}(2015)}]{giomi_geometry_2015}%
  \BibitemOpen
  \bibfield  {author} {\bibinfo {author} {\bibfnamefont {L.}~\bibnamefont {Giomi}},\ }\bibfield  {title} {\bibinfo {title} {Geometry and {{Topology}} of {{Turbulence}} in {{Active Nematics}}},\ }\href {https://doi.org/10.1103/PhysRevX.5.031003} {\bibfield  {journal} {\bibinfo  {journal} {Physical Review X}\ }\textbf {\bibinfo {volume} {5}},\ \bibinfo {pages} {031003} (\bibinfo {year} {2015})}\BibitemShut {NoStop}%
\bibitem [{\citenamefont {Doostmohammadi}\ \emph {et~al.}(2016)\citenamefont {Doostmohammadi}, \citenamefont {Adamer}, \citenamefont {Thampi},\ and\ \citenamefont {Yeomans}}]{doostmohammadi_stabilization_2016}%
  \BibitemOpen
  \bibfield  {author} {\bibinfo {author} {\bibfnamefont {A.}~\bibnamefont {Doostmohammadi}}, \bibinfo {author} {\bibfnamefont {M.~F.}\ \bibnamefont {Adamer}}, \bibinfo {author} {\bibfnamefont {S.~P.}\ \bibnamefont {Thampi}},\ and\ \bibinfo {author} {\bibfnamefont {J.~M.}\ \bibnamefont {Yeomans}},\ }\bibfield  {title} {\bibinfo {title} {Stabilization of active matter by flow-vortex lattices and defect ordering},\ }\href {https://doi.org/10.1038/ncomms10557} {\bibfield  {journal} {\bibinfo  {journal} {Nature Communications}\ }\textbf {\bibinfo {volume} {7}},\ \bibinfo {pages} {10557} (\bibinfo {year} {2016})}\BibitemShut {NoStop}%
\bibitem [{\citenamefont {Spera}\ \emph {et~al.}(2025)\citenamefont {Spera}, \citenamefont {Yeomans},\ and\ \citenamefont {Thampi}}]{spera_low-pass_2025}%
  \BibitemOpen
  \bibfield  {author} {\bibinfo {author} {\bibfnamefont {G.}~\bibnamefont {Spera}}, \bibinfo {author} {\bibfnamefont {J.~M.}\ \bibnamefont {Yeomans}},\ and\ \bibinfo {author} {\bibfnamefont {S.~P.}\ \bibnamefont {Thampi}},\ }\href {https://doi.org/10.48550/arXiv.2511.22701} {\bibinfo {title} {Low-{{Pass Filtering}} of {{Active Turbulent Flows}} to {{Liquid Substrates}}}} (\bibinfo {year} {2025}),\ \Eprint {https://arxiv.org/abs/2511.22701} {arXiv:2511.22701 [cond-mat]} \BibitemShut {NoStop}%
\bibitem [{\citenamefont {Ariel}\ \emph {et~al.}(2015)\citenamefont {Ariel}, \citenamefont {Rabani}, \citenamefont {Benisty}, \citenamefont {Partridge}, \citenamefont {Harshey},\ and\ \citenamefont {Be'er}}]{ariel_swarming_2015}%
  \BibitemOpen
  \bibfield  {author} {\bibinfo {author} {\bibfnamefont {G.}~\bibnamefont {Ariel}}, \bibinfo {author} {\bibfnamefont {A.}~\bibnamefont {Rabani}}, \bibinfo {author} {\bibfnamefont {S.}~\bibnamefont {Benisty}}, \bibinfo {author} {\bibfnamefont {J.~D.}\ \bibnamefont {Partridge}}, \bibinfo {author} {\bibfnamefont {R.~M.}\ \bibnamefont {Harshey}},\ and\ \bibinfo {author} {\bibfnamefont {A.}~\bibnamefont {Be'er}},\ }\bibfield  {title} {\bibinfo {title} {Swarming bacteria migrate by {{L\'evy Walk}}},\ }\href {https://doi.org/10.1038/ncomms9396} {\bibfield  {journal} {\bibinfo  {journal} {Nature Communications}\ }\textbf {\bibinfo {volume} {6}},\ \bibinfo {pages} {8396} (\bibinfo {year} {2015})}\BibitemShut {NoStop}%
\bibitem [{\citenamefont {Vicsek}\ \emph {et~al.}(1995)\citenamefont {Vicsek}, \citenamefont {Czir{\'o}k}, \citenamefont {{Ben-Jacob}}, \citenamefont {Cohen},\ and\ \citenamefont {Shochet}}]{vicsek_novel_1995}%
  \BibitemOpen
  \bibfield  {author} {\bibinfo {author} {\bibfnamefont {T.}~\bibnamefont {Vicsek}}, \bibinfo {author} {\bibfnamefont {A.}~\bibnamefont {Czir{\'o}k}}, \bibinfo {author} {\bibfnamefont {E.}~\bibnamefont {{Ben-Jacob}}}, \bibinfo {author} {\bibfnamefont {I.}~\bibnamefont {Cohen}},\ and\ \bibinfo {author} {\bibfnamefont {O.}~\bibnamefont {Shochet}},\ }\bibfield  {title} {\bibinfo {title} {Novel {{Type}} of {{Phase Transition}} in a {{System}} of {{Self-Driven Particles}}},\ }\href {https://doi.org/10.1103/PhysRevLett.75.1226} {\bibfield  {journal} {\bibinfo  {journal} {Physical Review Letters}\ }\textbf {\bibinfo {volume} {75}},\ \bibinfo {pages} {1226} (\bibinfo {year} {1995})}\BibitemShut {NoStop}%
\bibitem [{\citenamefont {Marchetti}\ \emph {et~al.}(2013)\citenamefont {Marchetti}, \citenamefont {Joanny}, \citenamefont {Ramaswamy}, \citenamefont {Liverpool}, \citenamefont {Prost}, \citenamefont {Rao},\ and\ \citenamefont {Simha}}]{marchetti_hydrodynamics_2013}%
  \BibitemOpen
  \bibfield  {author} {\bibinfo {author} {\bibfnamefont {M.~C.}\ \bibnamefont {Marchetti}}, \bibinfo {author} {\bibfnamefont {J.~F.}\ \bibnamefont {Joanny}}, \bibinfo {author} {\bibfnamefont {S.}~\bibnamefont {Ramaswamy}}, \bibinfo {author} {\bibfnamefont {T.~B.}\ \bibnamefont {Liverpool}}, \bibinfo {author} {\bibfnamefont {J.}~\bibnamefont {Prost}}, \bibinfo {author} {\bibfnamefont {M.}~\bibnamefont {Rao}},\ and\ \bibinfo {author} {\bibfnamefont {R.~A.}\ \bibnamefont {Simha}},\ }\bibfield  {title} {\bibinfo {title} {Hydrodynamics of soft active matter},\ }\href {https://doi.org/10.1103/RevModPhys.85.1143} {\bibfield  {journal} {\bibinfo  {journal} {Reviews of Modern Physics}\ }\textbf {\bibinfo {volume} {85}},\ \bibinfo {pages} {1143} (\bibinfo {year} {2013})}\BibitemShut {NoStop}%
\bibitem [{\citenamefont {Poderoso}\ \emph {et~al.}(2011)\citenamefont {Poderoso}, \citenamefont {Arenzon},\ and\ \citenamefont {Levin}}]{poderoso_new_2011}%
  \BibitemOpen
  \bibfield  {author} {\bibinfo {author} {\bibfnamefont {F.~C.}\ \bibnamefont {Poderoso}}, \bibinfo {author} {\bibfnamefont {J.~J.}\ \bibnamefont {Arenzon}},\ and\ \bibinfo {author} {\bibfnamefont {Y.}~\bibnamefont {Levin}},\ }\bibfield  {title} {\bibinfo {title} {New {{Ordered Phases}} in a {{Class}} of {{Generalized}} \${{XY}}\$ {{Models}}},\ }\href {https://doi.org/10.1103/PhysRevLett.106.067202} {\bibfield  {journal} {\bibinfo  {journal} {Physical Review Letters}\ }\textbf {\bibinfo {volume} {106}},\ \bibinfo {pages} {067202} (\bibinfo {year} {2011})}\BibitemShut {NoStop}%
\bibitem [{\citenamefont {O'Keeffe}\ \emph {et~al.}(2017)\citenamefont {O'Keeffe}, \citenamefont {Hong},\ and\ \citenamefont {Strogatz}}]{okeeffe_oscillators_2017}%
  \BibitemOpen
  \bibfield  {author} {\bibinfo {author} {\bibfnamefont {K.~P.}\ \bibnamefont {O'Keeffe}}, \bibinfo {author} {\bibfnamefont {H.}~\bibnamefont {Hong}},\ and\ \bibinfo {author} {\bibfnamefont {S.~H.}\ \bibnamefont {Strogatz}},\ }\bibfield  {title} {\bibinfo {title} {Oscillators that sync and swarm},\ }\href {https://doi.org/10.1038/s41467-017-01190-3} {\bibfield  {journal} {\bibinfo  {journal} {Nature Communications}\ }\textbf {\bibinfo {volume} {8}},\ \bibinfo {pages} {1504} (\bibinfo {year} {2017})}\BibitemShut {NoStop}%
\bibitem [{\citenamefont {Hong}(2018)}]{hong_active_2018}%
  \BibitemOpen
  \bibfield  {author} {\bibinfo {author} {\bibfnamefont {H.}~\bibnamefont {Hong}},\ }\bibfield  {title} {\bibinfo {title} {Active phase wave in the system of swarmalators with attractive phase coupling},\ }\href {https://doi.org/10.1063/1.5039564} {\bibfield  {journal} {\bibinfo  {journal} {Chaos: An Interdisciplinary Journal of Nonlinear Science}\ }\textbf {\bibinfo {volume} {28}},\ \bibinfo {pages} {103112} (\bibinfo {year} {2018})}\BibitemShut {NoStop}%
\bibitem [{\citenamefont {Boccelli}\ \emph {et~al.}(2025)\citenamefont {Boccelli}, \citenamefont {Martal{\`o}},\ and\ \citenamefont {Travaglini}}]{boccelli_turing_2025}%
  \BibitemOpen
  \bibfield  {author} {\bibinfo {author} {\bibfnamefont {S.}~\bibnamefont {Boccelli}}, \bibinfo {author} {\bibfnamefont {G.}~\bibnamefont {Martal{\`o}}},\ and\ \bibinfo {author} {\bibfnamefont {R.}~\bibnamefont {Travaglini}},\ }\href {https://doi.org/10.48550/arXiv.2509.20268} {\bibinfo {title} {Turing instability and 2-{{D}} pattern formation in reaction-diffusion systems derived from kinetic theory}} (\bibinfo {year} {2025}),\ \Eprint {https://arxiv.org/abs/2509.20268} {arXiv:2509.20268 [math-ph]} \BibitemShut {NoStop}%
\bibitem [{\citenamefont {Hamming}\ and\ \citenamefont {Hamming}(1986)}]{hamming_numerical_1986}%
  \BibitemOpen
  \bibfield  {author} {\bibinfo {author} {\bibfnamefont {R.~W.}\ \bibnamefont {Hamming}}\ and\ \bibinfo {author} {\bibfnamefont {R.~W.}\ \bibnamefont {Hamming}},\ }\href@noop {} {\emph {\bibinfo {title} {Numerical {{Methods}} for {{Scientists}} and {{Engineers}}}}}\ (\bibinfo  {publisher} {Courier Corporation},\ \bibinfo {year} {1986})\BibitemShut {NoStop}%
\bibitem [{\citenamefont {Pope}(2000)}]{pope_turbulent_2000}%
  \BibitemOpen
  \bibfield  {author} {\bibinfo {author} {\bibfnamefont {S.~B.}\ \bibnamefont {Pope}},\ }\href@noop {} {\emph {\bibinfo {title} {Turbulent {{Flows}}}}}\ (\bibinfo  {publisher} {Cambridge: Cambridge University Press},\ \bibinfo {year} {2000})\BibitemShut {NoStop}%
\bibitem [{\citenamefont {Takatori}\ \emph {et~al.}(2014)\citenamefont {Takatori}, \citenamefont {Yan},\ and\ \citenamefont {Brady}}]{takatori_swim_2014}%
  \BibitemOpen
  \bibfield  {author} {\bibinfo {author} {\bibfnamefont {S.~C.}\ \bibnamefont {Takatori}}, \bibinfo {author} {\bibfnamefont {W.}~\bibnamefont {Yan}},\ and\ \bibinfo {author} {\bibfnamefont {J.~F.}\ \bibnamefont {Brady}},\ }\bibfield  {title} {\bibinfo {title} {Swim {{Pressure}}: {{Stress Generation}} in {{Active Matter}}},\ }\href {https://doi.org/10.1103/PhysRevLett.113.028103} {\bibfield  {journal} {\bibinfo  {journal} {Physical Review Letters}\ }\textbf {\bibinfo {volume} {113}},\ \bibinfo {pages} {028103} (\bibinfo {year} {2014})}\BibitemShut {NoStop}%
\bibitem [{\citenamefont {Tan}(1992)}]{tan_shallow_1992}%
  \BibitemOpen
  \bibfield  {author} {\bibinfo {author} {\bibfnamefont {W.~Y.}\ \bibnamefont {Tan}},\ }\href@noop {} {\emph {\bibinfo {title} {Shallow {{Water Hydrodynamics}}: {{Mathematical Theory}} and {{Numerical Solution}} for a {{Two-dimensional System}} of {{Shallow-water Equations}}}}}\ (\bibinfo  {publisher} {Elsevier},\ \bibinfo {year} {1992})\BibitemShut {NoStop}%
\bibitem [{\citenamefont {Salmon}(1998)}]{salmon_lectures_1998}%
  \BibitemOpen
  \bibfield  {author} {\bibinfo {author} {\bibfnamefont {R.}~\bibnamefont {Salmon}},\ }\href@noop {} {\emph {\bibinfo {title} {Lectures on Geophysical Fluid Dynamics}}}\ (\bibinfo  {publisher} {Oxford University Press},\ \bibinfo {year} {1998})\BibitemShut {NoStop}%
\bibitem [{\citenamefont {Landau}\ and\ \citenamefont {Lifshitz}(2013)}]{landauCourseTheoreticalPhysics2013}%
  \BibitemOpen
  \bibfield  {author} {\bibinfo {author} {\bibfnamefont {L.~D.}\ \bibnamefont {Landau}}\ and\ \bibinfo {author} {\bibfnamefont {E.~M.}\ \bibnamefont {Lifshitz}},\ }\bibfield  {title} {\bibinfo {title} {Chapter {{IX}} "{{Shock Waves}}"},\ }in\ \href@noop {} {\emph {\bibinfo {booktitle} {Course of Theoretical Physics}}}\ (\bibinfo  {publisher} {Elsevier},\ \bibinfo {year} {2013})\BibitemShut {NoStop}%
\bibitem [{\citenamefont {Treumann}(2009)}]{treumannFundamentalsCollisionlessShocks2009}%
  \BibitemOpen
  \bibfield  {author} {\bibinfo {author} {\bibfnamefont {R.~A.}\ \bibnamefont {Treumann}},\ }\bibfield  {title} {\bibinfo {title} {Fundamentals of collisionless shocks for astrophysical application, 1. {{Non-relativistic}} shocks},\ }\href {https://doi.org/10.1007/s00159-009-0024-2} {\bibfield  {journal} {\bibinfo  {journal} {The Astronomy and Astrophysics Review}\ }\textbf {\bibinfo {volume} {17}},\ \bibinfo {pages} {409} (\bibinfo {year} {2009})}\BibitemShut {NoStop}%
\end{thebibliography}


\end{document}